\documentclass[a4paper,fleqn,usenatbib]{mnras}

\usepackage{newtxtext,newtxmath}

\usepackage[T1]{fontenc}
\usepackage{ae,aecompl}

\usepackage{graphicx}	
\usepackage{amsmath}	
\usepackage{amssymb}	

\newcommand{\hst}{{\it HST\/}}       
\newcommand{\herschel}{{\it Herschel\/}}       
\newcommand{\chandra}{{\it Chandra\/}}  
  
\newcommand{\rosat}{{\it ROSAT\/}}

\newcommand{\spitzer}{{\it Spitzer\/}}
\newcommand{\galex}{{\it GALEX\/}}
\newcommand{\swift}{{\it Swift\/}}

\newcommand{\xray}{\hbox{X-ray}}  
\newcommand{\cdfs}{\hbox{CDF-S}}
\newcommand{\cdfn}{\hbox{CDF-N}}
\newcommand{\goodss}{\hbox{GOODS-S}}
\newcommand{\goodsn}{\hbox{GOODS-N}}

\newcommand{\lx}{L_{\rm X}}
\newcommand{\nh}{N_{\rm H}}
\newcommand{\slx}{L_{\rm SX}}
\newcommand{\kbol}{k_{\rm bol}}
\newcommand{\LsunMsun}{L_\odot\ M_\odot^{-1}}
\newcommand{\lbol}{L_{\rm bol}}
\newcommand{\lam}{\lambda_{\rm Edd}}
\newcommand{\mbh}{M_{\rm BH}}

\newcommand{\lxbar}{\overline{L_{\rm X}}}
\newcommand{\bharbar}{\overline{\rm BHAR}}
\newcommand{\sfrbar}{\overline{\rm SFR}}
\newcommand{\mstar}{M_{\star}}
\newcommand{\mbulge}{M_{\rm bulge}}
\newcommand{\mhalo}{M_{\rm halo}}

\newcommand{\ks}{K_{\rm S}}

\title[$\bharbar$ dependence on $\mstar$ and $z$]{Linking 
black-hole growth with host galaxies: 
The accretion-stellar mass relation and its cosmic evolution}

\author[G. Yang et al.]{
G. Yang,$^{1,2}$\thanks{E-mail: gxy909@psu.edu (GY)}
W. N. Brandt,$^{1,2,3}$
F. Vito,$^{1,2}$
C.-T. J. Chen,$^{1,2}$
J. R. Trump,$^4$\newauthor
B. Luo,$^5$
M.~Y. Sun,$^{6,7}$
Y.~Q. Xue,$^{6,7}$
A.~M. Koekemoer,$^{8}$
D.~P. Schneider,$^{1,2}$\newauthor
C. Vignali,$^{9}$
and J.-X. Wang$^{6,7}$
\\
$^{1}$Department of Astronomy and Astrophysics, 525 Davey Lab, The Pennsylvania State University, University Park, PA 16802, USA\\
$^{2}$Institute for Gravitation and the Cosmos, The Pennsylvania State University, University Park, PA 16802, USA\\
$^{3}$Department of Physics, 104 Davey Laboratory, The Pennsylvania State University, University Park, PA 16802, USA\\
$^{4}$Department of Physics, 2152 Hillside Road, U-3046, University of Connecticut, Storrs, CT 06269, USA\\
$^{5}$School of Astronomy \& Space Science, Nanjing University, Nanjing 210093, China\\
$^{6}$CAS Key Laboratory for Research in Galaxies and Cosmology, Department of Astronomy, \\University of Science and Technology of China, Hefei 230026, China\\
$^{7}$School of Astronomy and Space Science, University of Science and Technology of China, Hefei 230026, China\\
$^{8}$Space Telescope Science Institute 3700 San Martin Drive, Baltimore MD 21218, USA\\
$^{9}$Universita di Bologn\'{a}, Via Ranzani 1, Bologna, Italy\\
}

\date{Accepted XXX. Received YYY; in original form ZZZ}

\pubyear{2017}

\begin{document}
\label{firstpage}
\pagerange{\pageref{firstpage}--\pageref{lastpage}}
\maketitle

\begin{abstract}
Previous studies suggest that the growth of supermassive 
black holes (SMBHs) may be fundamentally related to  
host-galaxy stellar mass ($\mstar$). 
To investigate this SMBH growth-$\mstar$ relation in detail, 
we calculate long-term SMBH accretion rate 
as a function of $\mstar$ and redshift [$\bharbar(\mstar, z)$]
over ranges of $\log(\mstar/M_\odot) = \text{9.5--12}$\ 
and $z = \text{0.4--4}$. 
Our $\bharbar(\mstar, z)$ is constrained by high-quality 
survey data (\hbox{GOODS-South}, \hbox{GOODS-North}, and 
COSMOS), and by the stellar mass function and the 
\xray\ luminosity function.
At a given $\mstar$, $\bharbar$ is higher at high redshift.
This redshift dependence is stronger in more massive systems
(for $\log(\mstar/M_\odot)\approx 11.5$, $\bharbar$ is 
three decades higher at $z=4$ than at $z=0.5$), 
possibly due to AGN feedback. 
Our results indicate that the ratio between $\bharbar$ 
and average star formation rate ($\sfrbar$) rises toward high 
$\mstar$ at a given redshift.
This $\bharbar/\sfrbar$ dependence on $\mstar$  
does not support the scenario that SMBH and galaxy growth 
are in lockstep. 
We calculate SMBH mass history [$\mbh(z)$] based on 
our $\bharbar(\mstar, z)$ and the $\mstar(z)$ from 
the literature, and find that the $\mbh$-$\mstar$ 
relation has weak redshift evolution since $z\approx 2$.
The $\mbh/\mstar$ ratio is higher toward massive galaxies:
it rises from $\approx 1/5000$ at $\log\mstar\lesssim 10.5$ 
to $\approx 1/500$ at $\log\mstar \gtrsim 11.2$.
Our predicted $\mbh/\mstar$ ratio at high $\mstar$ is similar 
to that observed in local giant ellipticals, suggesting that 
SMBH growth from mergers is unlikely to dominate over growth 
from accretion.
\end{abstract}




\section{Introduction}\label{sec:intro}
The connection between supermassive black holes (SMBHs)
and host galaxies is a central topic in extragalactic 
studies. 
Observations of the local universe reveal that 
black-hole mass ($\mbh$) is tightly related to 
host-galaxy bulge mass ($\mbulge$) and bulge 
velocity dispersion ($\sigma$) \citep[e.g.,][]{ferrarese00, 
gebhardt00, haring04, kormendy13}.
This $\mbh\text{--}\mbulge$ relation suggests that SMBHs 
and host galaxies evolve in a coordinated way over cosmic 
history, i.e., the so-called ``SMBH-galaxy coevolution'' 
scenario. 

The black-hole accretion rate (BHAR) can be inferred from 
\xray\ emission, thanks to the near-universality
and penetrating nature of \hbox{X-rays} generated by the 
SMBH accretion process (e.g., \citealt{brandt15, xue17}).
However, since orders-of-magnitude variability is likely 
common among active galactic nuclei (AGNs) on timescales 
of $\approx 10^2-10^7$~yr (e.g., \citealt{novak11}; 
\mbox{\citealt{hickox14}} and references therein), 
the instantaneous \xray\ luminosity ($\lx$) cannot be used 
to study the long-term behavior of black-hole accretion. 
The sample-mean BHAR of a galaxy population is often 
adopted as a proxy for the long-term average BHAR 
($\bharbar$, i.e., $\overline{{dM}_{\rm BH}/dt}$). 

Much work has focused on relations between $\bharbar$ 
(i.e., the activity of AGNs) 
and host-galaxy properties such as star formation rate 
(SFR) and stellar mass ($\mstar$) in the distant universe.
Some studies have found a positive correlation between 
sample-mean BHAR and SFR, and interpret this result as a 
reflection of an intrinsic $\bharbar$-SFR relation for 
individual galaxies 
\citep[e.g.,][]{mullaney12, chen13, hickox14, delvecchio15}.
In addition, observations have demonstrated that the 
AGN fraction above a given $\lx$\ threshold rises steeply 
toward high $\mstar$, suggesting a connection between 
$\bharbar$\ and $\mstar$\ \citep[e.g.,][]{xue10, lusso11, 
wang17}. 

The apparent $\bharbar$-SFR and $\bharbar$-$\mstar$\ 
relations might not both be fundamental, 
because SFR and $\mstar$\ 
are positively correlated through the ``star-formation main 
sequence'' for star-forming galaxies. 
To address this SFR-$\mstar$\ degeneracy, \citet{yang17}
studied the sample-mean BHAR dependence on both SFR and $\mstar$,
and concluded via partial-correlation analyses that 
$\bharbar$\ is primarily related to $\mstar$\ rather 
than SFR in general.
This result is consistent with some observations suggesting
AGN host galaxies have similar SFR as normal galaxies at a given
$\mstar$ \citep[e.g.,][]{rosario13b, xu15, stanley17, suh17}.
Therefore, the $\bharbar$-$\mstar$\ relation might be a key to 
addressing questions about SMBH-galaxy coevolution.

\citet{yang17} fitted the $\bharbar$-$\mstar$\ relation with a 
log-linear model and found that the slope is consistent with
unity (see their Fig.~5). 
They did not find significantly different 
$\bharbar$-$\mstar$\ relations for the two redshift 
ranges ($0.5 \leq z < 1.3$\ and $1.3 \leq z < 2.0$) 
used in their analyses.
However, the constraints on cosmic evolution were not 
tight due to the limited sample sizes and large
redshift bin widths.
Strong cosmic evolution could plausibly exist, 
since the cosmic BHAR density ($\rho_{\rm BHAR}$) 
significantly depends on redshift 
(e.g., \citealt{ueda14, aird15}).
\citet{yang17} only probed the $\bharbar$-$\mstar$\ 
relation for galaxies with low and intermediate $\mstar$\ 
[$\log(\mstar/M_\odot) \lesssim 10.5$]. 
However, massive galaxies may have a very different 
BHAR-$\mstar$ correlation, due to stronger AGN feedback 
governing SMBH growth 
\citep[e.g.,][]{mc_namara07, anderson15}.

In this paper, we investigate the $\bharbar$-$\mstar$\ relation
and its cosmic evolution [$\bharbar(\mstar,z)$] over 
ranges of $\log(\mstar/M_\odot) = \text{9.5--12}$
and $z = \text{0.4--4}$. 
We do not discuss the $\bharbar$-$\mbh$ relation.
The $\bharbar$-$\mbh$ relation might be more physically relevant 
than the $\bharbar$-$\mstar$ relation, as high-$\mbh$ SMBHs 
should be more capable of accreting surrounding material. 
It is even possible that the $\bharbar$-$\mstar$ relation 
ultimately originates from a $\bharbar$-$\mbh$ relation
(or other similar relations), since massive galaxies tend 
to host high-$\mbh$ SMBHs. 
However, unlike $\mstar$, $\mbh$ is not a property of host 
galaxies, and thus the $\bharbar$-$\mbh$ relation does 
not directly provide clues about SMBH-galaxy coevolution. 
Also, $\mbh$ measurements are often limited to rare luminous 
broad-line (BL) quasars and subject to large uncertainties 
\citep[e.g.,][]{shen13}.
On the other hand, $\mstar$ is an observable for most 
systems and its measurement techniques are 
mature \citep[e.g..,][]{santini15, davidzon17}.

Our $\bharbar(\mstar,z)$ is derived from the 
probability distribution of specific \xray\ luminosity 
[$P(\slx|\mstar,z)$, where $\slx=\lx/\mstar$].
The definition of $\slx$\ allows it to serve as a rough proxy 
for the Eddington ratio, i.e., $\lam=\lbol/L_{\rm Edd}$, 
where $L_{\rm Edd}=1.26\times 10^{38} 
(\mbh/M_\odot)$~erg~s$^{-1}$ is the Eddington luminosity 
and $\lbol$ is the bolometric luminosity, 
as $\lbol$ and $\mbh$\ are 
broadly correlated with $\lx$\ and $\mstar$, respectively 
\citep[e.g.,][]{lusso12, kormendy13}.
For our purpose, choosing the mathematical format of $\slx$  
is only to follow the convention of previous work 
\citep[e.g.,][]{aird12, bongiorno12}, and the choice of 
whether to derive $P(\slx|\mstar,z)$ or $P(\lx|\mstar,z)$ 
does not affect the final $\bharbar(\mstar,z)$.
Previous work on $P(\lx|\mstar,z)$ 
\citep[e.g.,][]{aird17, georgakakis17} focuses on 
interpreting the physical cause of $P(\lx|\mstar,z)$ 
itself (e.g., its shape and normalization).
Our work utilizes $P(\lx|\mstar,z)$ as a method to derive 
$\bharbar(\mstar,z)$, and study SMBH-galaxy coevolution 
based on $\bharbar(\mstar,z)$.
Therefore, our scientific goals are different from 
those works. 
 

At low and moderate $\slx$ ($\lesssim 1\ L_\odot\ M_\odot^{-1}$),
$P(\slx|\mstar,z)$ has a power-law shape with a slope of 
$\approx 0.4-0.6$\
(e.g., \citealt{aird12, bongiorno12, jones16, wang17}).
Some recent work identified a sharp drop toward high $\slx$
($\gtrsim 1\ L_\odot\ M_\odot^{-1}$; e.g., 
\citealt{bongiorno16, aird17, georgakakis17}).
The break at high $\slx$ is physically expected, because
otherwise the average \xray\ luminosity ($\lxbar$), 
which is the integral of 
$P(\slx|\mstar,z)\times \mstar \times \slx$, would diverge. 

In this work, we derive $P(\slx|\mstar,z)$\ utilizing the 
data from CANDELS survey \citep{grogin11, koekemoer11}, 
in particular the GOODS-South and GOODS-North fields 
(hereafter \goodss\ and \goodsn), as well as the 
data from COSMOS survey \citep{scoville07}, 
in particular the UltraVISTA field \citep{mccracken12}.
All these fields have superb multiwavelength coverage. 
The UV-to-mid-IR (MIR) data allow accurate photometric redshift 
and $\mstar$\ measurements \citep[e.g.,][]{santini15, laigle16}.
All fields have \chandra\ \xray\ observations, allowing us 
to derive $\lx$\ for the AGNs. 
Thanks to the excellent positional accuracy of \chandra\ 
($\approx 0.5\arcsec$), the matching between \xray\  
sources and optical/near-IR (NIR) sources is highly 
reliable, with typical false matching rate less than
a few percent \citep[e.g.,][]{xue16,luo17}. 
\goodss\ and \goodsn\ are small fields 
(\hbox{$\approx 170$~arcmin$^2$}) but are covered by the deepest 
\xray\ surveys, 
the 7~Ms \chandra\ Deep Field-South and 
2~Ms \chandra\ Deep Field-North 
\citep[\cdfs\ and \cdfn;][]{alexander03, xue16, luo17}.
These two fields provide constraints on 
$P(\slx|\mstar,z)$\ at low $\slx$.
COSMOS, covered by the COSMOS-Legacy survey, is a much larger 
field ($\approx 1.4$~deg$^2$ for the utilized UltraVISTA region) 
than \goodss\ and \goodsn, although the \xray\ data are much
shallower ($\approx 160$~ks; see \citealt{civano16}).
COSMOS generally constrains $P(\slx|\mstar,z)$\ at 
higher $\slx$ than \goodss\ and \goodsn.
Following \citet{bongiorno16}, we also constrain 
$P(\slx|\mstar,z)$\ utilizing the \xray\ luminosity function (XLF)
and stellar mass function (SMF) from the literature.
This technique can constrain $P(\slx|\mstar,z)$ for 
rare high-$\lx$ AGNs which are included in the XLF.

We do not calculate $\bharbar(\mstar,z)$ via 
averaging the instantaneous BHAR from samples of galaxies 
in surveys \citep[e.g.,][]{yang17}.
Survey data often have small-area effects; e.g., accretion 
power from rare luminous AGNs is not included due to  
the limited size of the survey area, and this effect is difficult
to determine when performing sample averaging of BHAR. 
Also, it is challenging to combine different surveys 
when directly averaging the instantaneous BHAR, because 
different \xray\ data have large differences in 
sensitivity. 
However, these issues can be properly 
considered when modeling $P(\slx|\mstar,z)$ 
\citep[e.g.,][]{aird12, bongiorno16}.

This paper is structured as follows. 
In \S\ref{sec:sample}, we describe the survey data, SMF, and XLF.
We detail our analyses and results in \S\ref{sec:data_ana}.
In \S\ref{sec:discuss}, we discuss the implications of our 
measurements, and summarize our work in \S\ref{sec:summary}.

Throughout this paper, we assume a cosmology 
with $H_0=70$~km~s$^{-1}$~Mpc$^{-1}$, $\Omega_M=0.3$, 
and $\Omega_{\Lambda}=0.7$, and a Chabrier
initial mass function \citep[][]{chabrier03}.
Quoted uncertainties are at the $1\sigma$\ (68\%)
confidence level, unless otherwise stated. 
We express $\slx$\ in units of $\LsunMsun$, 
and $\mstar$ ($\mbh$) in units of $M_\odot$,
unless otherwise stated.
$\lx$\ indicates AGN intrinsic \xray\ luminosity at 
rest-frame \hbox{2--10 keV} and is in units of 
erg~s$^{-1}$. 

\section{Data and Sample}\label{sec:sample}
\subsection{Multiwavelength Surveys}\label{sec:surv}
In this work, we compile observational data from the 
\goodss, \goodsn, and COSMOS surveys. 

\subsubsection{\goodss\ and \goodsn}\label{sec:goodss}
We include all 34,779 and 34,651 galaxies in the 
\goodss\ and \goodsn\ catalogs, respectively 
\citep{grogin11, koekemoer11, guo13}. 
In the catalogs, 4,314 and 4,660 sources have 
$\log\mstar>9.5$ (our $\log\mstar$ threshold; see 
\S\ref{sec:m_cut}).
The basic information for these two fields is summarized 
in Tab.~\ref{tab:surv}.
These two catalogs are based on $H$-band detections
with $3\sigma$\ limiting magnitude of $H\approx28$.
The solid angle is $\approx 170$~arcmin$^2$ for 
both the \goodss\ and \goodsn\ surveys.
We utilize both fields to minimize the effects of cosmic 
variance. 
 
\begin{figure}
\includegraphics[width=\linewidth]{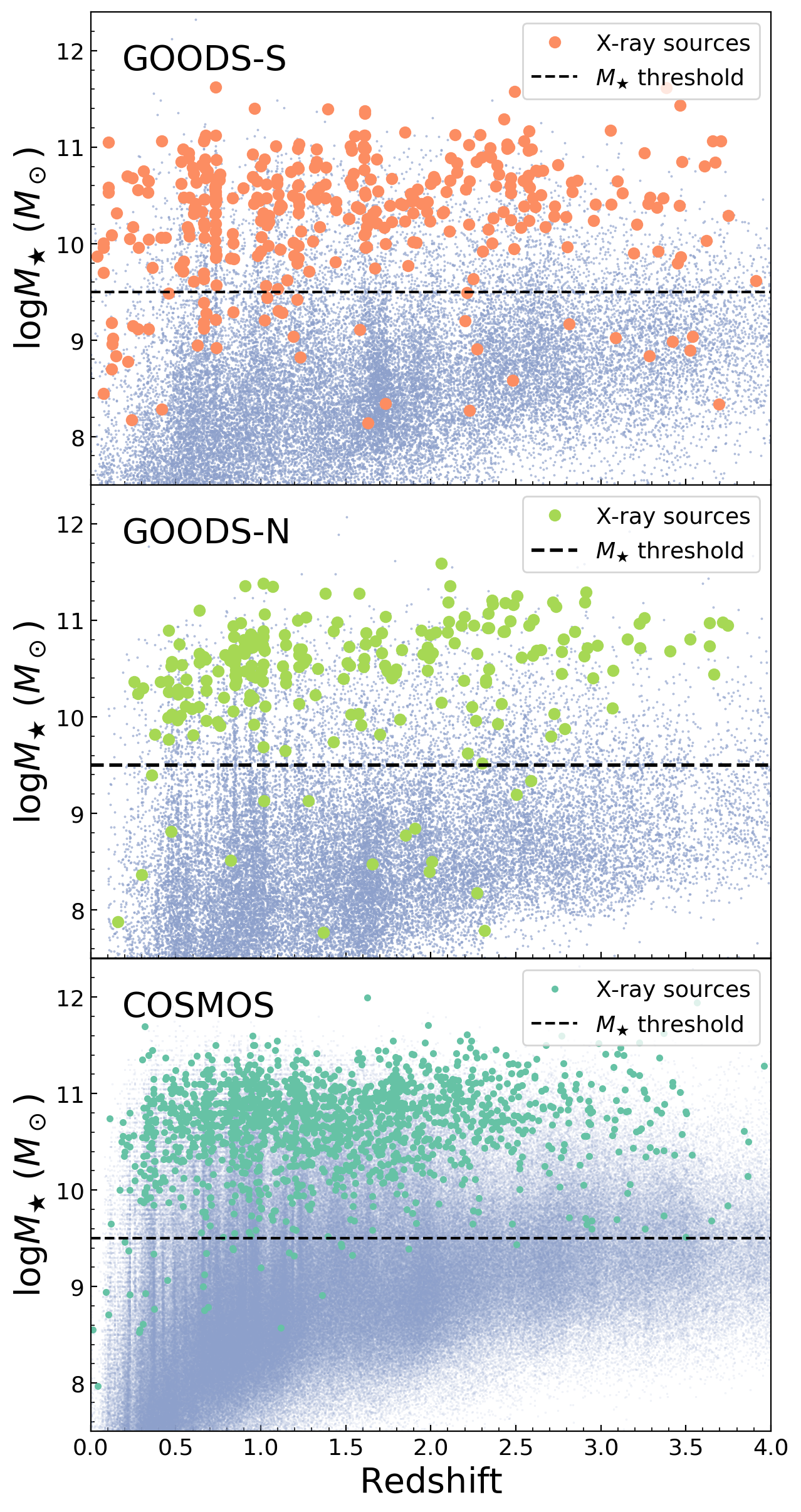}
\caption{$\mstar$\ as a function of redshift
for \goodss\ (top), \goodsn\ (middle), and COSMOS 
(bottom). 
The total numbers of galaxies in these three fields
are 34,779, 34,651, and 520,778, respectively.
The small blue points indicate galaxies; 
the larger points indicate \xray-detected sources.
The dashed horizontal line indicates our 
$\mstar$\ threshold (see \S\ref{sec:m_cut}).
}
\label{fig:M_vs_z}
\end{figure}

We adopt the galaxies' $\mstar$ and redshift from \citet{santini15} 
and Barro et al.\ (in prep.), respectively, for 
\goodss\ and \goodsn\ sources.
The $\mstar$ and redshift measurements in both fields
are based on spectral energy distribution (SED) 
fitting of broad-band photometry 
ranging from the $U$ band to \spitzer/IRAC bands (i.e., 
3.6, 4.5, 5.8, 8.0 $\mu$m).  
In both surveys, the $\mstar$\ estimations were
performed by several independent teams, some of whom 
used galaxy templates with nebular emission. 
The nebular emission is important at high redshift
($z\gtrsim 2$), where some strong emission lines enter
the wavelength ranges of the NIR bands. 
Since this work includes redshifts up to $z=4$, we 
adopt the median $\mstar$\ of teams using templates
with nebular emission; these median $\mstar$\ values 
are available in the catalogs of \citet{santini15}
and Barro et al.\ (in prep.). 
The SED-based $\mstar$ measurements have typical 
uncertainties of $\lesssim 0.3$~dex 
\citep[e.g.,][]{santini15}.
The redshifts are taken from spectroscopic measurements 
when available; otherwise, they are high-quality 
photometric redshifts.
These photometric redshifts in \goodss\ and \goodsn\ 
have median uncertainties
[$|z_{\rm phot}-z_{\rm spec}|/(1+z_{\rm spec})$]
of 0.02 and 0.004, respectively, and outlier fractions 
[$|z_{\rm phot}-z_{\rm spec}|/(1+z_{\rm spec})>0.15$] of 
9\% and 5\%, respectively.
The $\mstar$\ as a function of redshift is shown 
in Fig.~\ref{fig:M_vs_z} (top and middle).

We also obtain the SED-based SFR values  
from \citet{santini15} and Barro et al.\ (in prep.)
to classify galaxies as star-forming or quiescent. 
The SFRs from SED fitting have typical uncertainties of 
$\approx 0.5$~dex \citep[see, e.g., \S2.2 of][]{yang17}. 
Although the errors are relatively large, they 
are sufficient for our purpose 
\citep[star-forming vs.\ quiescent classification; 
e.g.,][]{aird17}.
We adopt a SFR threshold that is 1.3~dex below the 
star-forming main sequence (Eq.~1 in \citealt{aird17}).
If a galaxy has SFR above (below) this threshold, we 
classify it as star-forming (quiescent).
We do not classify galaxies hosting a BL AGN, as 
the total UV flux of these galaxies might have a
significant contribution from the AGN thereby 
biasing the SED-based SFRs.

\begin{table}
\begin{center}
\caption{Summary of Survey Data}
\label{tab:surv}
\begin{tabular}{ccccc}\hline\hline
Name & Area (deg$^2$) & Galaxies & \xray\ Depth & AGNs \\ 
(1) & (2) & (3) & (4) & (5) \\  \hline
\goodss & 0.05 & 4,314 & 7~Ms & 264 \\
\goodsn & 0.05 & 4,660 & 2~Ms & 195  \\
COSMOS  & 1.38 & 141,004 & 160~ks & 1,448 \\
\hline 
\end{tabular}
\end{center}
{\sc Note.} ---
(1) Field name.
(2) Area of the field in units of deg$^2$. 
(3) Number of galaxies above $\log\mstar=9.5$.
(4) \chandra\ effective exposure time. 
(5) Number of AGNs with $\log \slx>-2$.  
\end{table}

We cross-correlate these galaxies with the \xray\
sources in the \cdfs\ \citep{luo17} and \cdfn\ 
\citep{xue16} surveys using the method described in 
\citet{yang17}. 
\cdfs\ and \cdfn\ fully cover \goodss\ and \goodsn,
respectively. 
They are the deepest \xray\ surveys with 7~Ms and 2~Ms 
\chandra\ observations, respectively. 
We only use hard-band (observed-frame \hbox{2--7~keV}) 
detected sources, taking advantage of the fact that 
the hard band is less affected by obscuration than
the soft band (observed-frame \hbox{0.5--2~keV}). 
Soft-band detections are biased to less-obscured 
sources, and it is difficult to account for this bias 
with our methodology (see \S\ref{sec:like_surv}). 
Also, for a source detected in the soft band but not in
the hard band, it is not feasible to obtain their 
absorption-corrected $\lx$ which is necessary in 
our analyses (\S\ref{sec:like_surv}).

\begin{figure}
\begin{center}
\includegraphics[width=\linewidth]{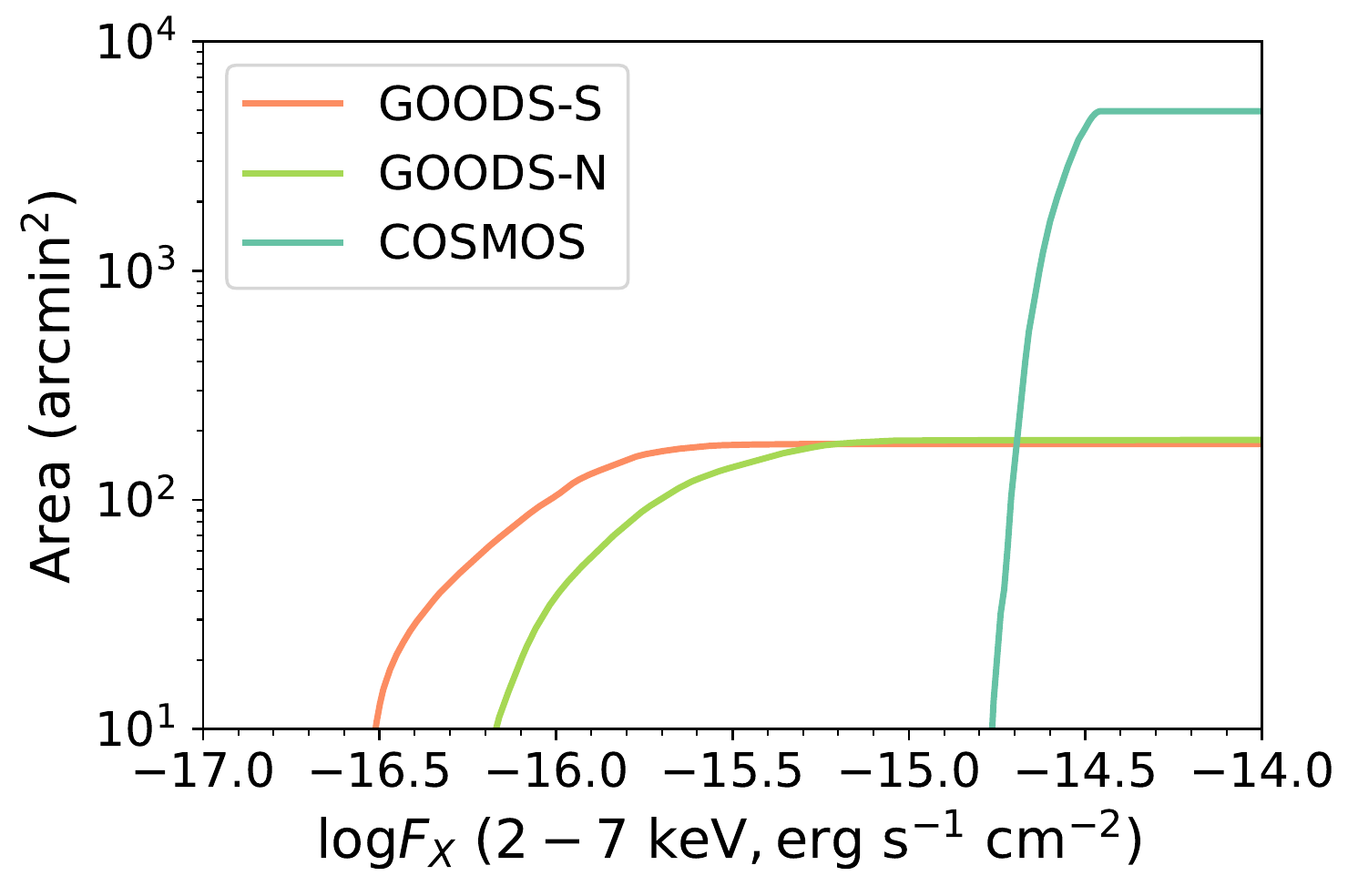}
\caption{\goodss, \goodsn, and COSMOS 
sky coverages as functions of hard-band 
(\hbox{2--7 keV}) \xray\ flux. 
Different colors indicate different surveys.
\goodss\ and \goodsn\ are quite deep but cover 
small areas; COSMOS covers a larger area but is 
shallower than \goodss\ and \goodsn. 
}
\label{fig:skycov}
\end{center}
\end{figure}

\begin{figure}
\includegraphics[width=\linewidth]{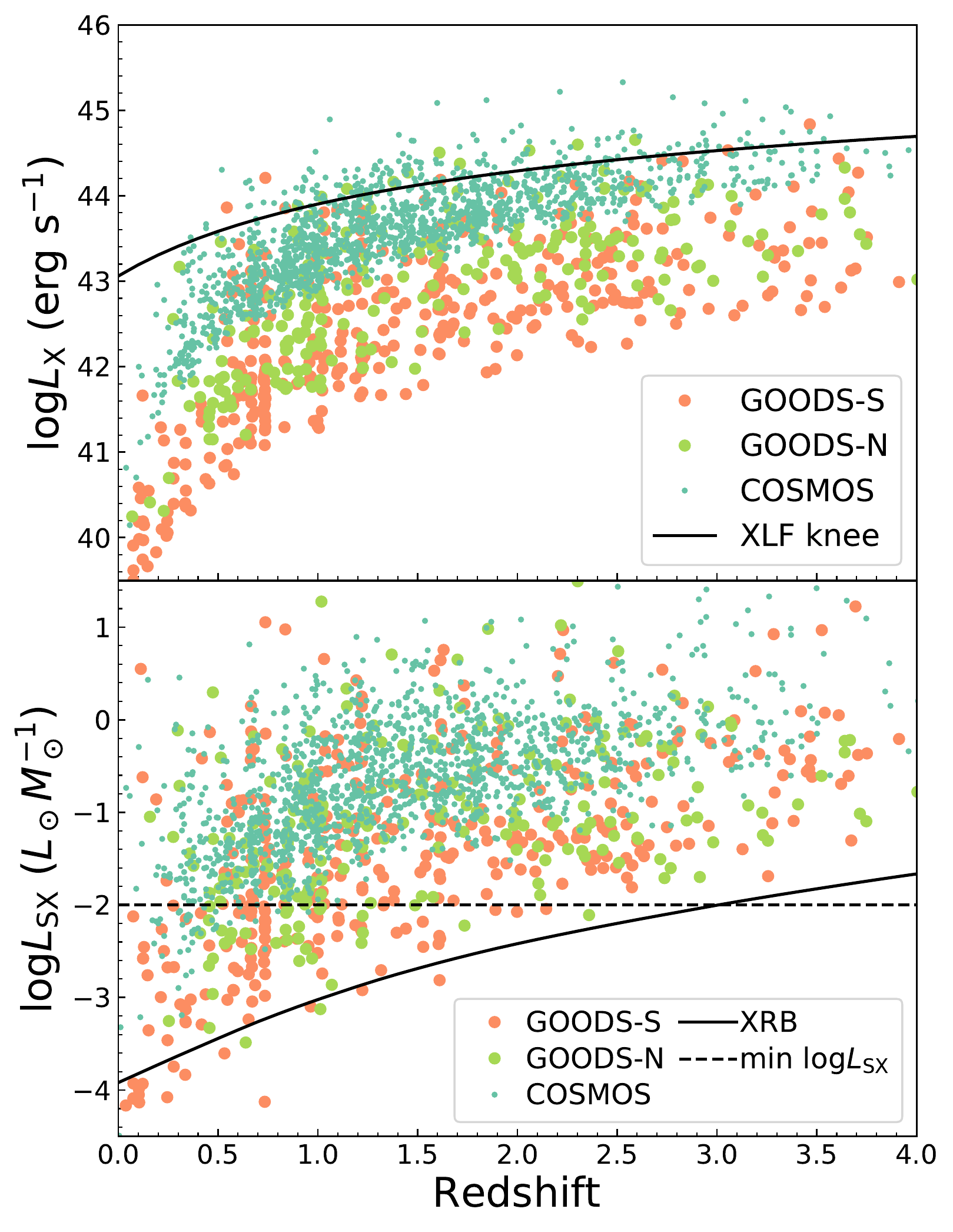}
\caption{Top panel: 
$\lx$\ vs.\ $z$\ for the \xray-detected 
sources in our sample. 
Different colors indicate sources from 
different surveys.
The black solid curve indicates the XLF knee 
luminosity from \citet{aird15}, who derived 
the knee luminosities for obscured and 
unobscured populations, respectively. 
Here, we adopt the values for the obscured AGNs,
which are the majority population.
The \goodss\ and \goodsn\ sources are generally less 
luminous than COSMOS sources at a given redshift.
Bottom panel: $\slx$\ vs.\ $z$. 
The black solid curve represents the $\slx$\ for 
\xray\ binaries of galaxies on the star-forming 
main sequence (see \S\ref{sec:xrb_cont}). 
The black dashed horizontal line indicates our 
$\slx$\ cut, i.e., $10^{-2}\ \LsunMsun$\ 
(see~\ref{sec:Lsx_cut}).
}
\label{fig:Lx_vs_z}
\end{figure}

We obtain hard-band fluxes from \citet{xue16} and 
\citet{luo17} and convert them to $\lx$\ 
assuming a power-law spectral shape with a photon 
index of $\Gamma=1.6$ and that absorption only has 
minor effects on the observed hard-band flux 
\citep[e.g.,][]{liu17, yang17}.
We justify these assumptions in \S\ref{sec:xray_obs}
and discuss the contamination from \xray\ binaries 
(XRBs) in \S\ref{sec:xrb_cont}.
We adopt $\Gamma=1.6$ instead of 
$\Gamma=1.8\text{--}1.9$, mainly because the former 
can produce $\lx$ values that agree better with 
those from \xray\ spectral fitting (see 
\S\ref{sec:xray_obs} for more details).
Although the observed-frame soft band 
corresponds to rest-frame energies above 2~keV at $z>3$, 
we do not use the soft-band data for high-$z$
sources, as obscuration generally becomes stronger 
toward high redshift, and the soft band can still be 
affected by obscuration at high redshift 
\citep[e.g.,][]{vito14a, liu17}.
Fig.~\ref{fig:skycov} displays the sky coverages
as a function of hard-band flux for \goodss\ and \goodsn, 
and Fig.~\ref{fig:Lx_vs_z} 
shows $\lx$ and $\slx$ as functions of redshift for \xray\ 
detected sources in these two fields. 
There are 264 and 195 \xray\ AGNs with $\log \slx > -2$
(our $\slx$ cut; see \S\ref{sec:Lsx_cut})
in \goodss\ and \goodsn, respectively. 

\subsubsection{COSMOS}\label{sec:cosmos}
We adopt the COSMOS2015 galaxy catalog for the COSMOS
field \citep{laigle16}; the basic information
for this field is summarized in Tab.~\ref{tab:surv}.
We only use sources within both the COSMOS and 
UltraVISTA regions, after removing the masked objects 
(1.38~deg$^2$; see Fig.~1 and Tab.~7 in \citealt{laigle16}).
The UltraVISTA region has deep NIR observations that
are critical in SED fitting. 
The COSMOS2015 catalog includes sources detected in 
a $\chi^2$ sum of $zYJHK_{\rm s}$ images 
(see \citealt{szalay99} for this detection technique). 
The $3 \sigma$\ limiting magnitude is $\ks\approx 24$.
We note that this limiting magnitude is only a rough 
estimation of the depth of the $\ks$ imaging data. 
The COSMOS2015 catalog is actually deeper than 
$\ks\approx 24$ due to the multiwavelength-based 
detection method used. 
In fact, $\approx 50\%$ of sources are fainter than 
$\ks\approx 24$.
The multiwavelength photometric data in the COSMOS2015
catalog include 18 broad bands from \galex\ NUV 
to \spitzer/IRAC, 12 medium optical bands, and 
2 narrow optical bands (see Tab.~1 in \citealt{laigle16}).

The refined catalog consists of 520,778 galaxies
with $\mstar$\ and photometric-redshift measurements,
among which 141,004 sources have $\log\mstar>9.5$.
Compared with the spectroscopic redshifts 
available \citep{lilly09},
the photometric redshifts have a median uncertainty 
of 0.006 and an outlier fraction of 3\%, thanks to 
the good coverage of the multiwavelength 
observations.\footnote{\label{foot:cosmos_photoz}\citet{laigle16} 
derived $\mstar$\ using the photometric redshifts. 
Thus, we do not adopt the spectroscopic redshifts 
for these objects even when these are available.
Since their photometric redshifts agree well with the
spectroscopic redshifts, the adoption of the former
instead of the latter should not affect our conclusions
materially.} 
Fig.~\ref{fig:M_vs_z} (bottom) shows $\mstar$\ vs.\ $z$\ 
for these sources. 
We also adopt the SED-based SFR measurements and 
perform star-forming vs.\ quiescent classifications 
as in \S\ref{sec:goodss}.

\citet{marchesi16} have matched these sources with 
the COSMOS-Legacy \xray\ survey, which has 160~ks
\chandra\ effective exposure time \citep{civano16}. 
We utilize hard-band flux to derive $\lx$\ for hard-band 
detected sources as in \S\ref{sec:goodss}.
The sky coverage as a function of hard-band flux is 
shown in Fig.~\ref{fig:skycov}. 
COSMOS covers an area $\approx 30$ times larger than 
\goodss/\goodsn\ but is substantially shallower. 
The $\lx$\ and $\slx$ as functions
of redshift are presented in Fig.~\ref{fig:Lx_vs_z}.
There are 1,448 \xray\ AGNs with $\log \slx > -2$ 
(see \S\ref{sec:Lsx_cut}).

The ultradeep fields of \goodss\ and \goodsn\ 
and the medium-deep field COSMOS are complementary 
in sampling a wide range of $\lx$.
At a given redshift, the most luminous COSMOS sources 
often have $\lx$ a few times higher than the knee 
luminosity;
the faintest GOODS sources have $\lx \approx 2$ 
decades below the knee luminosity. 
This wide range of sampled $\lx$ typically includes 
$\gtrsim 70\%$ of the total $\lx$ from the integral 
of the XLF. 
In this sense, our survey data sample contains most 
of cosmic accretion power.

\subsubsection{Broad-Line AGNs}
\label{sec:fit_blagn}
The photometric redshifts and $\mstar$ in the survey 
catalogs (\S\ref{sec:surv}) are obtained from SED fitting 
with galaxy templates.
This is appropriate for normal galaxies and non-BL
AGNs where rest-frame UV to NIR SEDs are 
dominated by stellar light (see, e.g., \citealt{luo10, brandt15}; 
\S2.2 of \citealt{yang17}). 
However, BL AGNs can be responsible for a large fraction 
of the total UV-to-NIR SED.
When they are present, AGN components should be 
considered properly in SED fitting, and thus 
the redshift and $\mstar$ measurements in the survey 
catalogs may not be reliable. 
There are 359 BL AGNs in our survey data identified 
from spectroscopic observations 
(\goodss: 22 sources from \citealt{silverman10}; 
\goodsn: 15 sources from \citealt{barger03};
COSMOS: 322 sources from \citealt{marchesi16}).
These observations provide reliable spectroscopic 
redshifts for these BL AGNs. 

Additional BL AGNs might be missed due to the lack of 
full spectroscopic coverage. 
Most of the missed BL AGNs should be in COSMOS due 
to its relatively large area.
Now we estimate the number of possibly missed BL AGNs 
for COSMOS. 
For the $\approx 710$ \xray\ AGNs with high-quality 
spectra \citep{marchesi16},
we find that BL AGNs are generally brighter and bluer than
non-BL AGNs, and these two populations are separated in 
the color-magnitude diagram by $B-r=0.8$ and $r=24$ 
(observed AB magnitude; from \citealt{laigle16}).
Applying this criterion ($B-r<0.8$ and $r<24$) to the 
$\approx 740$ \xray\ sources without high-quality spectra,
we estimate that the missed BL AGN number in COSMOS
is $\approx 70$, significantly smaller than the number of 
spectroscopically identified BL AGNs (322).
We caution that there might be additional BL AGNs not 
satisfying the empirical criterion.
However, those sources are faint ($r\geq 24$) and/or 
red ($B-r\geq0.8$), and their AGN SED components are 
less likely to dominate over the host-galaxy components.
Therefore, the missed BL AGNs in our sample should not 
affect our analyses significantly. 


To obtain reliable $\mstar$ for BL AGNs, we perform SED 
decomposition utilizing a minimum-$\chi^2$ method 
implemented in {\sc cigale} \citep{noll09, 
serra11}.\footnote{http://cigale.lam.fr/} 
{\sc cigale} can perform SED fitting for a source 
based on a set of photometric data at a given redshift,
and output physical parameters like $\mstar$ and SFR. 
We use the photometric data from the survey catalogs
(\S\ref{sec:surv}) and redshifts from spectroscopic
observations.
The AGN model in \citet{fritz06} is implemented 
in {\sc cigale}, and we follow the settings 
for the typical BL AGN template in 
Tab.~1 of \citet{ciesla15}.
We allow \textit{frac}$_{\rm AGN}=0\text{--}1$ with a 
step of 0.05, where \textit{frac}$_{\rm AGN}$ is the 
AGN fractional contribution to the total IR luminosity.
For galaxy components, 
we assume a Chabrier initial mass function (IMF; 
\citealt{chabrier03}), which is also adopted to measure 
$\mstar$ for galaxies in the survey data.
We adopt a $\tau$ model of star formation history (SFH)
and allow values for $\log(\tau/\rm yr)$ ranging from 8 
to 10.5 with a step size of 0.5~dex. 
The stellar templates are from the model of 
\citet{bruzual03}. 
We allow several possible values of metallicity 
($Z=0.0001, 0.0004, 0.004, 0.008, 0.02, 0.05$, where
$Z$ is the mass fraction of metals).
We adopt the Calzetti extinction law \citep{calzetti00}
allowing $E(B-V)$ to vary from 0.0\text{--}1.0 with a 
step of 0.1. 
{\sc cigale} also includes nebular and dust emission 
\citep{noll09, draine07}.
The above settings for galaxy SED is similar to those
used to derive photometric redshifts and $\mstar$ 
in the survey catalogs (\citealt{santini15, laigle16};
Barro et al.\ in prep.).

\begin{figure}
\includegraphics[width=\linewidth]{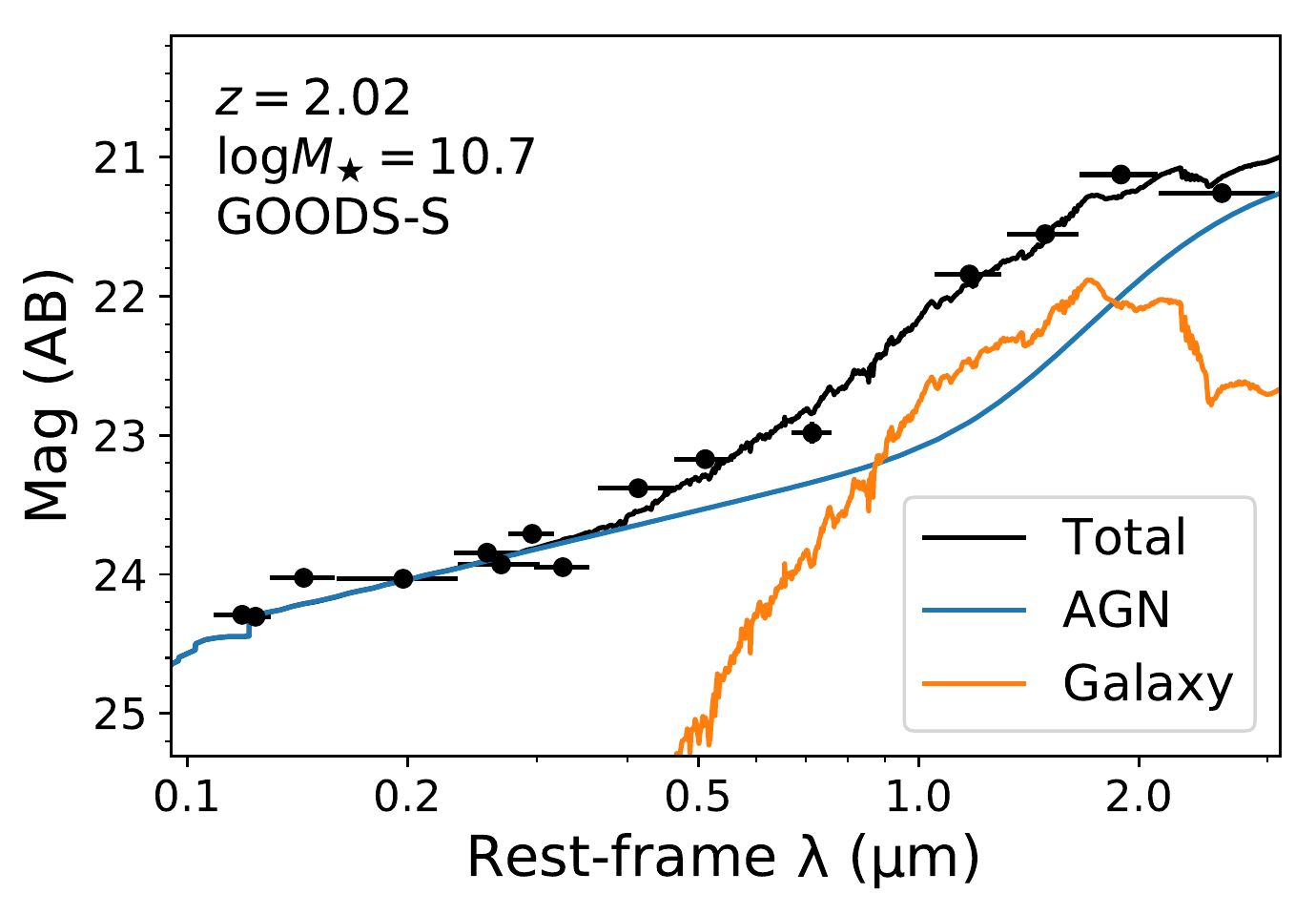}
\includegraphics[width=\linewidth]{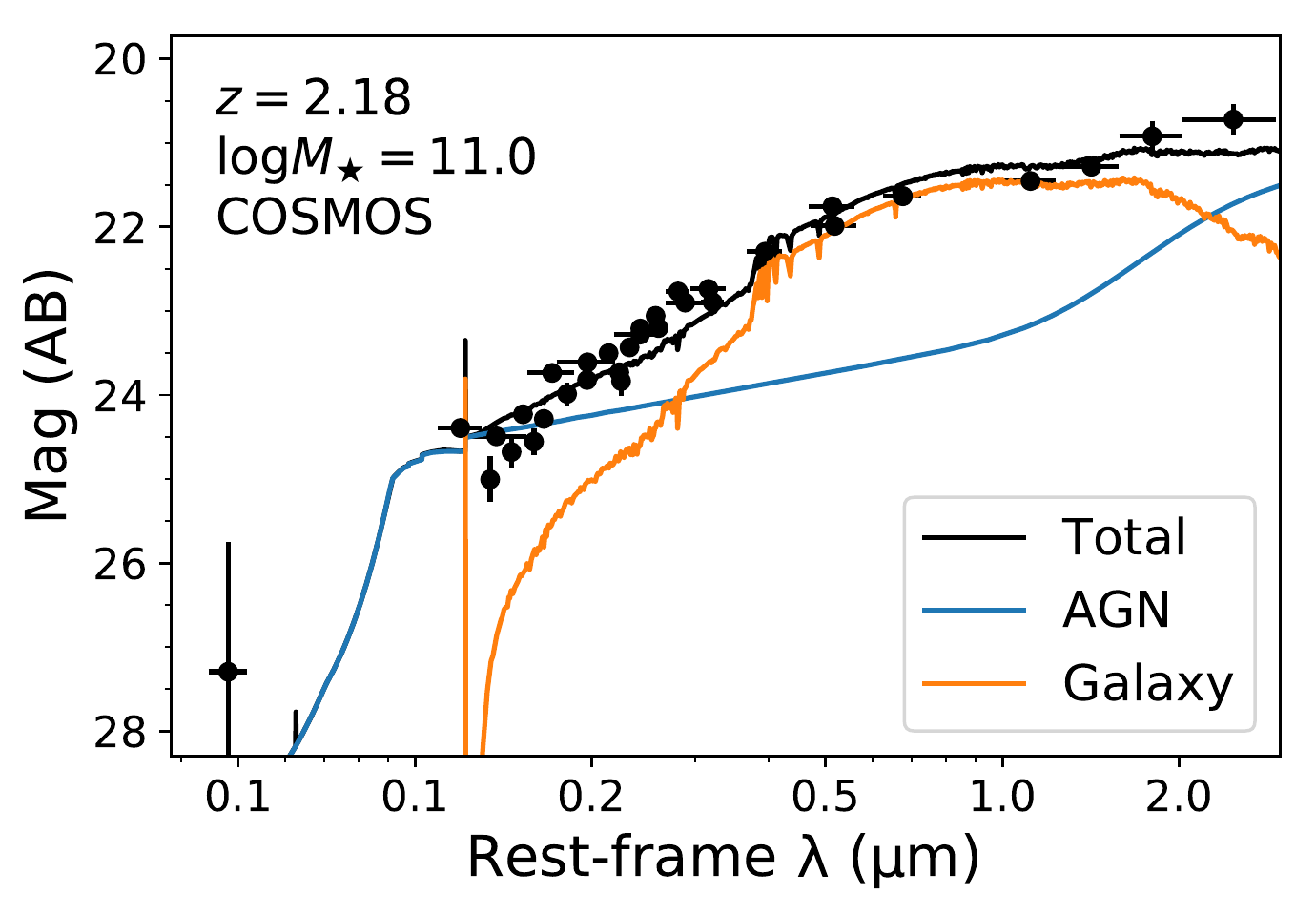}
\caption{Two examples of our SED decomposition for 
BL AGNs. 
The black data points are observed photometry, 
and the black line indicates our best-fit SED model. 
The blue and orange lines represent the AGN and 
galaxy components, respectively. 
At rest-frame $\approx 1\ \mu$m, the galaxy component 
often contributes significantly to the total SED. 
}
\label{fig:blagn_sed}
\end{figure}

Fig.~\ref{fig:blagn_sed} displays the SED fitting for 
two typical BL AGNs. 
Although the AGN component can be dominant 
over the galaxy component at rest-frame 
UV-to-optical wavelengths, 
the galaxy component often contributes
significantly to the total SED at 
NIR wavelengths ($\approx 1\ \mu$m, rest-frame). 
This effect arises because emission from 
the AGN accretion disk and the torus are both 
relatively weak in the NIR, but starlight 
often peaks in the NIR (see, e.g., Fig.~1 of 
\citealt{calistro_rivera16}).
The relatively high contrast of galaxy/AGN at NIR 
wavelengths ensures reliable $\mstar$ 
measurements, since NIR flux is critical 
in assessing $\mstar$ 
\citep[e.g.,][]{ciesla15, calistro_rivera16}. 

\begin{figure}
\includegraphics[width=\linewidth]{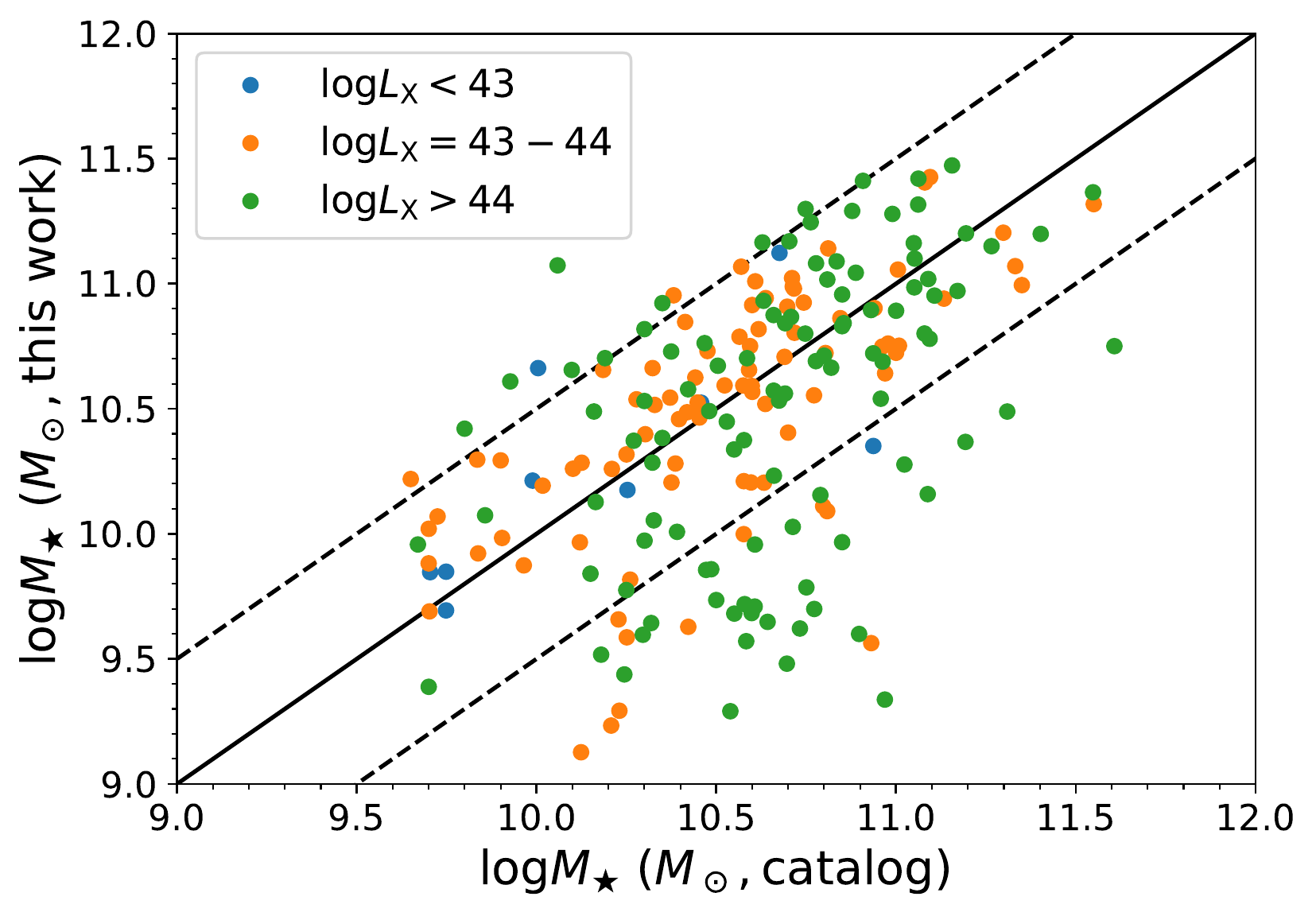}
\caption{Comparison between the $\mstar$ measurements 
from our SED fitting (with both BL AGN and galaxy 
templates) and those from the catalogs (with only galaxy
templates; see \S\ref{sec:surv}). 
Different colors indicate AGNs in different $\lx$\ 
regimes. 
The black solid line indicates the 1:1 relation between
the two $\mstar$ measurements;
the dashed lines indicate 0.5~dex offsets. 
A group of sources have our $\mstar$ measurements 
significantly lower than the catalog values. 
These sources tend to be luminous AGNs with 
$\log\lx\gtrsim 44$. 
}
\label{fig:MvsM}
\end{figure}

We compare our updated $\mstar$ values with those
from the survey catalogs in Fig.~\ref{fig:MvsM}.
Here, we only compare BL AGNs having spectroscopic 
redshifts consistent with the redshifts from the 
catalogs [$|\Delta z|/(1+z_{\rm spec})<0.15$; see 
Footnote~\ref{foot:cosmos_photoz}]. 
The systematics between these two measurements 
are low with the median of $\Delta\log\mstar$ 
being 0.01~dex. 
The scatter (median of $|\Delta\log\mstar$|) is 0.26. 
The two measurements agree within 0.5~dex for the
majority (73\%) of sources. 
From Fig.~\ref{fig:MvsM}, there is a group of 
sources with our $\mstar$ values being significantly 
lower than the corresponding catalog $\mstar$ values. 
These sources tend to be \xray\ luminous 
($\log\lx \gtrsim 44$; see Fig.~\ref{fig:MvsM}), 
and their AGN SED components are comparable to or even 
dominant over their galaxy components in the NIR.
Thus, the catalog measurements, which only utilize 
galaxy templates, significantly overestimate their 
$\mstar$.
There are also some sources for which our $\mstar$
measurements are $\approx 0.5$~dex higher than 
the catalog measurements. 
The reason is likely that the catalog measurements
underestimate the stellar age, resulting in small 
mass-to-light ratio. 
This is because when using galaxy templates 
only, the photometric data require young stellar 
age to account for the UV flux generated by the BL
AGN. 

Throughout the paper, for BL AGNs, we adopt the 
$\mstar$ values derived from our AGN-galaxy SED 
decomposition rather than the catalog values. 

We do not recalculate $\mstar$ for non-BL AGNs. 
However, these sources might have strong AGN 
emission in the rest-frame mid-IR, contributing 
significantly to the observed SED, especially 
in the IRAC bands.
To evaluate this effect, we compare our adopted
$\mstar$ values for non-BL AGNs in COSMOS with 
those from \citet{suh17}, who included AGN 
components in SED fitting. 
The two sets of $\mstar$ agree well: the offset
and scatter are 0.08~dex and 0.11~dex, 
respectively. 
We thus conclude that the presence of non-BL AGNs
does not bias our adopted $\mstar$ measurements.

\subsection{Stellar Mass Function and X-ray 
Luminosity Function}\label{sec:smf_xlf}
\begin{figure}
\includegraphics[width=\linewidth]{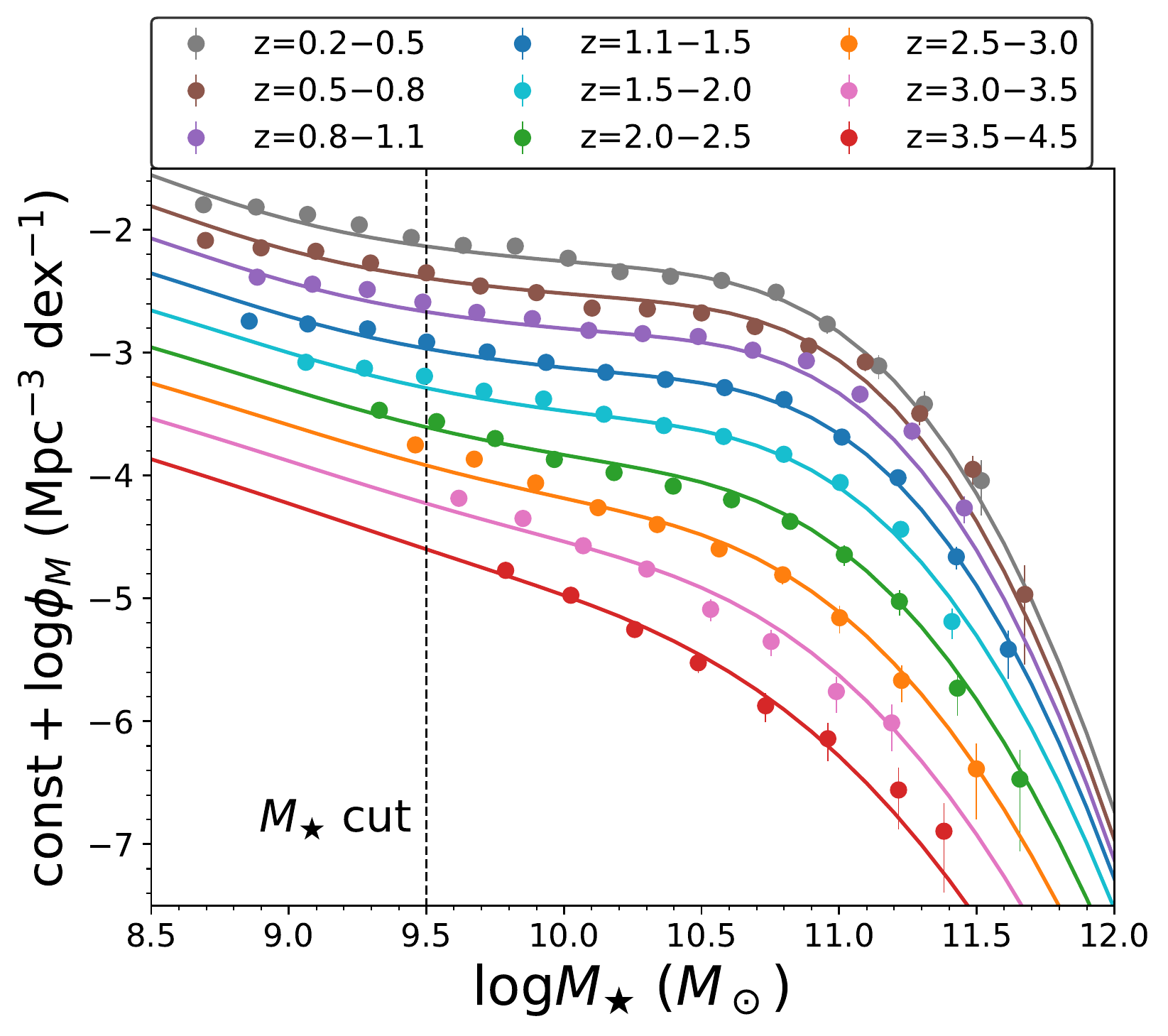}
\caption{Stellar mass function. 
The solid curves indicate the SMF model from 
\citet{behroozi13}. 
The data points are the observational results of 
\citet{davidzon17}. 
Different colors indicate different redshifts.
For display purposes only, the SMF (both the curve 
and data points) for a given redshift bin is shifted 
downward by 0.2~dex 
relative to the next lower redshift bin; the SMF 
for the lowest redshift bin is not shifted.  
The SMF model agrees well with the observations.}
\label{fig:smf}
\end{figure}

Our methodology also utilizes the SMF and XLF as inputs,
which provide additional constraints to $P(\slx|\mstar,z)$
besides the \goodss/\goodsn/COSMOS data set 
(see \S\ref{sec:like_xlf}), especially for 
luminous AGNs (see \S\ref{sec:like_xlf}).
We adopt the SMF from \citet{behroozi13}. 
Their SMF is based on a galaxy evolution model 
that considers both star formation and mergers
and is constrained by the observed SMF, specific 
SFRs ($\rm{sSFR}=\rm{SFR}/\mstar$), and 
$\rho_{\rm SFR}$. 
The SMF model covers ranges of 
$\log\mstar=7\text{--}12$ and $z=0\text{--}8$.
To evaluate the quality of this SMF, we compare it 
with the observational results of \citet{davidzon17}
(see Fig.~3 in \citealt{behroozi13} for comparison 
with some previous observations). 
Fig.~\ref{fig:smf} displays the results. 
The SMF from the \citet{behroozi13} model agrees well 
with the observed SMF values even at high redshift 
($z\approx 4$).

\begin{figure}
\includegraphics[width=\linewidth]{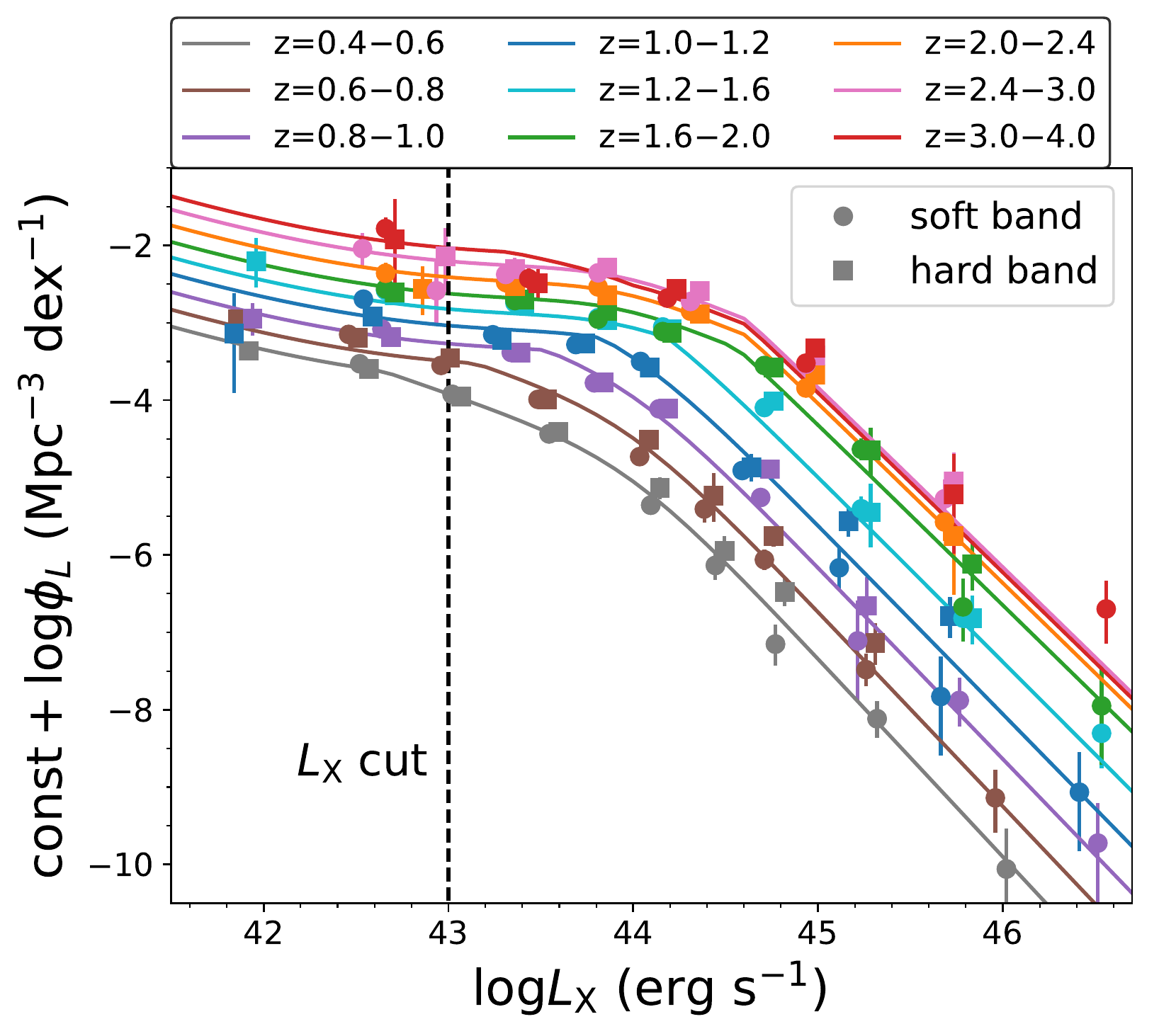}
\caption{\xray\ luminosity function. 
The circles and squares indicate the 
binned XLF values derived from the soft band and 
hard band, respectively, in \citet{ueda14}. 
The hard-band data are shifted slightly toward
the right for display purposes only.  
The solid curves are the best-fit XLF model in 
\citet{ueda14}. 
The soft-band XLF agrees well with the hard-band XLF.
Different colors indicate different redshifts. 
For display purposes only, the XLF (both the curve 
and data points) for a given redshift 
bin is shifted upward by 0.3~dex 
relative to the next lower redshift bin; the XLF 
for the lowest redshift bin is not shifted.
The vertical dashed solid line indicates 
our $\lx$ cut ($\log\lx=43$) for the XLF.
}
\label{fig:xlf}
\end{figure}

We adopt the binned soft-band XLF values of 
\citet{ueda14} (see their Fig.~10). 
Fig.~\ref{fig:xlf} compares the binned rest-frame 
\hbox{2--10~keV} XLF derived from soft-band and hard-band
observations.\footnote{In \citet{ueda14}, the hard-band 
data include various survey results above 2~keV (see their 
Tab.~1 for details).}
We choose the soft-band XLF instead of the hard-band 
XLF because the former extends to extremely high 
luminosity ($\log\lx \gtrsim 46$),
largely due to the wide-field surveys of \rosat.
The soft-band XLF is already corrected for 
obscuration and is thus consistent with the 
hard-band XLF when available (see Fig.~\ref{fig:xlf}). 
In practice, adopting the hard-band XLF would only
have minor effects on our results (\S\ref{sec:res}).  
We have also compared the XLFs from \citet{ueda14}
with those from \citet{aird15}, and found they 
agree well.  
Since \citet{ueda14} extend to higher $\lx$\ 
than \citet{aird15}, 
we adopt the XLF results from \citet{ueda14}. 
The XLF data points in \citet{ueda14} are derived 
from all \xray\ detected sources with $\log\lx \gtrsim 42$, 
and thus components of XRBs
might become important at the low-luminosity end.
To avoid this XRB contamination, we only adopt
the XLF at $\log\lx>43$. 
Also, we do not probe the low-$\slx$ regime ($\log\slx<-2$; 
see \S\ref{sec:Lsx_cut}) that could significantly contribute 
to the XLF at $\log\lx<43$. 
For example, at $\log\lx=42\text{--}42.5$, 
$32\%$ of the \xray\ sources in our sample have $\log\slx<-2$.
At $\log \lx<43$, our $P(\slx|\mstar,z)$ are constrained 
by the survey data (see \S\ref{sec:surv}) which have 
$\approx 420$ \xray\ AGNs with $\log \lx<43$. 

\subsection{Bolometric Correction}\label{sec:k_bol}
To obtain $\bharbar$ from the $P(\slx|\mstar,z)$, 
we require a bolometric correction 
($\kbol=\lbol/\lx$) which is a function of AGN 
luminosity \citep[e.g.,][see \S\ref{sec:res} for 
$\bharbar$ calculation]{steffen06, hopkins07, lusso12}.
Our $\kbol$ is based on the results of \citet{lusso12}, 
which are derived from the multiwavelength observations 
of \xray\ selected AGNs. 
We did not adopt the $\kbol$ from \citet{hopkins07}, because
they included the dust-reprocessed IR emission that is 
not directly produced by AGN accretion power. 
This inclusion results in a fraction of the 
accretion-powered radiation being accounted for twice. 
Thus, their $\kbol$ serves as an empirical bolometric correction
but is an overestimation for our purpose, i.e., to infer 
the accretion power of the SMBH. 

\begin{figure}
\includegraphics[width=\linewidth]{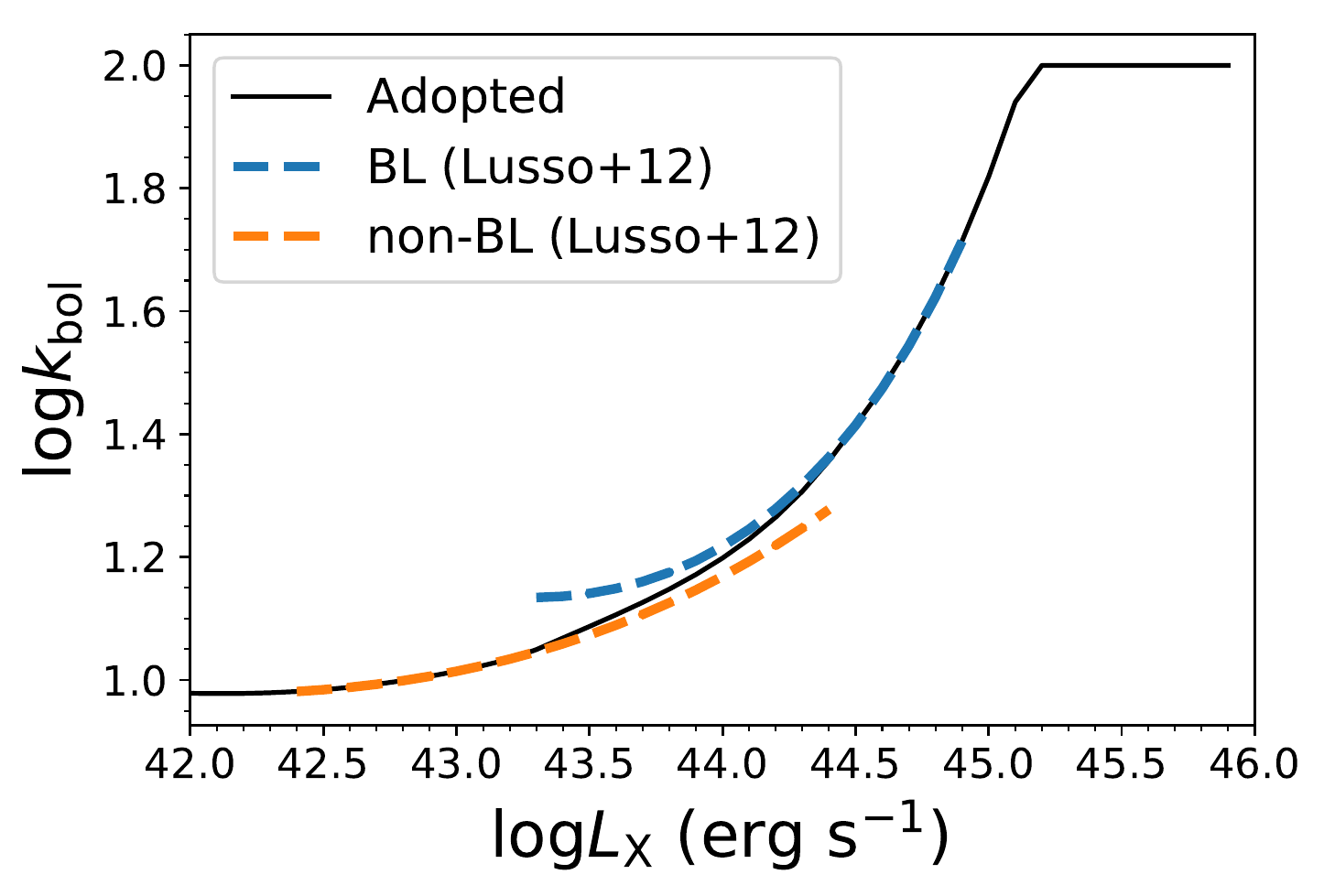}
\caption{Bolometric correction as a function of \xray\
luminosity.
The black solid line indicates the $\kbol$ adopted in
this work. 
The blue and orange dashed lines indicate $\kbol$ 
for BL and non-BL AGNs, respectively,
in \citet{lusso12}. 
$\kbol$ generally increases toward high $\lx$.
}
\label{fig:k_bol}
\end{figure}

\citet{lusso12} modeled $\kbol$ as polynomials of AGN $\lbol$ 
for BL ($k_{\rm bol,BL}$) and non-BL AGNs 
($k_{\rm bol,nBL}$), respectively (see their Tab.~2). 
We present their $k_{\rm bol,BL}$ and $k_{\rm bol,nBL}$ as a 
function of $\lx$\ in Fig.~\ref{fig:k_bol} 
(dashed curves).
Both $k_{\rm bol,BL}$ and $k_{\rm bol,nBL}$ are positively 
dependent on $\lx$.
The differences between $k_{\rm bol,BL}$ and $k_{\rm bol,nBL}$ 
are small ($\lesssim 0.1$~dex), consistent with the standard 
unified AGN model.
The BL AGNs are generally more luminous than the non-BL
AGNs in \citet{lusso12}.
Denoting the overlapping $\lx$ range of the BL and non-BL 
as $L_{\rm X,1}\text{--}L_{\rm X,2}$, we adopt $k_{\rm bol,nBL}$ 
if $\lx<L_{\rm X,1}$ and $k_{\rm bol,BL}$ if $\lx>L_{\rm X,2}$. 
For $L_{\rm X,1} \leq \lx \leq L_{\rm X,2}$, we adopt 
$\kbol$ as 
\begin{equation}
\kbol = k_{\rm bol,BL} \frac{\lx-L_{\rm X,1}}{L_{\rm X,2}-L_{\rm X,1}}
      + k_{\rm bol,nBL} \frac{L_{\rm X,2}-\lx}{L_{\rm X,2}-L_{\rm X,1}}.
\end{equation} 
This linear interpolation produces a smooth $\kbol$ as 
a function of $\lx$ (see Fig.~\ref{fig:k_bol}).
$k_{\rm bol,BL}$ diverges quickly toward high $\lx$ due
to the polynomial functional form of $\kbol$\ that 
\citet{lusso12} assumed. 
To avoid this divergence, we set an upper limit of 100
for the adopted $\kbol$, which is about the maximum 
value of $\kbol$ reported in the literature 
\citep[e.g.,][]{marconi04, hopkins07}.

Some studies suggest that $\kbol$ might also be 
related to $\lam$ (e.g., \citealt{vasudevan07}; but also 
see \citealt{shemmer08}).
While it is still under investigation whether $\kbol$ is 
more fundamentally related to AGN luminosity or 
$\lam$, we adopt the $\kbol$-$\lx$ relation because 
we are not able to obtain $\lam$ accurately. 
We discuss the effects of a $\kbol$-$\lam$ relation 
in \S\ref{sec:bhar_for_all}.

\section{Data Analyses}\label{sec:data_ana}
We derive $\bharbar(\mstar,z)$ from $P(\slx|\mstar,z)$
(\S\ref{sec:res}), which is constrained utilizing 
maximum-likelihood fitting (\S\ref{sec:ml_fit}). 
The fitting requires likelihood functions, 
and our likelihood functions are based on  
the survey data (\S\ref{sec:like_surv}) as well 
as the SMF and XLF (\S\ref{sec:like_xlf}). 
The survey data mostly constrain $P(\slx|\mstar,z)$\ for
low-to-moderate luminosity AGNs, while the SMF and XLF 
constrain $P(\slx|\mstar,z)$ for luminous AGNs.

\subsection{Likelihood Functions}\label{sec:like}
\subsubsection{The Likelihood Function of Survey Data}
\label{sec:like_surv}
With the survey data providing $\lx$, $\mstar$, and $z$\ 
for individual sources (\S\ref{sec:sample}), we can derive
the corresponding likelihood function,
which quantifies how well the input $P(\slx|\mstar,z)$ 
model matches the survey data.
The methodology is detailed in \S6.1 of 
\citet{aird12}; below we briefly summarize this technique. 

For each survey in \S\ref{sec:surv}, we denote the number 
of detected \xray\ AGNs as $N_{\rm AGN}$, where we only 
consider \xray\ sources with $\log \slx> -2$\ as AGNs 
(see \S\ref{sec:Lsx_cut}).
The log-likelihood function for this survey
can be derived from the unbinned 
Poisson probability \citep[e.g.,][]{loredo04} as 
\begin{equation}\label{eq:L_surv1}
\begin{split}
\ln L = -N_{\rm mdl} + 
	\sum\limits_{i=1}^{N_{\mathrm{AGN}}} 
	\ln P(L_{\mathrm{SX},i}|M_{\star,i},z_i),
\end{split}
\end{equation}
where $N_{\rm mdl}$ is the expected number 
of \xray\ detected AGNs from the input $P(\slx|\mstar,z)$ 
model.
In the survey area, there are $N_{\rm gal}$\ galaxies
within the redshift range, each with $\mstar$\ and 
redshift measurements. 
The quantity $N_{\rm mdl}$\ can be calculated as 
\begin{equation}\label{eq:N_mdl}
\begin{split}
N_{\rm mdl} & = \sum\limits_{i=1}^{N_{\mathrm{gal}}}
	\int_{-2}^{\infty} P(\slx|M_{\star,i},z_i) 
	p_{\mathrm{det}}(\lx, z_i) d \log \slx \\
	 & = \sum\limits_{i=1}^{N_{\mathrm{gal}}}
	\int_{-2}^{\infty} P(\slx|M_{\star,i},z_i) 
	p_{\mathrm{det}}(\slx M_{\star,i}, z_i) d \log \slx,
\end{split}
\end{equation}
where $p_{\mathrm{det}}$, as a function of $\lx$\ and $z$,
is the correction factor for survey sensitivity. 
The integral lower limit ($-2$) is the lowest $\slx$\ probed
in this work, and we discuss the effects of this $\slx$ cut
in \S\ref{sec:Lsx_cut}.
$p_{\mathrm{det}}(\lx, z)$\ is defined as 
\begin{equation}\label{eq:p_det}
\begin{split}
p_{\mathrm{det}}(\lx, z) = \frac{A[f(\lx, z)]}{A_{\rm tot}},
\end{split}
\end{equation}
where $f(\lx, z)$\ is the expected \xray\ flux in the 
detection band for a source with $\lx$\ at redshift $z$;
$A[f(\lx, z)]$\ is the sky coverage of the survey
sensitive to the flux $f(\lx, z)$ (Fig.~\ref{fig:skycov}); 
$A_{\rm tot}$\ is the total survey area.

We calculate the log-likelihood for each survey 
(\goodss, \goodsn, and COSMOS; see \S\ref{sec:surv}) 
independently according to Eq.~\ref{eq:L_surv1}, and 
combine them as 
\begin{equation}\label{eq:L_surv2}
\begin{split}
\ln L_{\rm surv} = \ln L_{\rm GOODS\text{-}S} + 
\ln L_{\rm GOODS\text{-}N} + \ln L_{\rm COSMOS}.
\end{split}
\end{equation}

Since our data only include \hbox{0.05--1.4~deg$^2$}
surveys, most of our \xray\ selected sources are 
low-to-moderate luminosity AGNs ($\log\lx \lesssim 44.5$; 
see Fig.~\ref{fig:Lx_vs_z}), and the number of 
luminous AGNs is limited.
However, this small-area effect does not introduce 
biases for these luminous AGNs, because the sizes
of the survey areas are already properly considered in 
the likelihood function through the term 
``$N_{\rm mdl}$'' (see Eq.~\ref{eq:N_mdl}).
Moreover, $P(\slx|\mstar,z)$ for luminous AGNs is 
constrained by our SMF-XLF methodology in 
\S\ref{sec:like_xlf}.

\subsubsection{The Likelihood Function of SMF and XLF}
\label{sec:like_xlf}
Besides the constraints from the survey data, we can also 
constrain $P(\slx|\mstar,z)$\ utilizing the SMF and XLF.
By definition, the SMF provides the comoving number density 
of galaxies at given $\mstar$\ and $z$. 
Thus, we can convolve the SMF with $P(\slx|\mstar, z)$\ 
at a given redshift to obtain the comoving number density 
of AGNs at given $\lx$\ and $z$\ (i.e., the XLF), 
\begin{equation}\label{eq:xlf}
\begin{split}
\phi_L(\lx|z) &= \int^{12}_{9.5} P(\slx|\mstar,z) 
		\phi_M(\mstar|z) d\log{\mstar} \\
   &=  \int^{12}_{9.5} P(\lx/\mstar|\mstar,z) \phi_M(\mstar|z) 
	d\log{\mstar},
\end{split}
\end{equation}
where $\phi_L$\ and $\phi_M$\ are the XLF and SMF, respectively.
Both $\phi_L$\ and $\phi_M$\ are in units of Mpc$^{-3}$~dex$^{-1}$. 
The integral range ($\log\mstar=9.5\text{--}12$) corresponds 
to the $\mstar$\ regime probed in our analyses, and we discuss
the effects of this $\mstar$\ cut in \S\ref{sec:m_cut}. 

To evaluate quantitatively how well a model 
$P(\slx|\mstar,z)$\ meets the constraints from the SMF 
and XLF (\S\ref{sec:smf_xlf}), we compare 
$N_{\rm mdl}$ and the observed numbers of \xray\ sources 
($N_{\rm obs}$) for each $\phi_L$\ bin in \citet{ueda14}. 
For any given model of $P(\slx|\mstar,z)$, 
we can obtain the corresponding model XLF 
($\phi_{L,\rm mdl}$) via Eq.~\ref{eq:xlf}.
We can convert $\phi_{L,\rm mdl}$ to $N_{\rm mdl}$\ as 
\begin{equation}\label{eq:Nmdl}
\begin{split}
N_{\rm mdl} &= \phi_{L,\rm mdl} 
		       \frac{N_{\rm obs}}{\phi_{L,\rm obs}},
\end{split}
\end{equation}
where ${N_{\rm obs}}/{\phi_{L,\rm obs}}$\ is the 
conversion factor between the number of sampled \xray\ 
sources and the XLF in \citet{ueda14}. 
${N_{\rm obs}}/{\phi_{L,\rm obs}}$ considers many
factors such as \xray\ survey area and sensitivity and is 
is provided by Y.\ Ueda (2017, private communication).
We can then obtain the log-likelihood function for the 
SMF and XLF as 
\begin{equation}\label{eq:L_xlf}
\begin{split}
\ln L_{\rm XLF} = \sum\limits_{i=1}^n 
\ln P(N_{\mathrm{obs},i}|\lambda=N_{\mathrm{mdl},i}),
\end{split}
\end{equation}
where $n$\ is the total number of $\lx$ and redshift 
bins available for the XLF, 
and $P(N_{\mathrm{obs},i}|\lambda=N_{\mathrm{mdl},i})$\
is the Poisson probability of $N_{\mathrm{obs}, i}$\ 
events for the rate parameter $\lambda=N_{\mathrm{mdl},i}$.

Since we apply an $\lx$ cut to the XLF ($\log \lx >43$; see 
\S\ref{sec:smf_xlf}), the SMF-XLF method does not constrain 
$P(\slx|\mstar,z)$ for low-luminosity AGNs. 
In fact, this method is also not very effective when assessing
moderate-luminosity AGNs with $43 < \log \lx \lesssim 44$.
For example, for an AGN with $\log \lx \approx 43.5$, its host 
galaxy can be either a moderate-$\mstar$ galaxy with relatively 
high $\slx$ (e.g., $\log \mstar \approx 10$ and 
$\log \slx \approx 0$) or a high-$\mstar$ galaxy with low $\slx$
(e.g., $\log \mstar \approx 11$ and $\log \slx \approx -1$). 
The SMF-XLF method cannot determine which case is more probable,
although it does require that the sum of the contributions from 
all galaxies must match the AGN number density. 
However, this degeneracy is weak in the high-$\lx$ regime.
For example, for an AGN with $\log\lx \approx 45$, 
its host galaxy is likely massive ($\log \mstar \gtrsim 11$)
with moderate-to-high $\slx$ 
($-0.5 \lesssim \log \slx \lesssim 0.5$);
otherwise, the corresponding $\slx$ would be too high and 
far beyond the high-$\slx$ cutoff ($\log \slx \approx 0$; 
see, e.g., \citealt{bongiorno16, aird17, georgakakis17}).
This is the reason why the SMF-XLF method is most effective for 
luminous AGNs.  

\subsection{Cuts in Parameter Space}
\subsubsection{Redshift Cuts}
\label{sec:z_cut}
We limit our study to the redshift range of 
$z=0.4\text{--}4$.
The primary reason for this is that we have a limited
number of \xray\ AGNs outside this redshift range.
Quantitatively, the number of AGNs with $z<0.4$ or $z>4$ 
is 89 which accounts for only $\approx5\%$ of the whole AGN 
sample.

\subsubsection{$\mstar$\ Cuts}
\label{sec:m_cut}
In this work, we only study galaxies having 
$\log \mstar=9.5\text{--}12$. 
The reason for the lower cut of $\mstar$\ is that
\xray\ AGNs are rarely detected in low-$\mstar$\ galaxies
(see Fig.~\ref{fig:M_vs_z});
in our sample (\S\ref{sec:surv}), only $\approx 5\%$ of
\xray\ AGNs are detected with $\log \mstar<9.5$.
The low detection rate of \xray\ AGNs for the low-$\mstar$\ 
systems makes it challenging to constrain effectively
$P(\slx|\mstar,z)$ for these sources.
We investigate galaxies with $\mstar$ up to $\log\mstar=12$.
In our sample, there are $\approx 190$ galaxies with 
$\log \mstar=11.5\text{--}12$ but only 7 galaxies 
above $\log \mstar = 12$.

These $\mstar$\ cuts have little effect on our methods in 
\S\ref{sec:like_xlf}, because the XLF is dominated by AGNs 
hosted in galaxies with $\log\mstar=9.5\text{--}12$.
At $\log\lx =43 \text{--} 43.5$
(the lowest-luminosity bin we adopt for the XLF; see 
\S\ref{sec:smf_xlf}), almost all (98\%) AGN-host galaxies 
have $\log \mstar = 9.5\text{--}12$ in our sample,
and the fraction is even larger toward higher luminosities.


\subsubsection{$\slx$\ Cut}
\label{sec:Lsx_cut}
We cannot probe low $\slx$\ values at high redshift due to 
survey sensitivity. 
For example, we only have two sources with 
$\log\slx < -2$\ at $z>2$. 
Also, at low $\slx$, it is difficult to disentangle 
\xray\ emission from AGNs and XRBs (see \S\ref{sec:xrb_cont}).
We thus limit our investigation to $\log \slx \geq -2$.

This $\slx$\ cut has little effect on our SMF-XLF method
(\S\ref{sec:like_xlf}), since galaxies with $\log \slx < -2$\ 
have almost no contribution to the XLF above 
$\log\lx=43$\ (the lowest $\lx$\ we adopt for the XLF; 
see \S\ref{sec:smf_xlf}).
For example, among the $\approx 1,500$\ AGNs with 
$\log \lx \geq43$ in our sample, only one source has 
$\log \slx < -2$.

Now we demonstrate that SMBH accretion below $\log \slx = -2$ 
does not contribute significantly to the growth of SMBHs 
over cosmic history. 
The Eddington ratio is defined as   
\begin{equation}\label{eq:lam_edd}
\begin{split}
\lam &= \frac{\lbol}{L_{\rm Edd}} \\
     &= \frac{\epsilon \mathrm{BHAR} c^2}{1.3 \times 10^{38}\ 
		\mathrm{erg\ s^{-1}}\ \mbh\ M_\odot^{-1}} \\
     &= \frac{M_\odot c^2}{1.3\times 10^{39}\ \mathrm{erg\ s^{-1}}} 
	\frac{\mathrm{BHAR}\epsilon_{0.1}}{\mbh} \\
     &= 0.044\ \mathrm{Gyr}\ 
	\frac{\mathrm{BHAR}\epsilon_{0.1}}{\mbh},
\end{split}
\end{equation}
where $c$\ is the speed of light, $\epsilon$ is radiation 
efficiency, and $\epsilon_{0.1}=\epsilon/0.1$.
$\lam$ can also be related to $\slx$ as 
\begin{equation}\label{eq:lam_edd_Lsx}
\begin{split}
\lam &= \frac{\lbol}{L_{\rm Edd}} \\
     &= \frac{\kbol\lx}{3.2\times 10^4 L_\odot \mbh M_\odot^{-1}} \\
     &= \frac{10\times5000}{3.2\times 10^4}
	\frac{\lx L_\odot^{-1}}{\mstar M_\odot^{-1}} 
	\frac{\kbol}{10} \frac{\mstar}{5000 \mbh} \\
     &= 1.6 \slx k_{10} m_{5},
\end{split}
\end{equation}
where $k_{10}=\kbol/10$ and $m_{5}=\mstar/(5000\mbh)$.
From Eqs.~\ref{eq:lam_edd} and \ref{eq:lam_edd_Lsx}, 
we can obtain the $e$-folding timescale of SMBH growth as
\begin{equation}\label{eq:t_e}
\begin{split}
t_e = \frac{\mbh}{\rm BHAR}
    = 0.028\ \mathrm{Gyr} 
      \frac{\epsilon_{0.1}}{\slx k_{10} m_{5}}.
\end{split}
\end{equation}
$\log \slx < -2$ corresponds to $\log \lx \lesssim 43.5$
with $k_{10} \approx 1$ (see Fig.~\ref{fig:k_bol}). 
Assuming the average $\slx$ for this low-$\slx$ 
($\log \slx < -2$) accretion is $\approx 10^{-2.5}$
and $\epsilon_{0.1}\approx 1$, then 
$t_e \approx 9 m_{5}^{-1}\ \mathrm{Gyr}$.
From our results in Sec.~\ref{sec:mbh_mstar}, $m_{5}$ 
ranges from $\approx 0.1$ to $1$ up to at least $z\approx 2$. 
Thus, the growth time for low-$\slx$ accretion should be 
$t_e \approx 9 \text{--} 90\ \mathrm{Gyr}$ comparable to or 
longer than the Hubble time.
Therefore, the accretion with $\log\slx<-2$\ is unlikely 
important in our overall understanding of SMBH growth. 

The above argument assumes $\epsilon\approx 0.1$, 
which is typical for a cold accretion disk 
\citep[e.g.,][]{shakura73, agol00}. 
For hot accretion flows, $\epsilon$ can be much lower 
($\epsilon \lesssim 0.01$). 
However, such low $\epsilon$ only arises when 
the accretion rate is low, i.e., 
\begin{equation}\label{eq:t_e2}
\begin{split}
\frac{0.1{\rm BHAR}c^2}{L_{\rm Edd}}\lesssim 10^{-3},
\end{split}
\end{equation}
(see Fig.~2 of \citealt{yuan14}).
Under this condition, the SMBH growth timescale 
is 
\begin{equation}\label{eq:t_e2}
\begin{split}
t_e &= \frac{\mbh}{\rm BHAR} \\
    &= \frac{L_{\rm Edd}}{1.3 \times 
 	    10^{38}\rm{BHAR}\ \rm{erg\ s^{-1}}\ \it M_\odot^{-1} } \\
    &= \frac{M_\odot c^2}{1.3 \times 10^{39}{\rm\ erg\ s^{-1}}}  
       \frac{L_{\rm Edd}}{0.1{\rm BHAR}c^2 } \\
    & \gtrsim \frac{M_\odot c^2}{1.3 \times 10^{39}{\rm\ erg\ s^{-1}}} 
	    \times 10^3 \\
    & \approx 44\ \rm{Gyr},
\end{split}
\end{equation}
which is also longer than Hubble time.
Therefore, low-$\epsilon$ accretion states should not 
contribute significantly to the overall SMBH growth 
across cosmic history.

Our argument above shows that low-$\slx$ accretion 
cannot increase $\mbh$ effectively over cosmic history.
However, low-$\slx$ accretion might still contribute more 
significantly to $\bharbar$ than high-$\slx$ accretion 
in specific limited ranges of redshift and $\mstar$.
In such regimes, our $\bharbar$ calculation, which does not 
account for low-$\slx$ accretion, will be inaccurate. 
We discuss this possibility in \S\ref{sec:contr_low_Lsx}.

\subsection{Maximum-Likelihood Fitting of $P(\slx|\mstar,z)$}
\label{sec:ml_fit}
The likelihood functions from survey data and 
the SMF-XLF data are the most effective in 
constraining $P(\slx|\mstar,z)$\
for AGNs in different $\lx$ regimes (see \S\ref{sec:like}).
Thus, they need to be combined to identify the 
best model for $P(\slx|\mstar,z)$. 
We obtain the final log-likelihood function as
\begin{equation}\label{eq:L_final}
\begin{split}
\ln L =  \ln L_{\rm surv} + \ln L_{\rm XLF},
\end{split}
\end{equation}
where $\ln L_{\rm surv}$ and $\ln L_{\rm XLF}$ 
are the log-likelihood functions in Eqs.~\ref{eq:L_surv2}
and \ref{eq:L_xlf}, respectively.
For a given model, we search for best-fit model parameters 
via maximum-likelihood fitting with 
{\sc iminuit}~v1.2.\footnote{See https://pypi.python.org/pypi/iminuit
for {\sc iminuit}.}

Following \citet{bongiorno16}, we model 
$P(\slx|\mstar, z)$\ as a smoothed double power law. 
We do not adopt a single power-law model or a Schechter
model \citep{schechter76} of $P(\slx|\mstar, z)$. 
A single power law would lead to an unphysical divergent 
$\lxbar$ (see \S\ref{sec:intro}).
A Schechter model results in an exponential 
decline toward high $\lx$ in the predicted XLF 
(see Eq.~\ref{eq:xlf}), inconsistent with the observed 
power-law decline \citep[e.g.,][]{ueda14, aird15}.
However, a smoothed double power law can produce a 
power-law decline in the XLF. 
The smoothed double power law at given 
$\mstar$\ and $z$ is written as
\begin{equation}
\begin{split}
P(\slx|\mstar, z) = 
	A \left[ \left(\frac{\slx}{L_c}\right)^{\gamma_1} + 
          \left(\frac{\slx}{L_\mathrm{c}}\right)^{\gamma_2} \right]^{-1}.
\end{split}
\end{equation}
We define $\gamma_1 \leq \gamma_2$, and thus they are the 
slopes in the low and high $\slx$\ regimes, respectively.
All of $\log A$, $\log L_\mathrm{c}$, $\gamma_1$, and 
$\gamma_2$\ are modeled in the general form of 
polynomial functions of $\log \mstar$\ and $\log(1+z)$, i.e., 
\begin{equation}\label{eq:poly_mdl}
\begin{split}
X &= X_0 + \alpha_0^X \log(1+z) + \alpha_1^X \log M_{10} \\
 &+ \beta_0^X[\log(1+z)]^2 + \beta_1^X\log(1+z)\log M_{10} 
   +\beta_2^X(\log M_{10})^2  \\
 & + ...
\end{split}
\end{equation}
where 
$M_{10} = \mstar/10^{10}M_\odot$, and $X$\ 
indicates $\log A$, $\log L_\mathrm{c}$, $\gamma_1$, or 
$\gamma_2$.

To determine the highest order of polynomial necessary 
for each $X$, we perform a detailed model selection 
presented in Appendix~\ref{sec:mdl_select}.
We find that $\log A$, $\log L_\mathrm{c}$, and $\gamma_2$ 
are consistent with 2nd-order, 2nd-order, and 
1st-order polynomials, respectively;
$\gamma_1$ is consistent with a constant value (0th-order 
polynomial).
The best-fit parameter values and their errors are listed 
in Tab.~\ref{tab:best_par}; these errors are estimated
from MCMC sampling (see Appendix~\ref{sec:mdl_select}).
Our conclusions (\S\ref{sec:discuss}) are robust under 
$>3\sigma$ confidence levels.
The fitting quality and comparison with previous work are 
presented in Appendix~\ref{sec:fit_qual}.

We are aware that there is an issue of 
``double-counting'' when combining the likelihood 
functions in Eq.~\ref{eq:L_final}, since there are some
AGNs being included in both survey and XLF data. 
This could lead to underestimation of our model 
uncertainties.
However, we expect that this issue only has 
minor effects on our results. 
For COSMOS AGNs, this issue does not exist because the 
XLF measurements do not include them (see Tab.~1 of 
\citealt{ueda14}). 
For \cdfs\ and \cdfn, only AGNs above $\log\lx=43$
(our $\lx$ cut for XLF; see \S\ref{sec:smf_xlf}) are 
counted doubly, and these objects account for only 
$\approx 50\%$ of the whole AGN population in these
two fields. 



\begin{table*}
\begin{center}
\caption{Best-Fit Model Parameters and $1\sigma$ Errors}
\label{tab:best_par}
\begin{tabular}{cccccccc}\hline\hline
 & $X_0$ & $\alpha^X_0$ & $\alpha^X_1$ & $\beta^X_0$ & $\beta^X_1$ & $\beta^X_2$  \\ \hline
$\log A$ & $-2.68^{+0.18}_{-0.19}$ & $0.81^{+0.68}_{-0.85}$ & $1.33^{+0.26}_{-0.20}$ & $-0.20^{+1.04}_{-0.90}$ & $1.00^{+0.43}_{-0.46}$ & $-0.76^{+0.09}_{-0.11}$ \\
$\log L_{\mathrm c}$ & $-0.57^{+0.19}_{-0.21}$ & $4.66^{+0.90}_{-0.62}$ & $-1.61^{+0.30}_{-0.39}$ & $-4.24^{+0.82}_{-1.07}$ & $0.90^{+0.62}_{-0.57}$ & $0.46^{+0.16}_{-0.12}$ \\
$\gamma_1$ & $0.43^{+0.04}_{-0.02}$ & -- & -- & -- & -- & -- \\
$\gamma_2$ & $2.45^{+0.27}_{-0.16}$ & $-0.88^{+0.72}_{-0.52}$ & $0.56^{+0.19}_{-0.36}$ & -- & -- & -- \\ \hline
\end{tabular}
\end{center}
\begin{flushleft}
{\sc Note.} --- The parameters are the polynomial coefficients in 
Eq.~\ref{eq:poly_mdl}. 
The symbol ``--'' indicates the corresponding parameter is 
not required, and hence fixed to zero in our modeling 
(see Appendix~\ref{sec:mdl_select}).
\end{flushleft}
\end{table*}

\subsection{The Results for $\bharbar(\mstar,z)$}
\label{sec:res}
The main goal of this paper is to characterize 
the overall SMBH growth for the majority galaxy population
at $z=\text{0.4--4}$.
Therefore, we derive the $\bharbar$ for all galaxies, 
including both star-forming and quiescent populations,
in \S\ref{sec:bhar_for_all}.
We also study the $\bharbar$ for star-forming galaxies in 
\S\ref{sec:bhar_SFonly}.
We do not investigate the $\bharbar$ for quiescent galaxies,
as explained in \S\ref{sec:bhar_quiescent}.

\subsubsection{All Galaxies}
\label{sec:bhar_for_all}
Due to the existence of strong AGN variability, ensemble-averaged
BHAR is often adopted as an approximation of long-term average 
BHAR (see \S\ref{sec:intro}). 
From our derived $\slx$\ distribution, we can obtain this
$\bharbar$ as a function of $\mstar$\ and redshift, i.e., 
\begin{equation}\label{eq:bharbar}
\begin{split}
& \bharbar(\mstar,z) \\
	& = \int^{\infty}_{-2} P(\slx|\mstar,z) 
	    \frac{(1-\epsilon)\kbol(\lx)\lx}{\epsilon c^2} 
	     d \log \slx \\
	& = \int^{\infty}_{-2} P(\slx|\mstar,z) 
	\frac{(1-\epsilon)\kbol(\mstar\slx)\mstar\slx}{\epsilon c^2} 
	d\log \slx.
\end{split}
\end{equation}
We use the $\kbol(\lx)$ function presented in \S\ref{sec:k_bol}.
Although theoretical studies suggest that $\epsilon$ depends
on factors like SMBH spin and the state of accretion flow 
\citep[e.g.,][]{shakura73, agol00, yuan14}, it is not feasible to 
determine $\epsilon$ accurately for individual AGNs from 
observations. 
The adopted $\epsilon=0.1$ is likely a typical value for the 
overall AGN population \citep[e.g.,][]{marconi04, davis11},
and has been widely adopted in previous work 
\citep[e.g.,][]{mullaney12, chen13, yang17}.
We caution that there might be uncertainties up to a factor of 
a few for this typical $\epsilon$.

The results are displayed in Fig.~\ref{fig:BHAR_vs_M}.
As explained in Appendix~\ref{sec:fit_qual} 
and \S\ref{sec:kbol_lam}, the 
model uncertainties at $z\lesssim 1$ and 
$\log\mstar\gtrsim 11.5$ might be underestimated due to 
limited AGN sample sizes and $\kbol$ uncertainties. 
Therefore, we mark $\bharbar$ in these ranges as dotted 
curves in Fig.~\ref{fig:BHAR_vs_M}.
Our derived $\bharbar(\mstar, z)$ shows that, at a given
redshift, $\bharbar$ generally increases toward high 
$\mstar$, although the $\mstar$ dependence is weaker 
toward the local universe. 
This positive $\bharbar$-$\mstar$ relation is also reported 
by \citet{yang17}. 
Their data are consistent with a linear $\bharbar$-$\mstar$
(in logarithmic space) relation with a slope of unity 
for sources at $z=0.5\text{--}2$ 
(the dashed line in Fig.~\ref{fig:BHAR_vs_M}). 

$\bharbar$ has stronger redshift evolution at 
higher $\mstar$.
For instance, for $\log\mstar=9.5$, $\bharbar$ 
at $z=4$ is $0.7\pm0.4$~dex higher than at $z=0.5$; 
for $\log\mstar=11.5$, $\bharbar$ rises by $3.0\pm0.3$~dex 
from $z=0.5$ to $z=4$.
This strong redshift dependence for massive galaxies 
is also found by recent works \citep[e.g.,][]{aird17, 
georgakakis17}.
\citet{yang17} compared the $\bharbar$-$\mstar$ relations 
for two broad redshift bins, $z=0.5\text{--}1.3$ and 
$z=1.3\text{--}2.0$, but did not find significant 
differences, possibly due to the limited sample size
of the \goodss\ field in their study (\S\ref{sec:goodss}).
Their relatively small sample lacks luminous AGNs 
(see Fig.~\ref{fig:Lx_vs_z}), 
and could lead to a generally underestimated $\bharbar$.
This underestimation should be stronger at higher redshifts,
where luminous AGNs contribute more significantly to 
$\rho_{\rm BHAR}$ than less-luminous AGNs 
\citep[e.g.,][]{ueda14}. 
In this work, we included the COSMOS field that probes  
more luminous AGNs than \goodss\ at a given redshift
(\S\ref{sec:cosmos}). 
Also, our methodology based on SMF and XLF provides 
further constraints on the luminous AGNs' contribution 
to $\bharbar$ (see \S\ref{sec:like_xlf}).

Since $\bharbar$ has a positive dependence on $\mstar$ in 
general, we expect more massive galaxies to have higher 
$\mbh$ in the local universe. 
However, the final $\mbh$ not only depends on $\mstar$ at 
$z=0$, but also on the stellar mass history 
[$\mstar(z)$].\footnote{\label{foot:Mstar_history}For example, 
consider two galaxies (dubbed ``A'' and ``B'') that both have 
$\log\mstar(z=0)=11.5$, but A forms earlier than B.
Assume at $z=2$, A and B have $\log\mstar=11$ and 10, 
respectively. 
Then A has much higher $\bharbar$ than B at $z=2$, 
according to Fig.~\ref{fig:BHAR_vs_M}. 
Therefore, the final $\mbh(z=0)$ of A will be higher than
that of B.}
Galaxy evolution is complex, e.g., high-$\mstar$ galaxies
tend to form earlier in cosmic history than low-$\mstar$ 
galaxies (``galaxy downsizing''; e.g., \citealt{cowie96}). 
In \S\ref{sec:mbh_mstar} and \S\ref{sec:mbh_mstar_z0}, we 
consider all these effects and predict typical $\mbh$-$\mstar$
relations at different redshifts.

\begin{figure}
\begin{center}
\includegraphics[width=\linewidth]{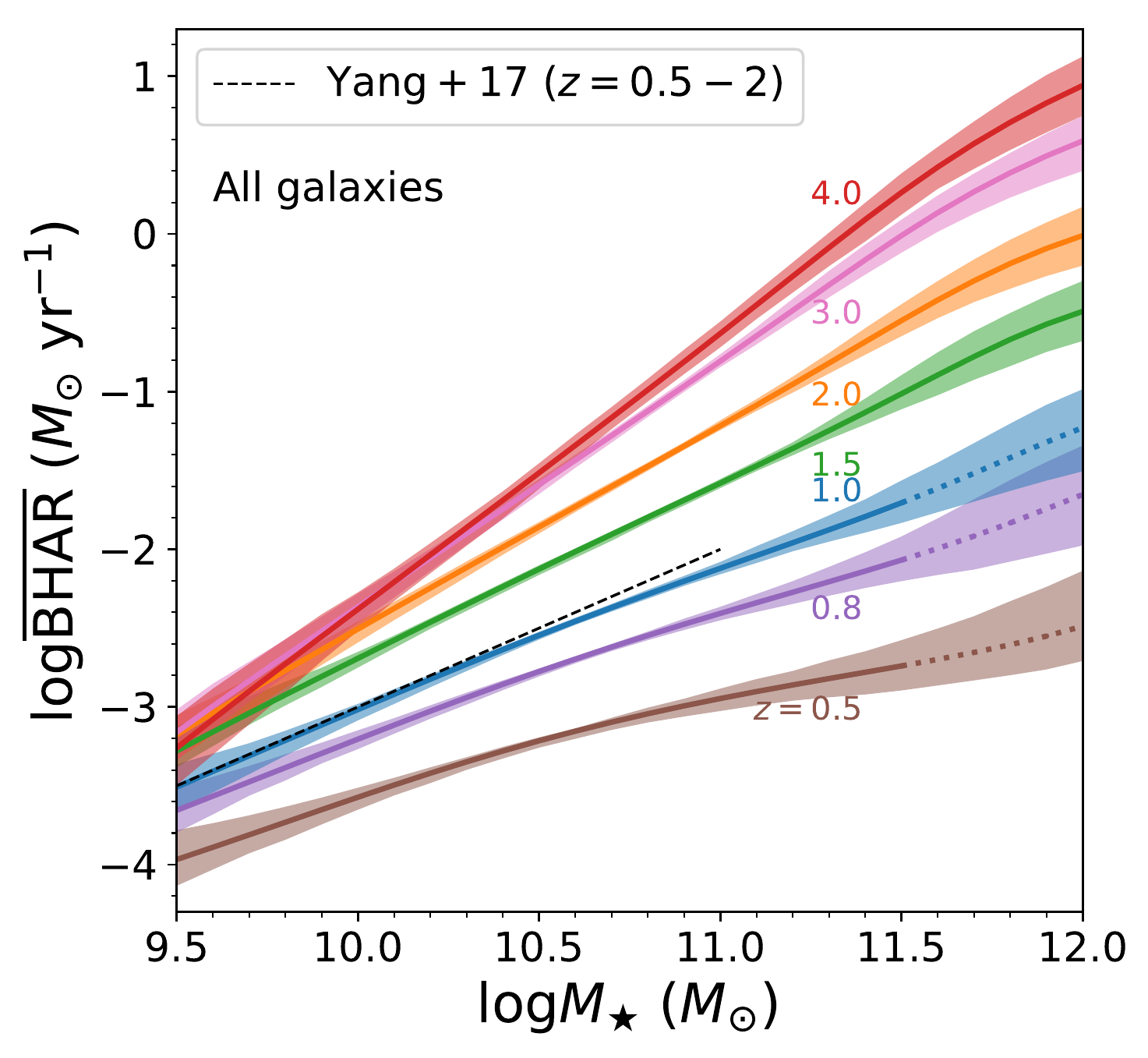}
\includegraphics[width=\linewidth]{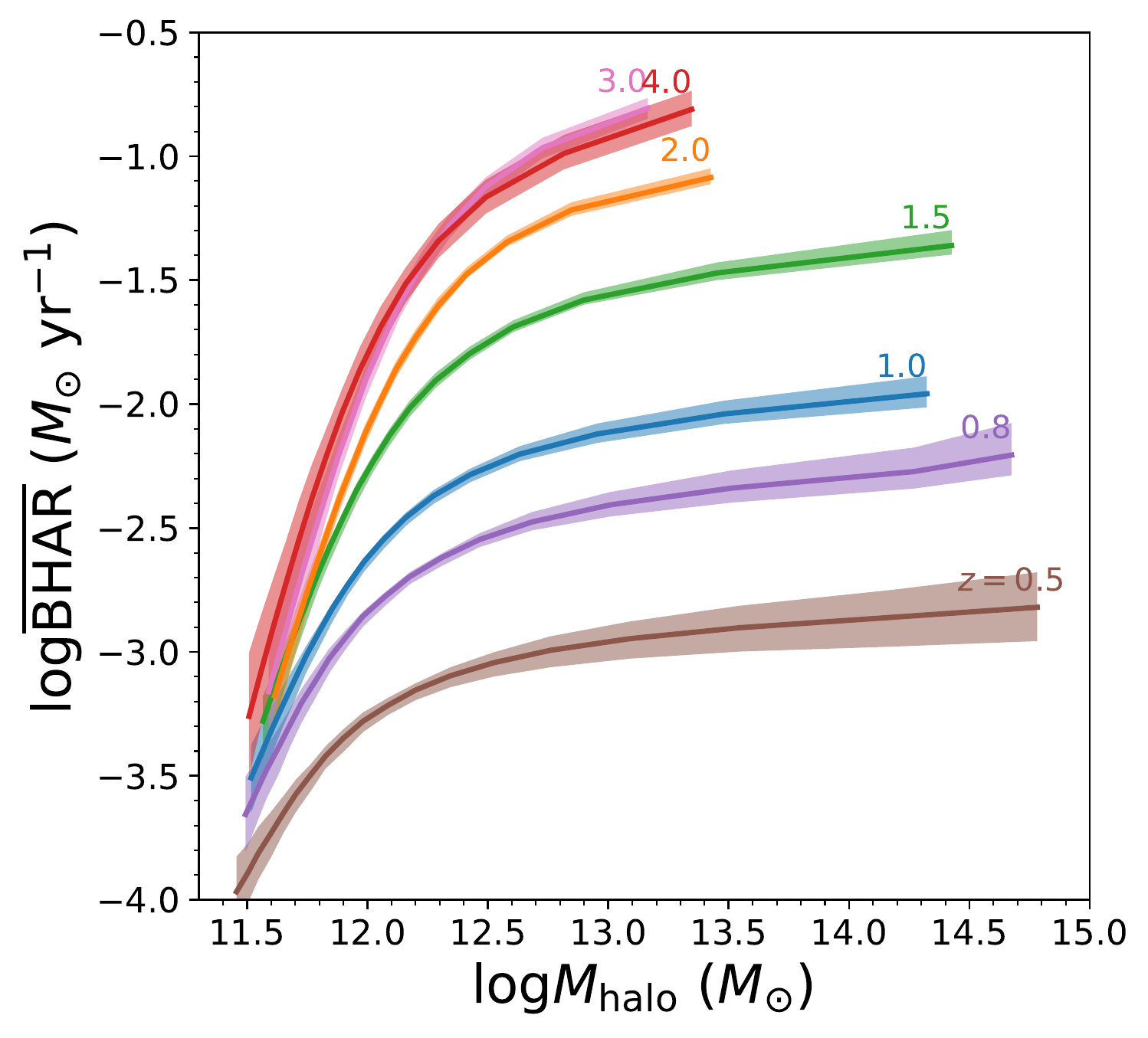}
\caption{Top panel: $\bharbar$ for all galaxies, 
derived from our best-fit $P(\slx|\mstar,z)$\ model, as 
a function of $\mstar$ and redshift. 
Different colors indicate different redshifts as 
labeled. 
The shaded regions indicate $1\sigma$ uncertainties 
derived from our MCMC sampling (see 
Appendix~\ref{sec:mdl_select}).
The dotted curves indicate the redshift and 
$\mstar$ regimes ($z\lesssim 1$ and 
$\log\mstar\gtrsim 11.5$) where $\bharbar$ might have larger 
uncertainties than estimated (see \S\ref{sec:kbol_lam} 
and Appendix~\ref{sec:fit_qual}).
The dashed black line indicates the best-fit relation 
from Eq.~6 of \citet{yang17} based on \goodss\ 
sources at $z=0.5\text{--}2.0$. 
Bottom panel: Same format as the top panel but for 
$\bharbar$ vs.\ $\mhalo$. 
$\bharbar$ is positively dependent on $\mhalo$ for 
$\log\mhalo \lesssim 12\text{--}13$, but the dependence 
becomes much weaker toward higher $\mhalo$.
There are no corresponding dotted curves 
(large $\bharbar$ uncertainties) in the bottom panel.
This is because, at $z\lesssim 1$, even for the highest 
$\mhalo$ shown, the corresponding $\log\mstar$ is below 
11.5.
}
\label{fig:BHAR_vs_M}
\end{center}
\end{figure} 

\subsubsection{Star-forming Galaxies}
\label{sec:bhar_SFonly}
The $\bharbar$-$\mstar$ slope in \S\ref{sec:bhar_for_all} 
becomes shallower toward low redshift.
This could be due to the fact that the quiescent galaxy 
population becomes increasingly significant at low redshifts 
in massive galaxies \citep[e.g.,][]{brammer11, tomczak14}, 
and that quiescent galaxies generally have weaker AGN 
activity compared to star-forming galaxies 
\citep[e.g.,][]{rosario13, aird17, wang17}.
To test this scenario, we derive $\bharbar(\mstar, z)$
for star-forming galaxies. 
We only use the survey-data constraints (\S\ref{sec:like_surv}), 
as the SMF-XLF constraints (\S\ref{sec:like_xlf}) 
are for the whole galaxy population rather than the 
star-forming subset.
 
\begin{figure}
\begin{center}
\includegraphics[width=\linewidth]{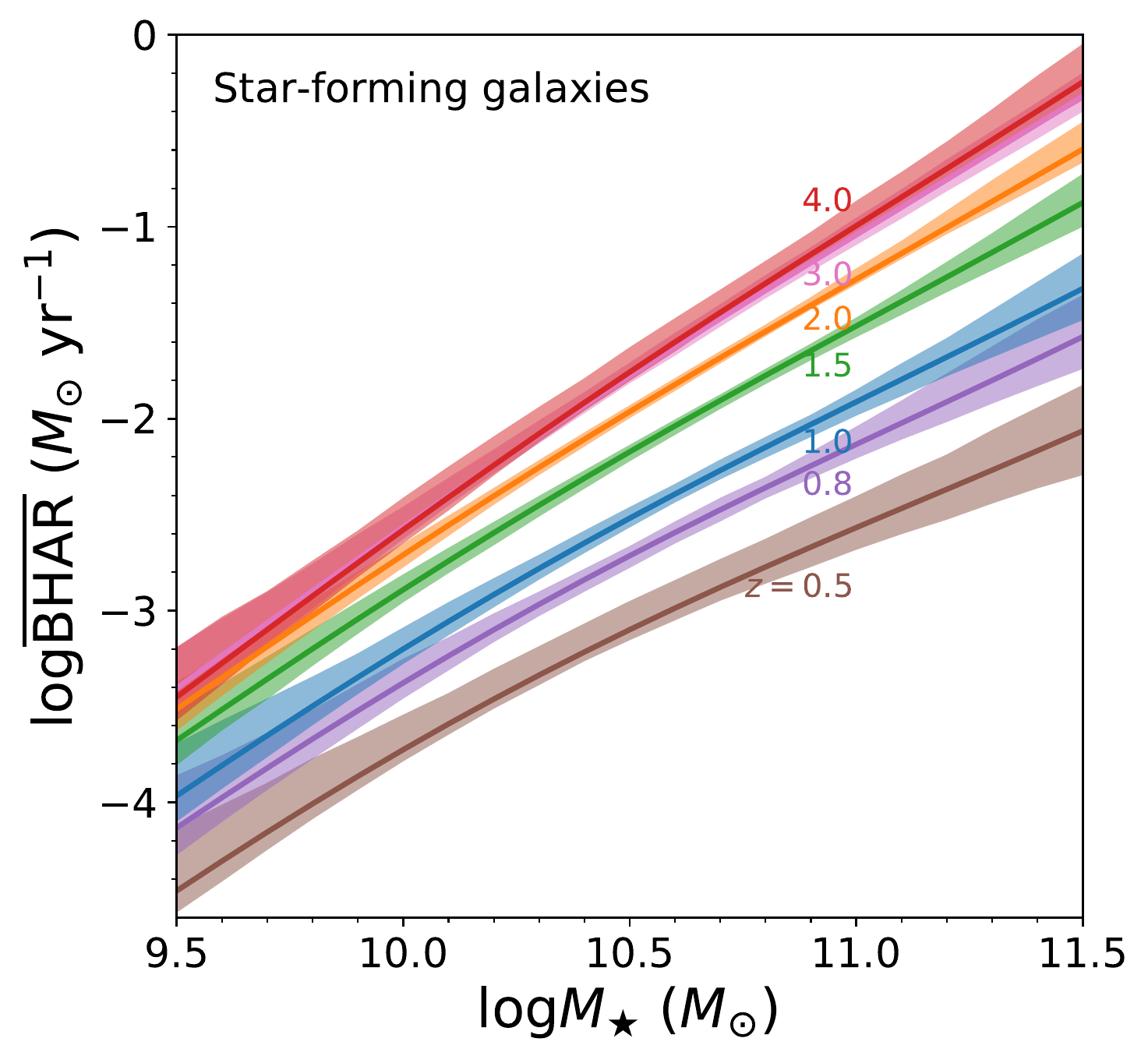}
\caption{Same format as Fig.~\ref{fig:BHAR_vs_M} (top) 
but for star-forming galaxies only.
We do not show $\log\mstar >11.5$ due to large 
uncertainties. 
The $\bharbar$ dependence on $\mstar$ 
does not become significantly weaker toward low 
redshift.
}
\label{fig:BHAR_vs_M_SFonly}
\end{center}
\end{figure}

We select star-forming galaxies using
the classifications in \S\ref{sec:surv}, which
are not applied to BL AGNs. 
We now discuss the star-forming/quiescent types for 
BL AGNs.
The star-forming population is dominant over the 
quiescent population for 
$\log\mstar \lesssim 10.5$ or $z\gtrsim 2$.
Therefore, we assign a BL AGN to be star-forming if it 
satisfies $\log\mstar \lesssim 10.5$ or $z\gtrsim 2$,
and this accounts for the majority population 
($\approx 70\%$) of the BL AGNs. 
For the other 30\% of BL AGNs, we include them in
our $\bharbar(\mstar, z)$ derivation considering two 
extreme scenarios.
The first scenario is that these 30\% of BL AGNs
are all star-forming \citep[e.g.,][]{zhang16}; the 
second is that they are all quiescent. 
The resulting $\bharbar(\mstar, z)$ for the two 
scenarios are similar at $\log\mstar \lesssim 11.5$
($\bharbar$ differences $\lesssim 0.3$~dex).
We thus only discuss the $\bharbar(\mstar, z)$ for 
star-forming galaxies at $\log\mstar \lesssim 11.5$. 
The results for the first scenario are displayed 
in Fig.~\ref{fig:BHAR_vs_M_SFonly}.

The $\bharbar$ dependence on $\mstar$ 
does not become significantly weaker toward low 
redshift.
Therefore, it is likely that the shallow slope of the
$\bharbar$-$\mstar$ relation for all galaxies in 
\S\ref{sec:bhar_for_all} is due to the increasing 
fraction of quiescent galaxies toward low redshifts 
and high $\mstar$.
However, one caveat is that our star-forming vs.\ quiescent 
classification scheme is largely empirical (\S\ref{sec:surv}).
Previous studies adopt various criteria for identifying 
star-forming galaxies
\citep[e.g.,][]{elbaz11, laigle16}, 
while the physical meanings of these bimodal classification 
schemes are still under debate \citep[e.g.,][]{feldmann17}.
Different classification methods can yield different 
$\bharbar$-$\mstar$ relations in Fig.~\ref{fig:BHAR_vs_M_SFonly}.

\subsubsection{Quiescent Galaxies}
\label{sec:bhar_quiescent}
We have tested deriving $\bharbar$ for quiescent 
galaxies. 
However, the uncertainties are generally large,
and thus we do not show the results. 
The large uncertainties are due to the limited sample 
size of quiescent galaxies, especially in the high-$z$
and low-$\mstar$ regime ($z\gtrsim 2$ and 
$\log\mstar\lesssim10.5$; see \S\ref{sec:bhar_SFonly}). 
At low redshifts and high $\mstar$, $\bharbar$ is also
significantly affected by whether to include  
BL AGNs as quiescent galaxies (see 
\S\ref{sec:bhar_SFonly}).

\subsection{Reliability Checks}

\subsubsection{Effects of \xray\ Obscuration}
\label{sec:xray_obs}
In \S\ref{sec:surv}, we derived $\lx$\ from observed-frame 
hard-band fluxes. 
However, even these measurements might suffer from obscuration, 
and thus the inferred $\lx$\ might be underestimated.
Also, we assume a photon index of 1.6 when deriving $\lx$,
and this choice could introduce additional uncertainties. 

To evaluate the accuracy of our $\lx$\ estimations, we 
compare them with those derived from XSPEC spectral 
fitting \citep[e.g.,][]{arnaud96}. 
This exercise is performed for \goodss\ sources for which 
we have $\lx$ from \xray\ spectral fitting (see \S2.3 in 
\citealt{yang17} for details of the spectral fitting). 
The intrinsic obscuration column density and photon index 
are free parameters in the fitting.
The $\lx$\ values derived from fluxes agree 
well with those derived from spectral fitting:
the systematic offset is $0.01$~dex, and 
the typical offset (the median of $|\Delta \log \lx|$)
is $0.09$~dex.
We do not find significant redshift or luminosity 
dependence of the systematic offset, indicating 
that our adopted $\lx$ values should also be reliable 
for COSMOS AGNs which are generally more luminous 
than \goodss\ AGNs. 

At $z>2$, assuming $\Gamma=1.8$ instead of 
$\Gamma=1.6$ leads to a systematic overestimation 
($\approx 0.1$~dex) of $\lx$ compared to that derived from 
spectral fitting, likely because $\Gamma$ appears generally 
lower at high redshift \citep[e.g.,][]{marchesi16b}.
At low redshift where the $k$ correction is smaller, the 
flux-based $\lx$ is only weakly dependent on $\Gamma$; 
for example, at $z=1$, the difference between $\lx$
produced by $\Gamma=1.6$ and $\Gamma=1.8$ is only 
0.04~dex.
The relatively low apparent value of $\Gamma$ (1.6) at high 
redshift might result from Compton reflection.  
At $z\gtrsim2$, the \xray\ observations can cover energies
above 20~keV where the Compton reflection component 
is strong. 
The reflection component is hard, and thus may cause 
the ``apparent'' $\Gamma$ value to be lower than the 
``intrinsic'' value.
However, due to the limited number of counts, it not feasible
to model properly the reflection component and obtain the
``intrinsic'' $\Gamma$. 
Our adopted $\Gamma=1.6$ is thus a practical approximation  
for the total \xray\ spectra rather than necessarily the 
intrinsic photon index for the transmission component. 

Most of the distant AGNs detected in \xray\ surveys are likely 
Compton-thin (CTN; $\nh<10^{24}$~cm$^{-2}$; e.g., \citealt{liu17}),
and the XLF we adopt does not include Compton-thick (CTK) AGNs.
Therefore, our derived $\bharbar(\mstar,z)$ does not account
for accretion contributed by CTK AGNs. 
Previous studies indicate that the CTK population 
is unlikely to be dominant over other AGNs 
and there is no evidence that the CTK fraction 
strongly depends on $\mstar$\ and/or $z$.
\citep[e.g.,][]{gilli07, treister09a, buchner15, ricci15, 
akylas16, baronchelli17}.
Thus, our $\bharbar(\mstar,z)$ should not be significantly 
affected by CTK AGNs. 

%

\subsubsection{Contamination from X-ray Binaries}
\label{sec:xrb_cont}
In the low-$\slx$\ regime, the \xray\ sources 
have a significant contribution of \xray\ emission 
from XRBs rather than AGNs. 
\xray\ emission from XRBs can be modeled 
as $L_{\rm X,XRB}=\alpha \mstar + \beta \mathrm{SFR}$,
where $\alpha$\ and $\beta$\ are functions of redshift. 
We adopt $\alpha$ and $\beta$ from model 269 of 
\citet{fragos13} which is preferred by the observations 
of \citet{lehmer16}.
Therefore, the $\slx$\ for XRBs can be written as
\begin{equation}\label{eq:xrb}
\begin{split}
L_{\rm SX,XRB} &= \frac{L_{\rm X,XRB}}{\mstar} \\
	       &= \frac{\alpha \mstar + \beta \mathrm{SFR}}{\mstar} \\
	       &= \alpha + \beta \mathrm{sSFR}.
\end{split}
\end{equation}

Since most star-forming galaxies have similar sSFR
at a given redshift (i.e., the star-forming ``main 
sequence''), the right-hand side of Eq.~\ref{eq:xrb}
only depends on redshift for star-forming galaxies.
For quiescent galaxies, the $L_{\rm SX,XRB}$\ 
is even lower than that of star-forming galaxies. 
We display the $L_{\rm SX,XRB}$\ for the main-sequence
galaxies as a function of redshift in 
Fig.~\ref{fig:Lx_vs_z} (bottom). 
The sSFR is from Eq.~13 of \citet{elbaz11}. 
The $L_{\rm SX,XRB}$\ increases toward higher redshift. 
Above our $\slx$\ threshold ($\log \slx=-2$; 
see \S\ref{sec:Lsx_cut}),
all the sources have $\slx$\ larger than $L_{\rm SX,XRB}$.
Even for the \goodss\ data (the deepest), the typical 
$\slx$ of our \xray\ AGNs is higher than $L_{\rm SX,XRB}$\
by an order of magnitude at any redshift. 
Also, almost all of our \xray\ sources above $\log \slx=-2$ 
in \goodss\ (98\%) and \goodsn\ (99\%) are classified as 
``AGN'' instead of ``galaxy'' by \citet{luo17} and \citet{xue16}, 
respectively. 
Therefore, our analyses should not be significantly 
affected by \xray\ emission from XRBs.

\subsubsection{Sample $\mstar$\ Completeness}
\label{sec:completeness}
\begin{figure}
\includegraphics[width=\linewidth]{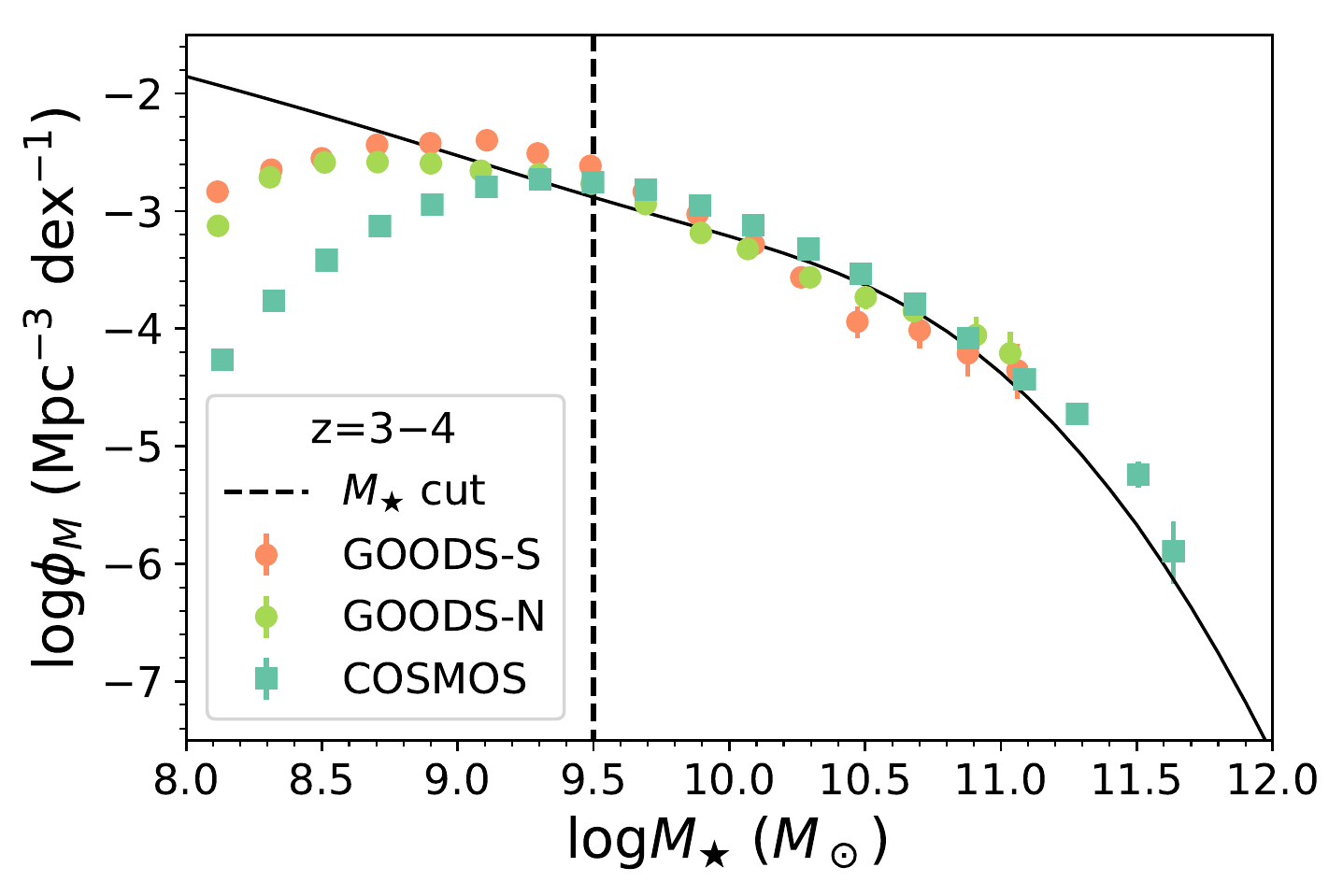}
\caption{Galaxy comoving number density as a function 
of $\mstar$ at $z=3.0\text{--}4.0$. 
The data points indicate values derived from our 
sample of different fields. 
We do not plot bins where only one source is available 
due to large uncertainties.  
The black solid curve represents the SMF from 
\citet{behroozi13}.
The vertical black dashed line indicates our 
$\mstar$ cut (see \ref{sec:m_cut}).
Our sources in all of the three fields are complete 
above our $\mstar$ cut.
}
\label{fig:smf_survey}
\end{figure}

To evaluate potential issues of $\mstar$\ completeness,
we derive the galaxy comoving number density as a function
of $\mstar$ for each field and compare with the SMF from 
\citet[][see \S\ref{sec:smf_xlf}]{behroozi13}.
The number density is calculated as the observed number of 
galaxies divided by the corresponding comoving volume. 
Fig.~\ref{fig:smf_survey} displays the results for 
$z=3\text{--}4$ (the highest redshift range probed in
this work). 
The number densities for all the three fields agree with 
the SMF above $\log\mstar=9.5$ (our $\mstar$\ cut; 
see \S\ref{sec:m_cut}), 
indicating our sample is complete for the $\mstar$ regime
probed in this work. 
The number density for COSMOS deviates from the SMF below 
$\log\mstar \approx 9$.
The number densities for \goodss\ and \goodsn\ both deviate 
from the SMF below $\log\mstar \approx 8.5$, as expected 
from their deeper imaging data than those of COSMOS. 

\subsubsection{Contribution to $\bharbar$ from Low-$\slx$ Accretion}
\label{sec:contr_low_Lsx}
Our derivation of $\bharbar$ does not include the 
contribution from $\log\slx<-2$ (see \S\ref{sec:Lsx_cut} 
and \S\ref{sec:bhar_for_all}). 
Although we have shown that this low-$\slx$ accretion cannot 
contribute significantly to the overall SMBH growth 
(see \S\ref{sec:Lsx_cut}), 
it can still affect the $\bharbar$-$\mstar$ relation, 
especially at low redshifts when the $\bharbar$ from 
high-$\slx$ accretion is low.

We here consider one extreme case in which 
$P(\slx|\mstar,z)$ is a narrow lognormal distribution 
centered at $\log\slx=-2.2$ with a width of $\pm 0.1$~dex;
then the $\bharbar$ contribution from $\log\slx<-2$ will 
be dominant over that from $\log\slx\geq-2$. 
This extreme case can be used to estimate the upper limit 
on $\bharbar$ contributed by $\log\slx<-2$. 
Quantitatively, we can estimate this upper limit as
\begin{equation}\label{eq:bhar_lowLsx}
\begin{split}
\bharbar &= \frac{\lx\kbol(1-\epsilon)}{\epsilon c^2} \\
	 &= \frac{\slx\kbol(1-\epsilon)}{\epsilon c^2} \mstar \\
	 &\approx 10^{-3.5} \frac{\mstar}{10^{10}M_\odot},
\end{split}
\end{equation}
where we adopt $\epsilon=0.1$ and $\kbol=10$ (see 
\S\ref{sec:Lsx_cut}). 
Only at low redshifts ($z\lesssim 0.8$), can our $\bharbar$ 
(see Fig.~\ref{fig:BHAR_vs_M}) be below this upper limit. 
Therefore, low-$\slx$ accretion can only be important at 
low redshifts.  

To examine whether low-$\slx$ accretion is actually 
dominant over high-$\slx$ accretion at low redshifts, we 
utilize \goodss\ where the deepest \xray\ observations and 
stacking data are available to probe low-$\slx$ accretion
\cite[e.g.,][]{vito16, yang17}.
We study sources in three $\mstar$ bins 
($\log\mstar=9.5\text{--}10$, 10--10.5, and 10.5--11) at 
$z=0.4\text{--}0.7$.
We do not probe higher $\mstar$ due to the limited number 
of galaxies (only 14 available). 
We use total $\lx$ to approximate total BHAR for each 
$\mstar$ bin.
For each $\mstar$ bin, we obtain the total $\lx$ for all 
sources in this bin, including both \xray\ detected 
and undetected sources (via stacking);
we also derive the contributions to total $\lx$ from 
$\log\slx\geq-2$ sources by summing the $\lx$ of 
detected \xray\ sources above $\log\slx=-2$. 
We find that, for all of the three $\mstar$ bins, the 
contributions from $\log\slx\geq-2$ sources 
account for $\gtrsim 80\text{--}90\%$ of the total $\lx$.
If we only consider the $\lx$ from AGNs by subtracting the 
expected XRB component (see Eq.~1 of \citealt{yang17}), 
almost all ($\gtrsim 99\%$) of the total $\lx$ from AGNs is 
from $\log\slx\geq-2$ sources.
The dominance of accretion at $\log\slx\geq-2$ indicates that 
the extreme lognormal $P(\slx|\mstar,z)$ distribution above
is unlikely to be physical, which is also suggested by some 
other studies \citep[e.g.,][]{aird17, georgakakis17}.

Therefore, we conclude that accretion from sources 
with $\log\slx<-2$, which is not included in our $\bharbar$
calculation, should not significantly affect our 
$\bharbar(\mstar, z)$ at all redshifts probed (i.e., 
$z=0.4\text{--}4$).

\subsubsection{Bolometric Correction}
\label{sec:kbol_lam}
The analyses in \S\ref{sec:res} assume 
$\kbol$ is a function of $\lx$. 
This luminosity-dependent $\kbol$ is generally larger 
for more massive galaxies, as they typically host AGNs 
with higher $\lx$ \citep[e.g.,][]{aird17, yang17}. 
However, $\kbol$ might be related to $\lam$ 
(see \S\ref{sec:k_bol}).

If we assume a $\kbol$-$\lam$ relation instead, 
then the $\kbol$ dependence on $\mstar$ will be weaker, 
because, compared to $\lx$, $\lam$ is likely much less 
dependent on $\mstar$ \citep[e.g.,][]{aird12, lusso12, 
aird17}.
We test an extreme case such that $\kbol$ is a constant 
value of 22.4 (the median of $\kbol$ values in the local
AGN sample of \citealt{vasudevan07}).
The resulting $\bharbar(\mstar, z)$ is broadly similar to 
that in Fig.~\ref{fig:BHAR_vs_M}.
$\bharbar$ generally has positive dependence on $\mstar$
and redshift, and the redshift dependence is stronger 
in more massive systems. 
However, we find that the $\bharbar$-$\mstar$ relation 
becomes flat at $z\lesssim 1.0$ and 
$\log \mstar \gtrsim 11.5$. 
This redshift and $\mstar$ regime is marked in 
Fig.~\ref{fig:BHAR_vs_M} (see also Appendix 
\ref{sec:fit_qual}).

\section{Discussion}\label{sec:discuss}

\subsection{Physical Causes of the $\bharbar$-$\mstar$\ 
Relation and its Cosmic Evolution}\label{sec:bhar_mstar}
At a given $\mstar$, $\bharbar$ is positively correlated 
with redshift out to $z\approx4$ (see Fig.~\ref{fig:BHAR_vs_M}). 
The physical reason could be that cold gas which 
fuels SMBH growth becomes more abundant toward high 
redshift \citep[e.g.,][]{mullaney12, popping12, vito14}.
At a given redshift, $\bharbar$ is generally higher in 
massive galaxies. 
As discussed in \S4.2 of \citet{yang17}, this positive 
dependence of $\bharbar$ on $\mstar$\ might be due to,
e.g., deeper potential wells \citep[e.g.,][]{bellovary13} 
and/or higher SMBH occupation fractions at high $\mstar$
\citep[e.g.,][]{volonteri10}. 

The positive dependence of $\bharbar$ on $\mstar$ becomes 
significantly weaker at low redshift, i.e., the 
$\bharbar$-$\mstar$ slope becomes shallower. 
One possible cause of the shallower $\bharbar$-$\mstar$ 
slope at low redshifts is that the fraction of star-forming galaxies,
which have generally stronger AGN activity than quiescent 
galaxies, decreases toward low redshift and high-$\mstar$.
We show that when only including star-forming galaxies, 
the $\bharbar$-$\mstar$ slope does not become significantly 
shallower at low redshifts (see \S\ref{sec:bhar_SFonly}).

The low $\bharbar$ of massive galaxies at low redshift is 
understandable considering the cosmic evolution 
of the SMF and XLF.
Luminous quasars ($\log\lx\gtrsim 44$), which are likely 
responsible for most of the SMBH growth in massive galaxies 
\citep[e.g.,][]{marconi04}, become much rarer toward the 
local universe (e.g., \citealt{schmidt68, ueda14}).
The number density of massive galaxies, however, becomes 
higher toward low redshift (see, e.g., Fig.~15 of 
\citealt{davidzon17}).
Therefore, the average SMBH accretion power for the massive
galaxy population decreases sharply toward low redshift. 
The strong redshift dependence of $\bharbar$ for massive 
galaxies is also evident in our survey data. 
The detected AGN fraction ($\log \slx >-2$) is $\approx 9\%$
among massive galaxies ($\log\mstar > 11$) at $z>2$, 
while the fraction is $\approx 5\%$ at $z<2$. 
Considering the larger incompleteness at higher redshift,
the intrinsic AGN fractions for $z>2$ and $z<2$ should have 
even larger differences.

The physical cause of this strong $\bharbar$ redshift 
evolution might be AGN feedback that could regulate the 
growth of SMBHs in massive galaxies. 
At high redshift, effective quasar-mode accretion could launch 
powerful winds that expel the cold gas in host galaxies 
\citep[e.g.,][]{king15a}.
Due to the lack of cold gas, SMBH growth drops significantly 
toward low redshift and hot-gas accretion occurs. 
The hot accretion flow could produce jets that prevent the 
gas from cooling and thereby maintain a low accretion rate 
\citep[e.g.,][]{croton06, yuan14}.

\begin{figure}
\begin{center}
\includegraphics[width=\linewidth]{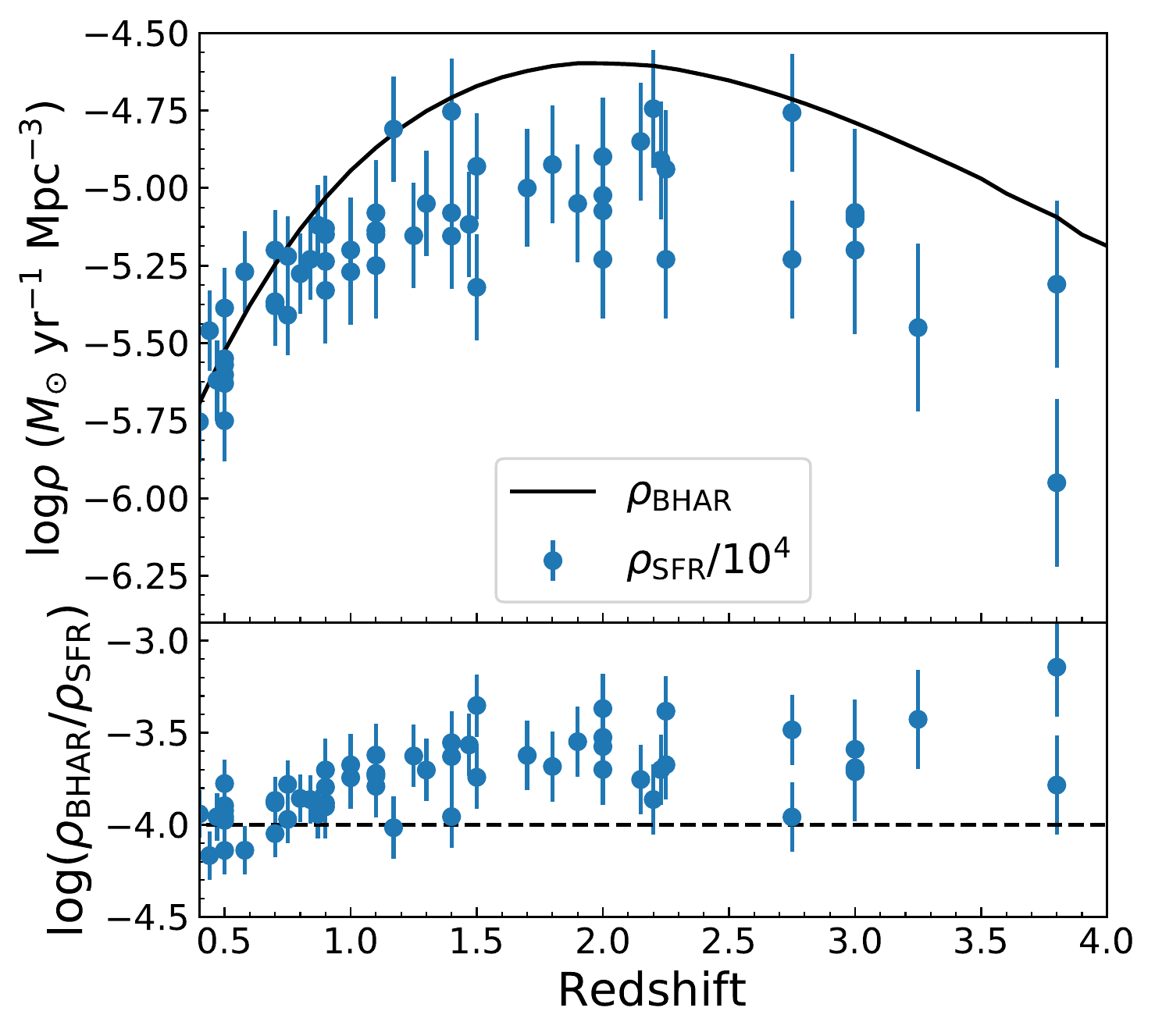}
\caption{Top panel: $\rho_{\rm BHAR}$
and $\rho_{\rm SFR}$ as a function of redshift. 
The black solid curve indicates $\rho_{\rm BHAR}$ derived 
from our best-fit model using Eq.~\ref{eq:rho_bhar};
the blue data points indicate $\rho_{\rm SFR}$ scaled by 
a factor of $10^{-4}$ \citep[][]{behroozi13}.
Bottom panel: the $\rho_{\rm BHAR}/\rho_{\rm SFR}$ ratio 
as a function of redshift. 
The black horizontal dashed line represents a constant 
ratio of $10^{-4}$.
$\rho_{\rm BHAR}/\rho_{\rm SFR}$ depends on redshift 
weakly. 
}
\label{fig:rho_bhar_sfr}
\end{center}
\end{figure}

The positive dependence of $\bharbar$ on redshift (especially 
at $z\gtrsim 2$) does not contradict the observational fact 
that $\rho_{\rm BHAR}$
peaks at $z\approx 2$ \citep[e.g.,][]{brandt15}. 
We calculate $\rho_{\rm BHAR}$ by convolving $\bharbar$ with
the SMF, i.e., 
\begin{equation}\label{eq:rho_bhar}
\begin{split}
\rho_{\rm BHAR}(z) &= \int^{12}_{9.5} \bharbar(\mstar,z)
		\phi_M(\mstar|z) d\log{\mstar}.
\end{split}
\end{equation}
Fig.~\ref{fig:rho_bhar_sfr} (top) displays the results. 
The $z\approx 2$ peak is successfully reproduced. 
$\bharbar$ describes the accretion power 
\textit{per galaxy} while $\rho_{\rm BHAR}$ characterizes 
the accretion power per comoving volume. 
Their different redshift evolution indicates that 
the drop of $\rho_{\rm BHAR}$ toward high redshift is 
driven by the evolution of the SMF:
the number density of galaxies with $\log\mstar\approx 
9.5\text{--}12$ decreases toward 
high redshift \citep[e.g.,][]{davidzon17}. 
A similar conclusion is also found by \citet{vito17}
who studied \xray\ AGNs at $z=3\text{--}6$.
Our $\rho_{\rm BHAR}$ curve has a similar shape as that of 
\citet[][see their Fig.~20]{ueda14}, but it is a factor of 
$\approx 3$ lower than theirs. 
The difference is primarily caused by the different 
$k_{\rm bol}$ adopted: our $k_{\rm bol}$ 
(from \citealt{lusso12}; 
\S\ref{sec:k_bol}) is $\approx 3$ times lower than their 
$k_{\rm bol}$ (from \citealt{hopkins07}) at 
$\log\lx \approx 44$ (the XLF break luminosity). 
For the reasons explained in \S\ref{sec:k_bol}, we believe
our adopted $k_{\rm bol}$ is more reliable than that 
from \citet{hopkins07}.

Some studies suggest that $\mbh$ might be fundamentally related
to the mass of the dark matter halo ($\mhalo$; e.g., 
\citealt{ferrarese02}; but also see \citealt{kormendy13}). 
We thus change the variable $\mstar$ to $\mhalo$ utilizing 
the typical redshift-dependent $\mhalo$-$\mstar$ relation 
from \citet[][see their Fig.~7]{behroozi13}, and convert 
$\bharbar(\mstar, z)$ to $\bharbar(\mhalo, z)$.
Fig.~\ref{fig:BHAR_vs_M} (bottom) displays the 
results.\footnote{\label{foot:behroozi}Due to the complicated 
evolution history of massive systems (e.g., frequent mergers), 
\citet{behroozi13} do not provide relations such as 
$\mhalo$-$\mstar$ and $\sfrbar$-$\mstar$ 
at very high $\mstar$ (P.\ Behroozi 2017, private 
communication).}
At $\log\mhalo\lesssim 12\text{--}13$, $\bharbar$ is 
positively dependent on $\mhalo$ and strongly so at
high redshifts.
At higher $\mhalo$, $\bharbar$ becomes relatively flat
as a function of $\mhalo$, indicating $\mhalo$ is not 
a good tracer 
of $\bharbar$ in massive systems \citep[e.g.,][]{kormendy13}.

To clarify the cause of this flatness, we express the 
slope of the $\bharbar$-$\mhalo$ relation as 
\begin{equation}
\label{eq:bhar_mhalo}
\frac{d\log\bharbar}{d\log\mhalo} = \frac{d\log\bharbar}{d\log\mstar} 
			    \frac{d\log\mstar}{d\log\mhalo}. 
\end{equation}
The slope of the $\mstar$-$\mhalo$ relation 
($d\log\mstar/d\log\mhalo$) 
is small ($\lesssim 0.3$) above 
$\log\mhalo\approx 12\text{--}13$ 
(see Fig.~7 of \citealt{behroozi13}), 
and this results in a small slope of
the $\bharbar$-$\mhalo$ relation ($d\log\bharbar/d\log\mhalo$),
even though the $\bharbar$-$\mstar$ relation can be steep 
(e.g., $d\log\bharbar/d\log\mstar\gtrsim 1$ at $z\gtrsim 1$; 
see Fig.~\ref{fig:BHAR_vs_M} top).
The physical cause of the flat $\mstar$-$\mhalo$ relation 
in massive systems is likely that gas cannot efficiently 
cool and collapse to form stars. 
This inefficient cooling might be due to AGN feedback 
and/or deep gravitational potential wells in massive halos 
(e.g., \citealt{rees77}; \S8.4 of \citealt{kormendy13}).

\subsection{BHAR vs.\ SFR}\label{sec:bhar_vs_sfr}
The redshift-evolution curves of $\rho_{\rm BHAR}$ and 
$\rho_{\rm SFR}$ are similar at least up to 
$z\approx2$ (e.g., \citealt{kormendy13, aird15}; 
Fig.~\ref{fig:rho_bhar_sfr}, which we will discuss in
detail below).
This similarity is considered a supporting point 
of the straightforward scenario that the long-term 
growth of SMBHs and their host galaxies are in 
lockstep. 
However, $\rho_{\rm BHAR}$ and $\rho_{\rm SFR}$ are quantities 
averaged over the whole galaxy population from low $\mstar$
to high $\mstar$, 
and their evolutionary similarity might not hold for galaxies
with different $\mstar$. 

\begin{figure}
\begin{center}
\includegraphics[width=\linewidth]{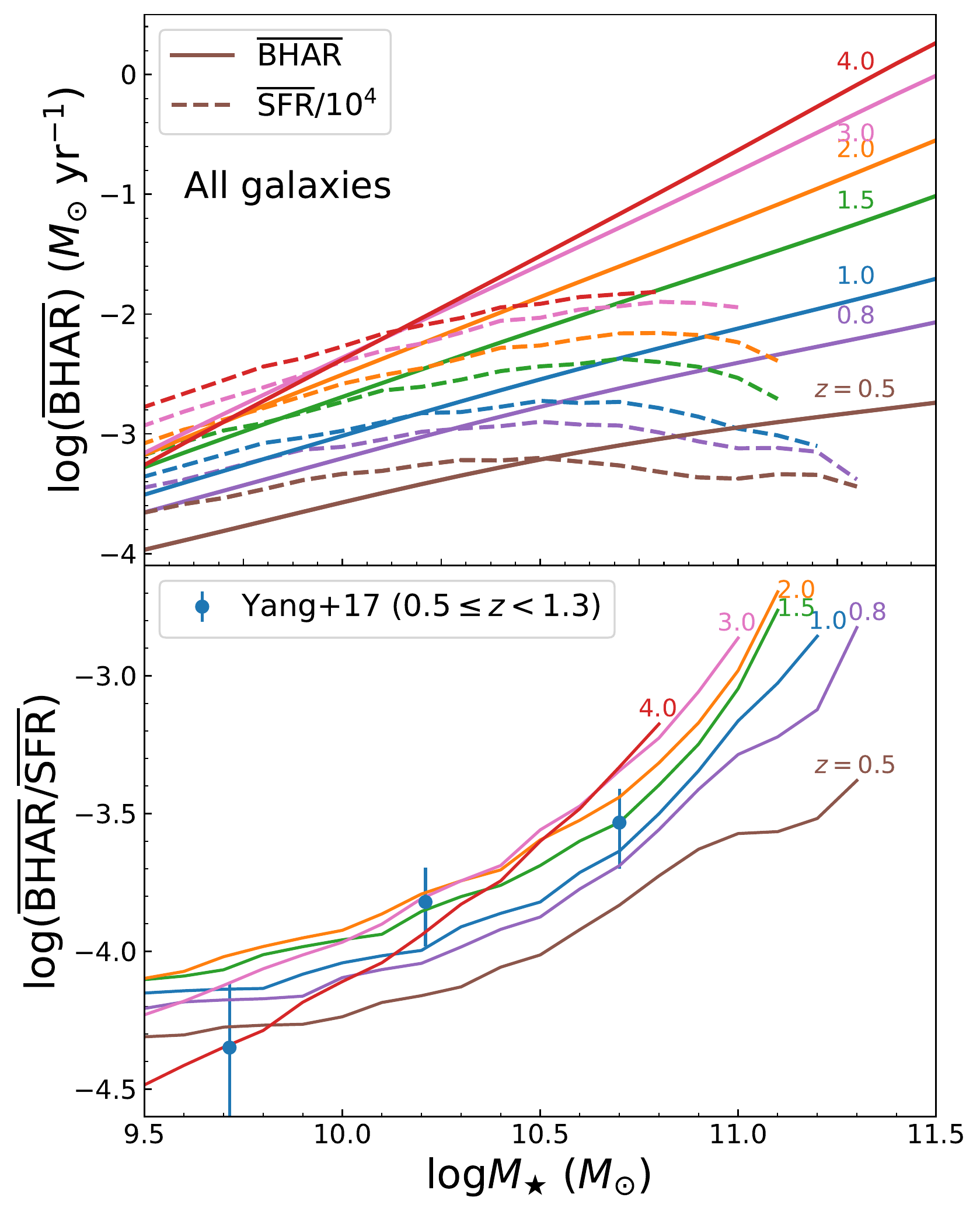}
\caption{Top panel: Comparison between $\bharbar$\ 
and $\sfrbar$\ as functions of $\mstar$. 
The solid lines indicate $\bharbar$; 
the dashed lines indicate $\sfrbar$ scaled by a factor
of $10^{-4}$.
Different colors indicate different redshifts as labeled.
Bottom panel: The ratio between $\bharbar$\ and $\sfrbar$ 
as functions of $\mstar$ at different redshifts. 
The purple data points indicate the measured 
$\bharbar/\sfrbar$ values from \citet{yang17}.
At $z\gtrsim 0.5$, $\bharbar/\sfrbar$ rises toward high
$\mstar$; it is flat as a function of $\mstar$ at lower 
redshift. 
}
\label{fig:BHAR_SFR_vs_M}
\end{center}
\end{figure}

To investigate this possibility, we compare the 
$\bharbar$-$\mstar$ relation with the $\sfrbar$-$\mstar$
relation from \citet{behroozi13} in 
Fig.~\ref{fig:BHAR_SFR_vs_M} (top).
The $\sfrbar$-$\mstar$ relation is truncated at high $\mstar$
due to the reason in Footnote~\ref{foot:behroozi}.
We note that the $\sfrbar$-$\mstar$ relation represents 
an ensemble-averaged property (the same as for the 
$\bharbar$-$\mstar$ relation), and thus it is not affected by 
the SFR variability of individual galaxies. 
We normalize the $\sfrbar$ values by a factor of $10^{-4}$
so that they become comparable with $\bharbar$. 
Fig.~\ref{fig:BHAR_SFR_vs_M} (bottom) presents the 
$\bharbar/\sfrbar$ ratio as a function of $\mstar$.

At a given $\mstar$, both $\bharbar$ and $\sfrbar$
rise toward high redshifts in general.
However, $\bharbar/\sfrbar$ has relatively weak redshift 
evolution, especially at $z\gtrsim0.8$. 
At a given redshift, $\bharbar$ rises more strongly 
as a function of $\mstar$ than $\sfrbar$.
As a result, the $\bharbar/\sfrbar$ ratio is positively 
correlated with $\mstar$ at a given redshift, broadly 
consistent with the findings of \citet[][the data points 
in Fig.~\ref{fig:BHAR_SFR_vs_M} bottom]{yang17}.
This positive dependence disfavors the straightforward 
coevolution model where SMBH and galaxy growth 
are in lockstep and $\bharbar/\sfrbar$ is a universal 
constant \citep[e.g.,][]{hickox14}. 
As explained in \S4.2 of \citet{yang17}, the larger 
$\bharbar/\sfrbar$ toward high $\mstar$ may indicate that
massive systems are more effective in feeding cold gas
to their central SMBHs due to, e.g., their deep potential 
wells.
Another possibility is that the SMBH occupation fraction
increases toward high $\mstar$. 
Our $\bharbar/\sfrbar$ agrees with the sample of 
\citet[][]{yang17} at $0.5 \leq z < 1.3$ (the red points 
in Fig.~\ref{fig:BHAR_SFR_vs_M}).
We do not compare with their high-redshift 
($1.3 \leq z < 2.0$) sample, because luminous AGNs 
($\log\lx\gtrsim 44$) are dominant at high redshift 
but almost absent in their sample. 
This rarity of luminous AGNs is due to the small area of 
\goodss\ used in \citet{yang17} and the removal of BL AGNs 
from their sample.

Fig.~\ref{fig:rho_bhar_sfr} compares our $\rho_{\rm BHAR}$ 
(\S\ref{sec:bhar_mstar}) with $\rho_{\rm SFR}$ from the 
observational data compiled by \citet{behroozi13}. 
The $\rho_{\rm BHAR}$ and $\rho_{\rm SFR}$ curves have similar
shapes consistent with previous work \citep[e.g.,][]{kormendy13, 
aird15}. 
The $\rho_{\rm BHAR}/\rho_{\rm SFR}$ ratio only rises slightly
($\approx 0.3$~dex) from $z=0$ to $z\gtrsim 3$. 
We note that this weak redshift dependence does not contradict 
our conclusion that $\bharbar/\sfrbar$ for massive galaxies 
($\log\mstar \gtrsim 11$) rises strongly toward high redshift 
(see Fig.~\ref{fig:BHAR_SFR_vs_M}).
This is because massive galaxies are relatively rare and their
weight in $\rho_{\rm BHAR}/\rho_{\rm SFR}$ is not dominant
over that of less-massive galaxies. 

\begin{figure}
\begin{center}
\includegraphics[width=\linewidth]{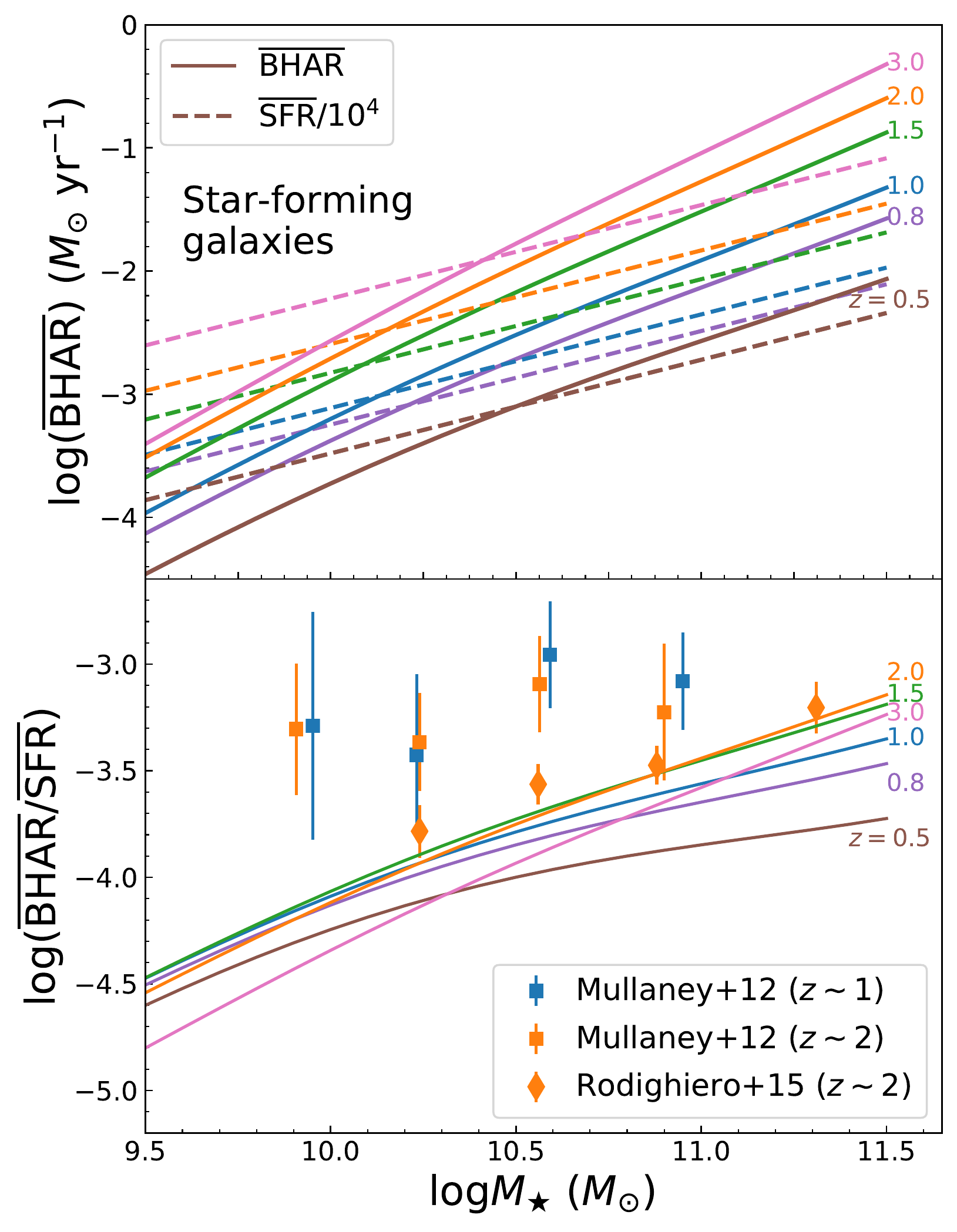}
\caption{Same format as Fig.~\ref{fig:BHAR_SFR_vs_M} 
but for star-forming galaxies only. 
The data points in the bottom panel indicate measurements 
from \citet{mullaney12, rodighiero15}.
}
\label{fig:BHAR_SFR_vs_M_SFonly}
\end{center}
\end{figure}

Fig.~\ref{fig:BHAR_SFR_vs_M_SFonly} compares the
$\bharbar$-$\mstar$ and $\sfrbar$-$\mstar$ relations for 
star-forming galaxies. 
The $\sfrbar$ vs.\ $\mstar$ are from the star-forming 
main-sequence relations (Eq.~6 of \citealt{aird17b}). 
We do not compare these relations above $z=3$, as the 
uncertainties of the main-sequence relations are large 
at $z\gtrsim 3$. 
At a given redshift, the $\sfrbar$-$\mstar$ relation
generally has a slope shallower than the $\bharbar$-$\mstar$ 
relation. 
This leads to $\bharbar/\sfrbar$ rising toward 
high $\mstar$ (Fig.~\ref{fig:BHAR_SFR_vs_M_SFonly} bottom),
similarly to the trend in Fig.~\ref{fig:BHAR_SFR_vs_M} (bottom).
Also similarly to Fig.~\ref{fig:BHAR_SFR_vs_M} (bottom), 
the $\bharbar/\sfrbar$ has weak redshift evolution at 
a given $\mstar$ above $z\gtrsim 0.8$.
The $\bharbar/\sfrbar$ curve in Fig.~\ref{fig:BHAR_SFR_vs_M_SFonly}
is concave down while that in Fig.~\ref{fig:BHAR_SFR_vs_M}
is concave up.

Fig.~\ref{fig:BHAR_SFR_vs_M_SFonly} also shows the 
$\bharbar/\sfrbar$ measurements from the literature. 
The $\bharbar/\sfrbar$ values from \citet{mullaney12} 
are higher than ours, possibly due to their large 
uncertainties. 
However, the observations of \citet{rodighiero15} 
at $z\sim 2$ have much smaller uncertainties and 
agree with our $\bharbar/\sfrbar$.
Since these measurements are for star-forming galaxies,
they are not displayed in Fig.~\ref{fig:BHAR_SFR_vs_M},
which is for all galaxies.

\subsection{Evolution of The $\mbh$-$\mstar$ Relation}\label{sec:mbh_mstar}
Our $\bharbar(\mstar, z)$\ describes the long-term average BHAR
for a galaxy at given $\mstar$\ and $z$. 
Based on $\bharbar(\mstar, z)$, we can calculate the  
cumulative accreted SMBH mass for galaxies with given 
stellar mass history 
[$\mstar(z)$], i.e., 
\begin{equation}
\label{eq:mbh_z}
\mbh(z) = \int^{z}_{4} \bharbar(\mstar(z'),z')
	   \frac{dt}{dz'} dz' + \mbh|_{z=4},
\end{equation}
where $t$\ is the cosmic time; the prime superscript in $z'$\ 
is to differentiate the redshift $z$ as the integral upper 
boundary.
We adopt the stellar mass history of \citet{behroozi13} 
who derived the average $\mstar(z)$\ for galaxies 
as a function of $\mstar(z=0)$\ up to 
$\log\mstar(z=0)\approx11.4$\ (see their Fig.~9).
To obtain the initial condition $\mbh|_{z=4}$, we test 
different $\mbh/\mstar|_{z=4}$ values and multiply them with
$\mstar(z=4)$ from \citet{behroozi13}. 

When $\mstar(z)$ lies below our limit of $\log\mstar=9.5$
in Eq.~\ref{eq:mbh_z} (i.e., our $M_\star$ cut; see 
\S\ref{sec:m_cut}), we assume their $\bharbar$ as 
$\bharbar(\log\mstar=9.5,z) \times (\mstar/10^{9.5})^\zeta$.
We set the power-law index $\zeta=1$, i.e., assuming 
$\bharbar$\ at $\log\mstar \leq 9.5$ is proportional to 
$\bharbar$ 
at $\log\mstar=9.5$ \citep[e.g., \S3.2 of ][]{yang17}.
To reduce the uncertainties of this approximation, we do not 
discuss the $\mbh$-$\mstar$ relation for 
$\log\mstar < 10$ at all redshifts.\footnote{We still need 
to assume $\bharbar$ for $\log \mstar \leq 9.5$, even though
we do not probe $\log\mstar < 10$. For example, a galaxy 
with $\log\mstar(z=0)=10.5$ has $\log\mstar<9.5$ in the early 
universe ($z>2$), and we need to account for its
SMBH accretion at $z=2\text{--}4$.}
In fact, we find that setting $\zeta$ to 0.5 or 2 
does not affect our conclusions.
For redshifts below our cut ($z=0.4$; 
\S\ref{sec:z_cut}), we use $\bharbar$ from model extrapolation. 
This should not bring noticeable uncertainties given 
that $\bharbar$ is low at $z<0.4$. 
In fact, we have tested two extreme scenarios:
$\bharbar(\mstar,z<0.4)=\bharbar(\mstar,z=0.4)$ 
(no redshift evolution at $z<0.4$)  
and $\bharbar=0$ (no SMBH growth at $z<0.4$), 
respectively.
Both of these cases produce almost the same
$\mbh$-$\mstar$ relation as that from model 
extrapolation. 

In Eq.~\ref{eq:mbh_z}, we assume long-term BHAR can be 
inferred from $\mstar$ and redshift. 
This assumption will not be strictly correct, if,
for example, $\bharbar$ also depends on SFR at 
given $\mstar$ and redshift. 
However, this SFR dependence is likely weak
based on previous work \citep[e.g.,][]{stanley17, 
yang17}, and significantly different $\bharbar$ might
only be found between galaxies with extremely 
different SFRs, e.g., star-forming vs.\ 
quiescent galaxies \citep[e.g.,][]{aird17, wang17}. 
In the case of weak SFR dependence, our ensemble 
analyses are still valid in characterizing the 
typical $\mbh$-$\mstar$ relation for the entire 
galaxy population.

\begin{figure*}
\begin{center}
\includegraphics[width=\linewidth]{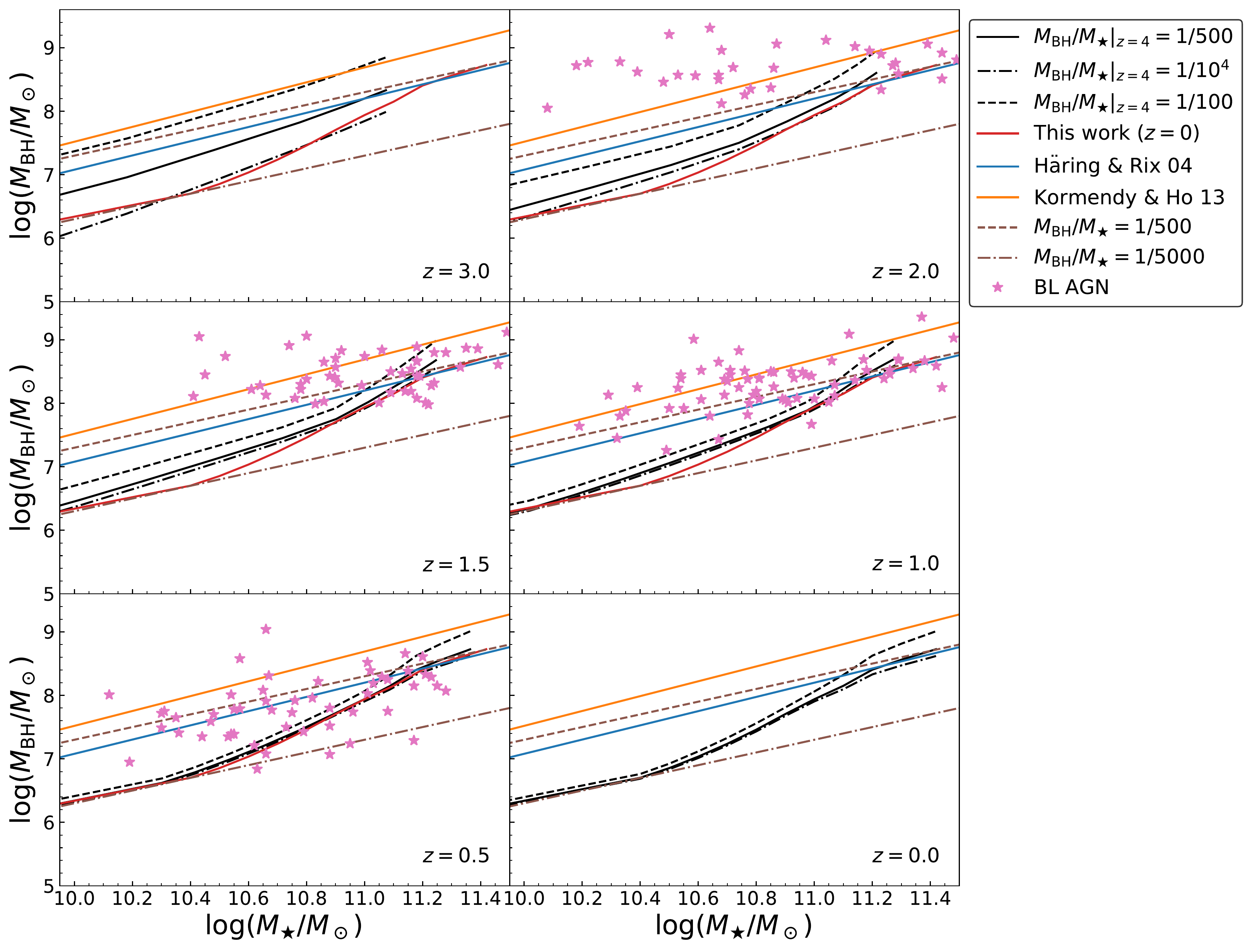}
\caption{SMBH mass vs.\ stellar mass at different redshifts. 
The black solid, dot-dashed, and dashed lines indicate our 
best-fit model assuming $\mbh/\mstar|_{z=4}=1/500$, $1/100$,
and $1/10^4$, respectively.
The MCMC 1$\sigma$ uncertainties are small ($\lesssim 0.1$~dex) 
and thus are not shown.
The red solid line represents our $\mbh\text{--}\mstar$ 
relation at $z=0$ assuming $\mbh/\mstar|_{z=4}=1/500$.
The brown dashed and dot-dashed lines are $\mbh/\mstar=1/500$
and $\mbh/\mstar=1/5000$, respectively;
the blue and orange lines indicate the 
$\mbh\text{--}\mbulge$\ relations in the local universe from 
\citet{haring04} and \citet{kormendy13}, respectively.
The pink stars identify BL AGNs from the literature
(see \S\ref{sec:mbh_mstar}). 
Our $\mbh$-$\mstar$ relation is not strongly affected
by the assumptions about $\mbh/\mstar|_{z=4}$ at $z\lesssim 2$.
The $\mbh$-$\mstar$ relation has weak redshift evolution 
since $z\approx 2$, and it is similar to the local
$\mbh\text{--}\mbulge$\ relations at high $\mstar$ where 
$\mstar \approx \mbulge$. At low $\mstar$, our $\mbh$-$\mstar$ 
relation deviates from the $\mbh\text{--}\mbulge$\ relations 
likely because the bulge is less dominant toward low $\mstar$. 
}
\label{fig:Mbh_vs_Mstar}
\end{center}
\end{figure*}

We show the resulting $\mbh$ as a function of $\mstar$ at 
different redshifts in Fig.~\ref{fig:Mbh_vs_Mstar}.
The $\mbh$-$\mstar$ curves are truncated at the highest 
$\mstar(z)$ values in \citet[][see 
Footnote~\ref{foot:behroozi}]{behroozi13}.
As expected, the effects of the $\mbh/\mstar|_{z=4}$ assumption
generally become weaker toward low redshift.
At $z\lesssim 2$, the $\mbh/\mstar|_{z=4}=1/100$ (black dashed)
and $\mbh/\mstar|_{z=4}=1/10^4$ (black dash-dotted) curves 
only differ from the $\mbh/\mstar|_{z=4}=1/500$ (black solid)
curve by $\lesssim 0.3$~dex in vertical direction. 
Thus, below we only discuss the $\mbh$-$\mstar$ relations at 
$z \leq 2$.

From Fig.~\ref{fig:Mbh_vs_Mstar}, the $\mbh$-$\mstar$ 
relation only has weak redshift evolution since 
$z \approx 2$ (solid black vs.\ red curves).
This result does not contradict Fig.~\ref{fig:BHAR_SFR_vs_M} 
which shows that $\bharbar$ and $\sfrbar$ have strong redshift 
dependence.
This is because $\bharbar$ and $\sfrbar$ are both low at 
$z \lesssim 0.8$ compared to at higher redshift. 
For example, at a given $\mstar$, the $\bharbar$ ($\sfrbar$) 
at $z=0.5$ is lower than $\bharbar$ ($\sfrbar$) at $z=3$ by 
a factor of $\approx 10$.
Therefore, the SMBH and galaxy growth at $z\lesssim 0.8$ 
are not important, and the $\mbh$-$\mstar$ relation is thus
primarily determined by $\bharbar/\sfrbar$ at higher redshift. 
At $z\gtrsim 0.8$, the $\bharbar/\sfrbar$ does not have 
strong redshift dependence at a given $\mstar$. 
The weak redshift dependence of the $\mbh$-$\mstar$ relation 
is consistent with the observations of \citet{schindler16} 
which show that the ratio between SMBH and stellar mass 
densities is flat as a function of redshift since $z\approx 5$.
Based on observations of BL AGNs, \citet{sun15} also 
found a redshift-independent $\mbh$-$\mstar$ relation 
(but also see \citealt{merloni10}).

Since massive galaxies tend to form at high redshifts 
(\S\ref{sec:res}), our result 
suggests that their SMBHs should follow a similar trend. 
The massive galaxies ($\log\mstar \gtrsim 11$) at $z \approx 2$
are likely the progenitors of local giant ellipticals for 
which $\mstar=\mbulge$.
Their $\mbh$ values are already at the level expected from 
the local $\mbh$-$\mbulge$ relation \citep[e.g.,][]{haring04,
kormendy13}. 
This conclusion disfavors the scenario that the local $\mbh$-$\mbulge$ 
relation for giant ellipticals mostly formed at low redshift,
as proposed in \S4.3 of \citet{yang17}.
Their scenario assumes their $\bharbar$-$\mstar$ relation 
(see Fig.~\ref{fig:BHAR_vs_M}) extends to $\log\mstar \gtrsim 11$
and has weak redshift evolution at $z\lesssim 2$, which 
is inconsistent with our results. 

The $\mbh$-$\mstar$ relations at $z\lesssim 2$ differ 
significantly from those expected from a constant 
$\mbh/\mstar$ ratio (the brown lines in 
Fig.~\ref{fig:Mbh_vs_Mstar}).
The $\mbh/\mstar$ ratio is positively related to $\mstar$:
it is $\sim 1/5000$ at $\log\mstar \lesssim 10.5$ and rises
to $\sim 1/500$ at $\log\mstar \gtrsim 11.2$.
This positive dependence is predicted by \citet[][see their 
\S4.3]{yang17} based on the fact that $\bharbar/\sfrbar$
is generally higher toward high $\mstar$ (see 
Fig.~\ref{fig:BHAR_SFR_vs_M}). 

Fig.~\ref{fig:Mbh_vs_Mstar} also compares our 
$\mbh$-$\mstar$ relation with the observations of BL AGNs
at $z \gtrsim 0.5$
\citep[][]{jahnke09, merloni10, bennert11, cisternas11, 
schramm13, sun15}.
At $\log\mstar \gtrsim 11$, these observations agree 
well with our $\mbh$-$\mstar$ relation. 
However, at lower $\mstar$, the $\mbh$ values for BL AGNs
are significantly higher than those predicted by our 
$\mbh$-$\mstar$ relation.
The discrepancies might be caused by selection biases
and/or $\mbh$ measurement uncertainties. 
Due to observational sensitivity, the BL AGNs with $\mbh$
measured tend to have massive SMBHs, and thus do not
well represent the whole galaxy population at given $\mstar$
\citep[e.g.,][]{lauer07}.
The measurements of $\mbh$ are based on single-epoch spectra,
and large uncertainties (both scatter and biases) exist 
in these measurements \citep[e.g.,][]{shen13}.
In fact, in the local universe where $\mbh$ can be measured 
for BL AGNs with lower luminosities and normal galaxies, 
the observations agree better with our $\mbh$-$\mstar$ 
relation (see \S\ref{sec:mbh_mstar_z0}).

We do not calculate $\mbh$ vs.\ $\mstar$ for 
star-forming galaxies due to the lack of their average 
stellar mass history (i.e., the needed $\mstar(z)$ is 
not provided by \citealt{behroozi13} or other works 
to our knowledge), and we only make qualitative 
arguments below. 
Since the cosmic evolution of $\bharbar/\sfrbar$ for
star-forming galaxies is weak at $z\gtrsim 0.8$ (see 
\S\ref{sec:bhar_vs_sfr}), their $\mbh$-$\mstar$ relation 
should also have weak dependence on redshift (as 
explained above for all galaxies).
The $\mbh/\mstar$ likely rises in more massive galaxies 
where $\bharbar/\sfrbar$ is higher. 
These qualitative conclusions for star-forming galaxies 
are similar to those for all galaxies 
(Fig.~\ref{fig:Mbh_vs_Mstar}).

\subsection{The $\mbh$-$\mstar$ Relation at $z=0$}
\label{sec:mbh_mstar_z0}
\begin{figure*}
\begin{center}
\includegraphics[width=\linewidth]{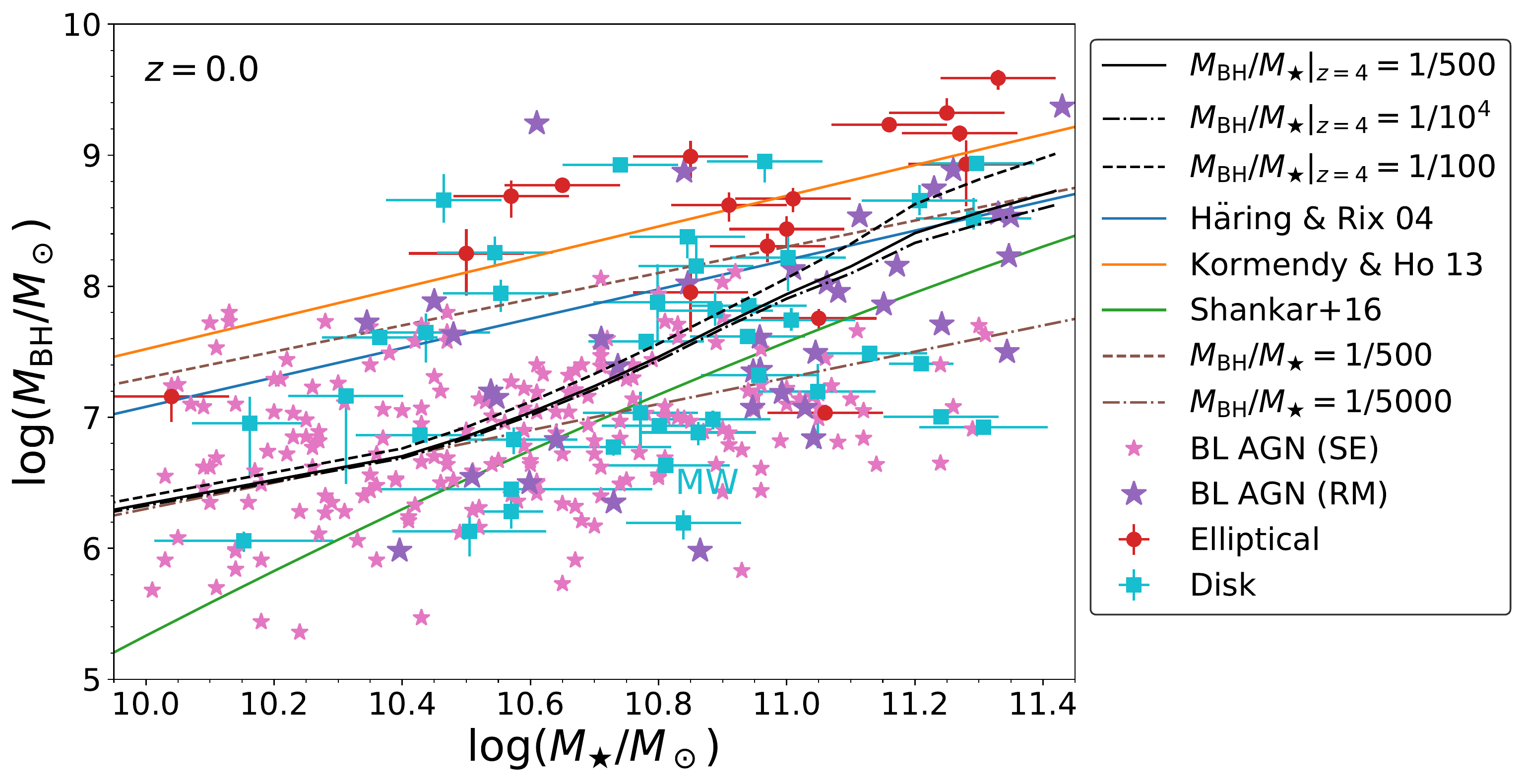}
\caption{Same format as the bottom-right panel in 
Fig.~\ref{fig:Mbh_vs_Mstar}, but plotted against data points
of local systems. 
The red points and blue squares indicate elliptical and 
disk galaxies, respectively, with $\mbh$ measurements 
from stellar dynamics or megamasers. 
The disk galaxies include S0 and spiral galaxies.
The data point representing the Milky Way is labeled as 
``MW'' at its lower right. 
The pink and purple stars represent local BL AGNs 
based on $\mbh$ measurements of single-epoch spectra and 
reverberation mapping, respectively. 
The blue and orange lines indicate the 
$\mbh\text{--}\mbulge$\ relations from \citet{haring04} and 
\citet{kormendy13}, respectively;
the green solid line represents the local ``unbiased''
$\mbh\text{--}\mstar$\ relation from \citet{shankar16}. 
}
\label{fig:Mbh_vs_Mstar_z0}
\end{center}
\end{figure*}

Fig.~\ref{fig:Mbh_vs_Mstar_z0} compares our $\mbh$-$\mstar$ 
relation at $z=0$ with observations.
At $\log\mstar \gtrsim 11$ where $\mstar \approx \mbulge$, 
our derived $\mbh$\ is similar to that expected from the 
$\mbh$-$\mbulge$ relations in
\citet{haring04}\ and \citet{kormendy13}. 
However, at lower $\mstar$, our $\mbh$ is significantly lower 
compared to that expected from the $\mbh$-$\mbulge$ relations.
This result is expected, because the bulge components become less 
dominant and $\mbulge/\mstar$ drops toward low $\mstar$ 
\citep[e.g.,][]{kelvin14}.

SMBH growth can occur via both accretion and mergers.
Some studies propose that mergers could be the main 
channel of SMBH growth for local giant ellipticals, 
and their $\mbh$-$\mstar$ relation might be a natural 
result of merger averaging (e.g., 
\citealt{peng07, jahnke11}; \S8.5 of \citealt{kormendy13}). 
However, the similarity between our $\mbh$-$\mstar$
relation and the $\mbh$-$\mbulge$ relations 
at high $\mstar$ suggests that accretion is sufficient
to produce the observed 
$\mbh$-$\mstar$\ relation in local giant ellipticals, 
since merger growth is not included in our $\mbh$ 
calculation (Eq.~\ref{eq:mbh_z}).\footnote{Our 
scheme of $\mbh$ calculation in Eq.~\ref{eq:mbh_z} 
is different from the So{\l}tan argument \citep{soltan82}.
The So{\l}tan argument considers $\mbh$ density which 
is not affected by mergers. 
However, our $\mbh$ calculation aims to 
characterize the average behavior for individual galaxies. 
Both accretion and mergers can contribute to the $\mbh$
for individual galaxies, although we only consider the 
accretion component in Eq.~\ref{eq:mbh_z}.}
Also, merger averaging of low-$\mstar$ galaxies can only 
lead to the final $\mbh/\mstar$ ratio being farther from 
that observed in giant ellipticals, because $\mbh/\mstar$ 
is lower toward low $\mstar$.  
This argument can also apply to the giant ellipticals 
beyond our probed $\mstar$ limit ($\log\mstar\approx11.4$).
Therefore, merger growth might not dominate over accretion 
growth for giant ellipticals in general.
We caution that our scheme describes the ensemble-averaged 
growth history for SMBHs and galaxies, and it is possible
that mergers are dominant in some individual systems. 

Fig.~\ref{fig:Mbh_vs_Mstar_z0} displays $\mbh$ and $\mstar$ values
for individual normal galaxies (cyan and red data points).
The $\mbh$ values are measured via reliable stellar dynamics or 
megamasers. 
These data points are mainly based on Tabs.~2 (elliptical galaxies) 
and 3 (disk galaxies) of \citet{kormendy13} where $\mbulge$
values are provided, and the unreliable $\mbh$ measurements
(the purple rows in these two tables) are discarded. 
For elliptical galaxies, $\mbulge$ equals $\mstar$; 
for disk galaxies, we divide $\mbulge$ by the $K_s$-band 
bulge-to-total luminosity ratio ($B/T$; also provided in 
\citealt{kormendy13}) to approximate $\mstar$. 
The data points also include the recent $\mbh$ measurements 
based on megamasers for three galaxies (Mrk~1029, J0437$+$2456, 
and UGC~6093; \hbox{\citealt{greene16}}). 
These data broadly agree with our $\mbh$-$\mstar$ relation:
the typical $\mbh/\mstar$ is low (one over several thousand) 
at $\log\mstar\lesssim10.5$ and becomes high (one over several 
hundred) at $\log\mstar\gtrsim 11$.
At $10.5\lesssim\log\mstar\lesssim11$, the observed $\mbh$
scatters around that expected from our $\mbh$-$\mstar$ relation,
although the scatter of $\mbh$ is large at a given $\mstar$.

Fig.~\ref{fig:Mbh_vs_Mstar_z0} also compares our $\mbh$-$\mstar$
relation with measurements for local BL AGNs based on 
single-epoch spectra or reverberation mapping 
\citep{reines15}.
At $\log\mstar \lesssim 10.5$, these measurements broadly agree
with our $\mbh$-$\mstar$ relation and the measurements 
for normal galaxies. 
However, at higher $\mstar$, the $\mbh$ for BL AGNs is 
systematically lower than the $\mbh$ for normal galaxies
at given $\mstar$.  
This discrepancy could result from large 
uncertainties in the $\mbh$ measurements for BL AGNs 
(\S\ref{sec:mbh_mstar}).
For example, these $\mbh$ measurements assume a single 
viral coefficient ``$f$'' for all sources, which might have 
systematic errors up to a factor of a few 
\citep[e.g.,][]{shen13}.
Another possibility is that the normal galaxies with 
$\mbh$ measurements may not be representative of the 
entire galaxy population due to selection bias. 
\citet{shankar16} proposed that observations tend to select 
massive SMBHs for which $\mbh$ measurements are achievable. 
They argued that this selection effect might strongly bias 
the observed $\mbh\text{--}\mstar$\ relation in the 
local universe, and they presented an intrinsic 
$\mbh\text{--}\mstar$\ relation (the green curve in 
Fig.~\ref{fig:Mbh_vs_Mstar_z0}).
However, our model $\mbh$\ does not suffer from this
selection bias but is generally higher than 
that of \citet{shankar16} at any given $\mstar$.
This discrepancy suggests that the selection bias 
might not be as strong as they proposed.

\begin{figure}
\begin{center}
\includegraphics[width=\linewidth]{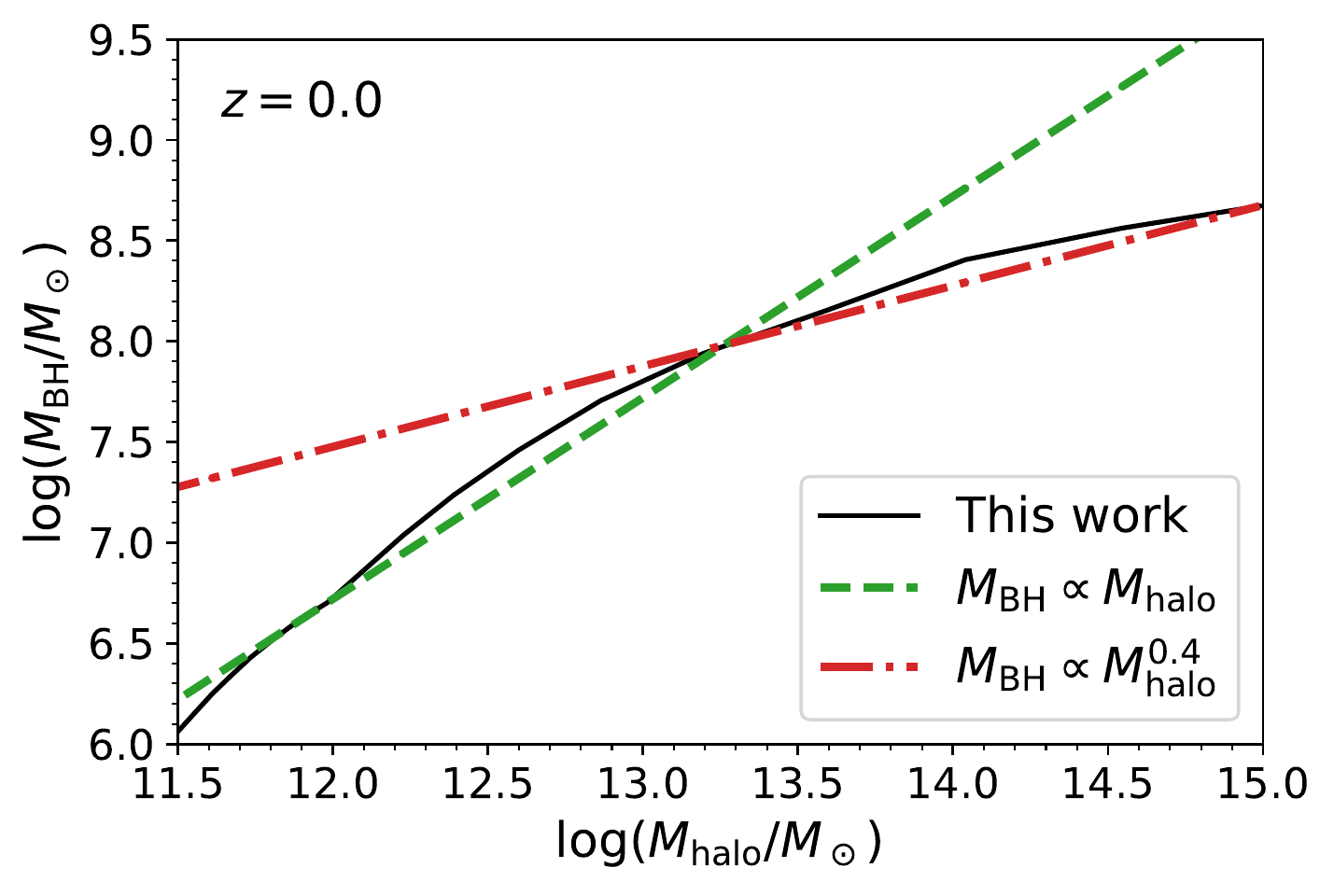}
\caption{The local $\mbh$-$\mhalo$ relation. 
The black solid curve indicates our best-fit model assuming
$\mbh/\mstar|_{z=4}=1/500$. 
The curve flattens toward high $\mhalo$.
The green dashed and red dash-dotted lines indicate 
$\mbh\propto\mhalo$ and $\mbh\propto\mhalo^{0.4}$, 
respectively; they are normalized at $\log\mhalo=12$ 
and 15, respectively.
These two lines are not from fitting but are just to
guide the eye.
}
\label{fig:Mbh_vs_Mhalo_z0}
\end{center}
\end{figure}

Fig.~\ref{fig:Mbh_vs_Mhalo_z0} shows the $\mbh$-$\mhalo$
relation converted from our $\mbh$-$\mstar$ relation 
(see \S\ref{sec:bhar_mstar} for the method). 
The relation is steep ($\mbh\propto\mhalo$; 
the green dashed line) at $\log\mhalo\lesssim 13$, 
but flattens at higher $\mhalo$ ($\mbh\propto\mhalo^{0.4}$; 
the red dash-dotted line). 
This behavior qualitatively agrees with 
\citet[][see their \S6.10.2]{kormendy13} who assumed 
$\mbulge\approx \mstar$.
The flattening is expected from the weak $\bharbar$ dependence
on $\mhalo$ in massive systems (see Fig.~\ref{fig:BHAR_vs_M}). 

\section{Summary and Future Prospects}\label{sec:summary}
We have derived $\bharbar$\ as a function of $\mstar$ and
redshift for galaxies with $\log\mstar=9.5\text{--}12$\ and 
$z=\text{0.4--4}$. 
Our method is based on survey data from \goodss, \goodsn,
and COSMOS, as well as the SMF and XLF from the literature.
Our main results are summarized below:
\begin{enumerate}

\item $\bharbar$ generally increases toward high 
$\mstar$ at a given redshift (see \S\ref{sec:bhar_for_all} 
and \ref{sec:bhar_mstar}). 
This $\mstar$ dependence becomes weaker toward the 
local universe. 
For example, from $\log\mstar=9.5$ to 
$\log\mstar=11.5$, $\bharbar$ increases by a factor
of $\sim 10^3$ at $z\approx4$ while the factor
is only $\sim 10$ at $z\approx0.5$.
This is probably due to the increasing 
fraction of quiescent galaxies at low redshift 
and high $\mstar$, as the $\bharbar$-$\mstar$ slope
does not become much shallower toward low redshift 
for star-forming galaxies (see \S\ref{sec:bhar_SFonly}).
At a given $\mstar$, $\bharbar$ is higher toward 
high redshift, and this redshift dependence is 
stronger for more massive systems, 
e.g., for $\log\mstar\approx 11.5$, $\bharbar$ is 
$\approx$ three decades higher at $z=4$ than at $z=0.5$.
This strong redshift evolution for massive galaxies 
is plausibly explained by AGN feedback. 

\item 
At a given $\mstar$, $\bharbar/\sfrbar$ has 
weak redshift evolution at $z\gtrsim 0.8$.
At a given redshift, $\bharbar/\sfrbar$ depends 
positively on $\mstar$. 
These features also exist when limiting the sample
to star-forming galaxies.
This positive dependence is inconsistent with the 
scenario that SMBH and galaxy growth are in lockstep.
We have converted the $\bharbar$-$\mstar$ relation 
to the $\bharbar$-$\mhalo$ relation based on the 
recipe of \citet{behroozi13}. 
$\bharbar$ positively depends on $\mhalo$ at 
$\log \mhalo \lesssim 12\text{--}13$
but becomes flat at higher $\mhalo$, indicating 
SMBH growth is not well traced by $\mhalo$ in 
massive systems.

\item Based on our $\bharbar(\mstar, z)$, $\mstar(z)$ 
from \citet{behroozi13}, and the assumption of 
$\mbh/\mstar|_{z=4}$, we have derived the $\mbh$-$\mstar$
relation at different redshifts (see \S\ref{sec:mbh_mstar}). 
At $z\lesssim 2$, our $\mbh$-$\mstar$ relation does  
not strongly depend on the $\mbh/\mstar|_{z=4}$ assumption.
The $\mbh/\mstar$ ratio is higher toward massive galaxies:
it rises from $\approx 1/5000$ at $\log\mstar\lesssim 10.5$ 
to $\approx 1/500$ at $\log\mstar \gtrsim 11.2$.
Our results suggest the $\mbh$-$\mstar$ relation is 
already largely in place at $z\approx 2$.
However, the $\mbh$ expected from our $\mbh$-$\mstar$ 
relation at $\log\mstar \lesssim 11$ is significantly 
lower than that from the observations of BL AGNs at
$z\gtrsim 0.5$. 
This discrepancy might be due to selection biases 
and/or inaccuracy of direct $\mbh$ measurements
of distant BL AGNs. 

\item Our predicted $\mbh$-$\mstar$ relation (at $z=0$) 
is similar to the local $\mbh$-$\mbulge$ relation at 
$\log\mstar\gtrsim 11$, suggesting that mergers do not 
dominate over accretion in the SMBH growth of giant 
ellipticals. 
At $\log\mstar \lesssim 10.5$, our $\mbh$-$\mstar$ 
relation broadly agrees with the observations of 
local normal galaxies and BL AGNs. 
At higher $\mstar$, our predicted $\mbh$ are broadly 
consistent with those of normal galaxies but are 
systematically higher than those of BL AGNs. 
We have also derived the $\mbh$-$\mhalo$ relation 
at $z=0$ using the $\mstar$-$\mhalo$ relation from
\citet{behroozi13}. 
Similar to the $\bharbar$-$\mhalo$ relation, 
$\mbh$-$\mhalo$ relation is steep ($\mbh\propto\mhalo$) 
at $\log\mhalo\lesssim 13$ but becomes relatively 
flat ($\mbh\propto\mhalo^{0.4}$) at higher $\mhalo$.


\end{enumerate}

In this work, we do not study SMBH growth at $z<0.4$
due to the limited sample size at low redshifts. 
Although SMBH accretion in the low-$z$ epoch should not affect 
the $\mbh$-$\mstar$ relation significantly (see 
\S\ref{sec:mbh_mstar}), it is critical for understanding 
how SMBH-galaxy coevolution develops from high redshifts 
to the local universe.
Future work could combine wide-field optical/NIR (e.g., SDSS,
UKIDSS, and 2MASS) and \xray\ (e.g., \swift/BAT, \chandra\ 
Source Catalog, and 3XMM) surveys to investigate the 
$\bharbar$-$\mstar$ relation at low redshift 
\citep[e.g.,][]{salim07, haggard10, shimizu15}. 
We also do not investigate SMBH growth for 
quiescent galaxies (see \S\ref{sec:bhar_quiescent});
future studies could derive their $\bharbar$-$\mstar$
relation.

The scatter of the observed local $\mbh$-$\mstar$ relation 
is large (Fig.~\ref{fig:Mbh_vs_Mstar_z0}). 
At least one origin of this scatter is related to $\mstar(z)$ 
(see Footnote~\ref{foot:Mstar_history} and Eq.~\ref{eq:mbh_z}).
We use average $\mstar(z)$ in this work while individual 
galaxies have different stellar mass histories. 
This aspect can be investigated based on the $\mstar(z)$ for
individual systems from numerical simulations such as 
Illustris \citep{genel14}.
Another origin of the large scatter may be that $\bharbar$ 
is dependent on other galaxy properties such as SFR, morphology, 
and cosmic environment in addition to $\mstar$ (see, e.g., 
\S\ref{sec:bhar_SFonly} and \citealt{martini09, 
kocevski12, aird17}). 
The detailed $\bharbar$ dependence on SFR and morphology can
be investigated using the CANDELS survey fields where 
deep \chandra, \herschel, and \hst\ observations are 
available.
The $\bharbar$ dependence on environment can be studied 
using surveys of several deg$^2$ such as COSMOS and 
\hbox{X-SERVS} (Chen et al.\ in prep.), where reliable 
environmental measurements can be performed 
\citep[e.g.,][]{darvish17}.

\section*{Acknowledgements}
We thank the referee (James Aird) for the constructive 
comments that improved the paper. 
We thank James Aird, Guillermo Barro, Peter Behroozi, 
Francesca Civano, Iary Davidzon, and Yoshihiro Ueda 
for providing relevant data.
We thank James Aird, Denis Burgarella, Andy Fabian, 
Andy Goulding, Jenny Greene, Clotilde Laigle, 
Jennifer Li, Amy Reines, Alexey Vikhlinin, and 
Feifan Zhu for helpful discussions. 
G.Y., W.N.B., F.V., and C.-T. J.C.\ acknowledge support
from Chandra X-ray Center grant GO4-15130A, the Penn
State ACIS Instrument Team Contract SV4-74018 (issued
by the Chandra \xray\ Center, which is operated by the
Smithsonian Astrophysical Observatory for and on behalf of
NASA under contract NAS8-03060), and the V.M. Willaman 
Endowment. 
B.L. acknowledges support from the National Natural Science 
Foundation of China grant 11673010 and the Ministry of 
Science and Technology of China grant 2016YFA0400702.
M.Y.S. and Y.Q.X. acknowledge the support from the 973 Program 
(2015CB857004), NSFC-11473026, NSFC-11421303, the CAS Frontier 
Science Key Research Program (QYZDJ-SSW-SLH006), and the 
Fundamental Research Funds for the Central Universities.
The Guaranteed Time Observations (GTO) for the \cdfn\ included 
here were selected by the ACIS Instrument Principal 
Investigator, Gordon P.\ Garmire, currently of the Huntingdon 
Institute for \xray\ Astronomy, LLC, which is under contract 
to the Smithsonian Astrophysical Observatory; Contract 
SV2-82024.
This project uses {\sc astropy} (a Python package; see 
\citealt{astropy}) and the SVO Filter Profile Service 
(http://svo2.cab.inta-csic.es/theory/fps/).




\bibliographystyle{mnras}
\bibliography{all.bib} 



\appendix

\section{Modeling of $P(\slx|\mstar, \lowercase{z})$}
\label{sec:mdl_select}
We describe $P(\slx|\mstar, z)$ at given $\mstar$\ and 
redshift with a smoothed double power law, and each 
parameter in this function 
($\log A$, $\log L_{\rm c}$, $\gamma_1$, and $\gamma_2$)
is modeled as a polynomial function of $\log \mstar$\ and 
$\log(1+z)$ (see \S\ref{sec:ml_fit}). 
To decide the orders of the polynomial functions,
we utilize the Akaike information criterion
\citep[AIC; e.g.,][]{burnham02}.
The AIC technique is applicable to any log-likelihood function, 
and it does not assume a specific distribution of parameter 
uncertainties.
For a given polynomial model, the AIC value is calculated
as $\mathrm{AIC}=-2\ln L_{\rm max}+ 2k$, where $k$\ is the number
of free model parameters and $\ln L_{\rm max}$\ is the maximum 
value of the likelihood function in Eq.~\ref{eq:L_final}.
$\ln L_{\rm max}$\ is calculated with the maximum-likelihood 
fitting method in \S\ref{sec:ml_fit}.
The AIC balances the complexity of the model and 
its efficiency in describing the data. 
In general, models with smaller AIC values are considered 
as more probable.

Assuming the currently favored model has an $\rm AIC$\ value 
of $\rm AIC_F$\ and a proposed model has an $\rm AIC$\ value 
of $\rm AIC_P$,
we accept the proposed model as the new favored model 
if $\Delta \rm AIC = AIC_P - AIC_F < \Delta AIC_{T}$. 
We choose the AIC threshold as $\rm \Delta AIC_{T}=-7$\
(e.g., Chapter~2.6 of \citealt{burnham02}).
This value corresponds to a $3\sigma$\ confidence 
level in the case when parameter uncertainties are 
Gaussian \citep[e.g.,][]{murtaugh14}. 

We first use a model of constant $\log A$, $\log L_\mathrm{c}$, 
$\gamma_1$, and $\gamma_2$ (i.e., 0th-order polynomials),
and obtain $\rm AIC_F=24998.7$. 
We then test the 1st-order polynomial models of $\log A$, i.e., 
\begin{equation}
\begin{split}\label{eq:poly1}
\centering
\log A = \log A_0 + \alpha_0^A \log(1+z) + \alpha_1^A \log M_{10}.
\end{split}
\end{equation}
The fitting of the 1st-order polynomial model produces
$\rm AIC_P=18164.2$, and thus the change of AIC is
$\Delta \rm AIC = AIC_P - AIC_F = -6834.5 < \Delta AIC_{T}$.
Therefore, we accept the 1st-order polynomial model of 
$\log A$, and with this model, we 
further test the 1st-order models of
$\log L_\mathrm{c}$, $\gamma_1$, and $\gamma_2$. 
We find that the 1st-order models of 
$\log L_\mathrm{c}$ and $\gamma_2$ significantly reduces 
the AIC value ($<-7$). 
Thus, $\gamma_1$ is consistent with a 
model that has no redshift or $\mstar$ dependence.

\begin{figure*}
\includegraphics[width=\linewidth]{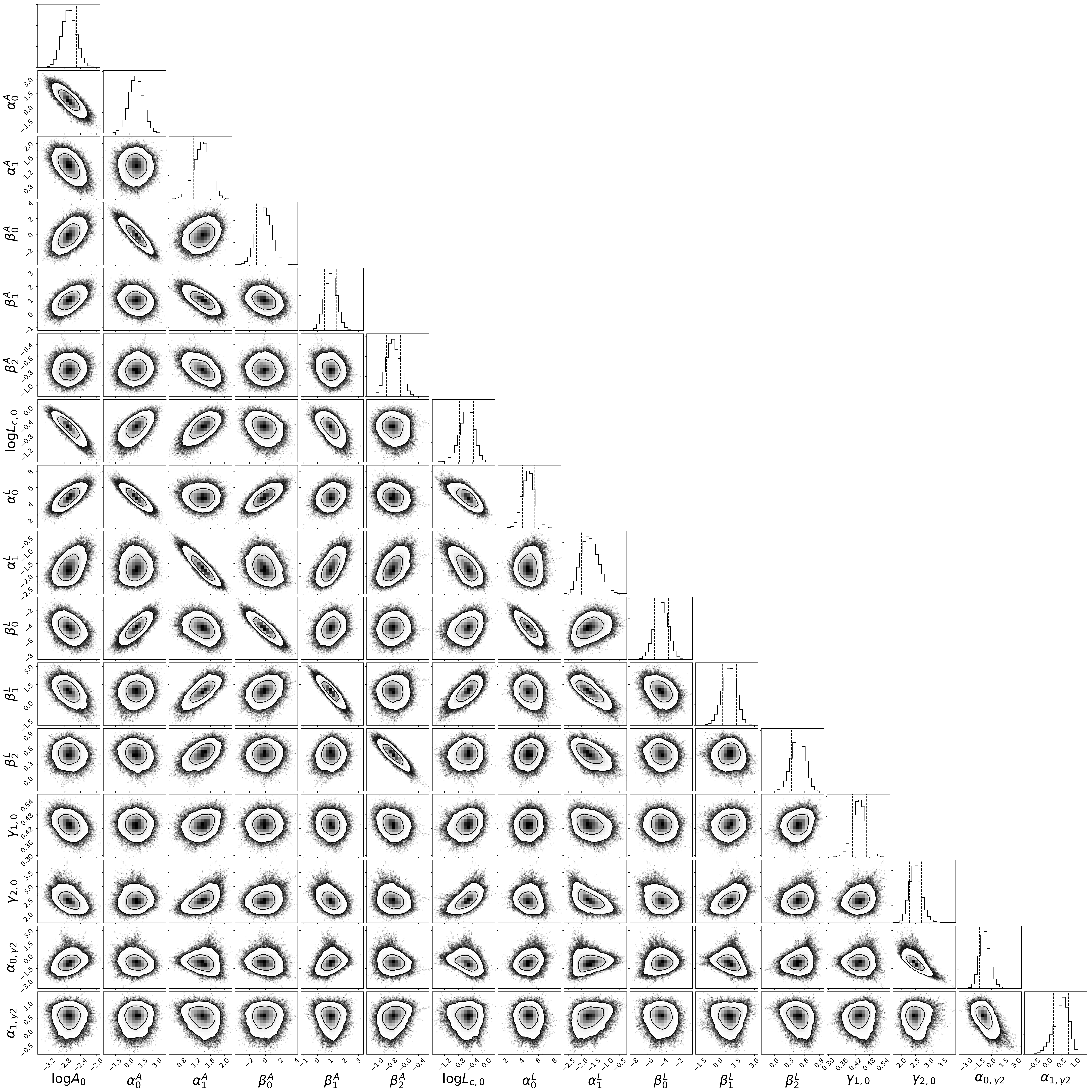}
\caption{2D contours and 1D histograms for our MCMC
sampling results.
The vertical dashed lines in the histograms 
indicate the $1\sigma$\ confidence intervals. 
The contour confidence levels are 1$\sigma$ (68.3\%) 
and 2$\sigma$ (95.5\%), respectively. 
The grayscale pixels inside the 1$\sigma$ contour 
denote probability with darken color indicating higher 
probability. 
The points outside the 2$\sigma$ contour represent 
individual MCMC sampling. 
This figure is plotted using {\sc cornor.py} 
\citep{foreman_mackey16}.
Some contours are highly irregular/tilted, indicating strongly 
correlated parametric uncertainties. 
}
\label{fig:mcmc_contour}
\end{figure*}

We further test higher-order polynomial models of $\log A$
$\log L_\mathrm{c}$, and $\gamma_2$.
The required polynomial orders for $\log A$, $\log L_\mathrm{c}$,
and $\gamma_2$ are 2nd, 2nd, and 1st, respectively. 
After determining the orders of the polynomials and obtaining 
the best-fit model parameters,
we estimate the parameter uncertainties utilizing Markov 
chain Monte Carlo (MCMC) sampling with {\sc emcee} 
\citep{foreman13}.\footnote{http://dan.iel.fm/emcee/current/}
The 2D contours and 1D histograms of the sampling results 
are shown in Fig.~\ref{fig:mcmc_contour}.
The $1\sigma$ confidence range of a parameter is calculated
as the 16\%--84\% percentile range of the corresponding 
1D histogram (shown as the vertical dashed lines in 
Fig.~\ref{fig:mcmc_contour}). 
We use the same method to derive uncertainties shown in 
Fig.~\ref{fig:BHAR_vs_M}.
The best-fit parameters and the MCMC uncertainties are 
presented in Tab.~\ref{tab:best_par}.
From Fig.~\ref{fig:mcmc_contour}, the contours can be highly
tilted and/or irregular (e.g., $\beta^L_{0}$ vs.\ $\alpha^L_0$),  
and thus the parametric uncertainties in Tab.~\ref{tab:best_par} 
can be strongly correlated. 

\section{$P(\slx|\mstar, \lowercase{z})$ 
Fitting Quality}\label{sec:fit_qual}
\begin{figure*}
\begin{center}
\includegraphics[width=\linewidth]{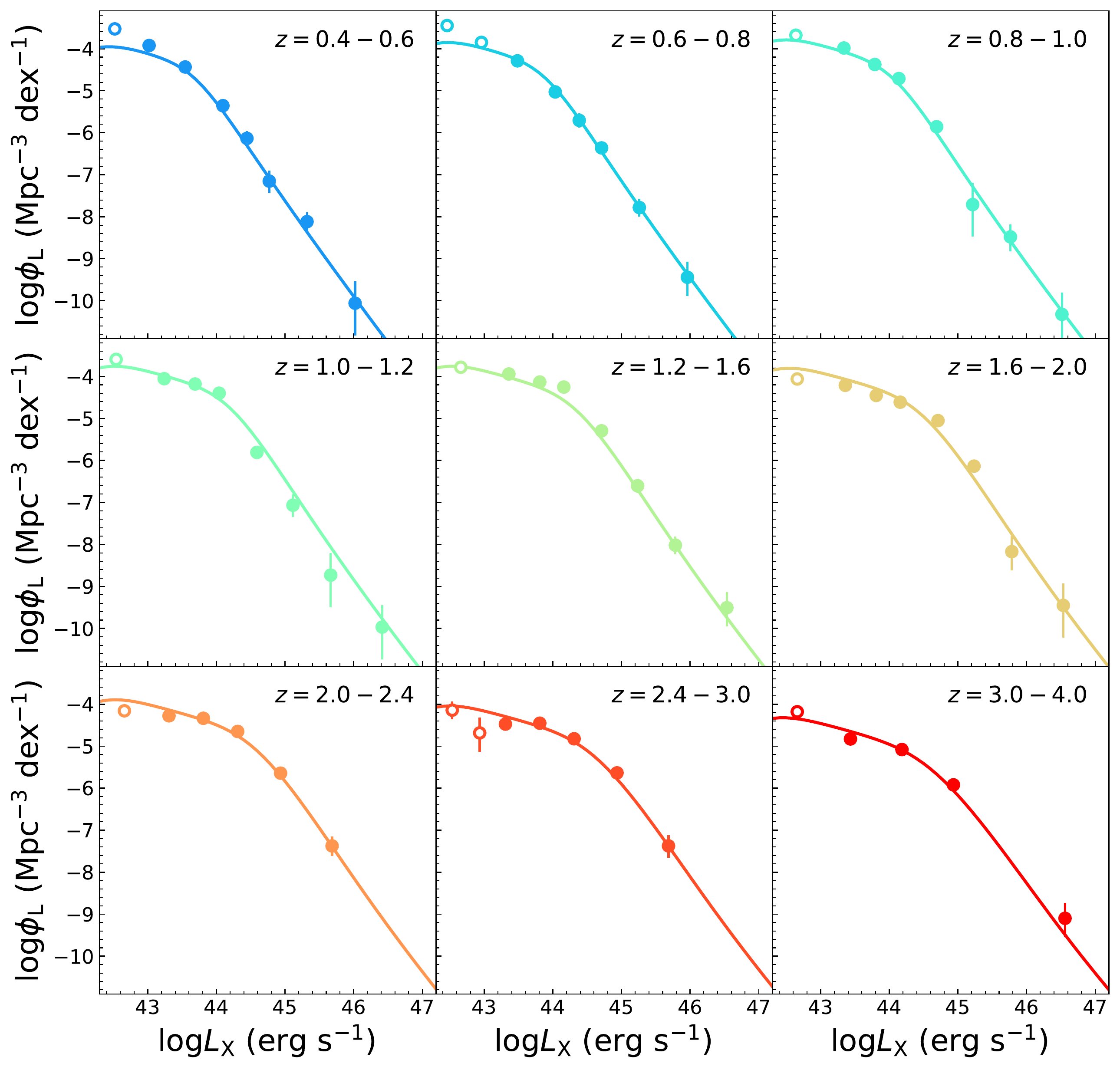}
\caption{The XLFs in redshift bins from $z=0.4\text{--}4$. 
The solid curves are the XLFs derived from our best-fit 
$P(\slx|\mstar,z)$\ model.
The data points indicate the soft-band XLFs from 
\citet{ueda14}.
The solid symbols represent the XLFs used in this work;
the open symbols represent the XLFs below our $\lx$ threshold
(see \S\ref{sec:smf_xlf}).
The XLFs from our model fit the observed data acceptably
in general. 
}
\label{fig:xlf_fit}
\end{center}
\end{figure*}

In Fig.~\ref{fig:xlf_fit}, we compare the observed XLFs
(\S\ref{sec:smf_xlf}) and the XLFs derived from our best-fit 
model (Eq.~\ref{eq:xlf}). 
The model XLFs generally agree well with the observed 
values. 
Fig.~\ref{fig:xlf_fit} also shows the observed 
XLFs that are not used to constrain $P(\slx|\mstar,z)$
($\log\lx<43$; open symbols). 
At $z\gtrsim 0.8$, our model XLFs are consistent with the 
observed values at $\log\lx<43$. 
At lower redshifts, our model underestimates the XLFs 
at $\log\lx \approx 42.5$.
This underestimation might be due to the fact that our 
model XLFs do not include low-$\slx$ AGNs and XRBs 
(see \ref{sec:smf_xlf}).

In Fig.~\ref{fig:P_Lsx_fit}, we compare the best-fit 
$P(\slx|\mstar,z)$\ with the binned values from our sample,
utilizing the $N_{\mathrm{obs}}/N_{\mathrm{mdl}}$\ method,
where $N_{\mathrm{obs}}$\ and $N_{\mathrm{mdl}}$\ are the
observed and model-expected numbers of sources in the 
bin, respectively \citep[e.g.,][]{aird12}.
Specifically, we derive the $N_{\mathrm{mdl}}$\ from the 
best-fit $P(\slx|\mstar,z)$\ using Eq.~\ref{eq:N_mdl}.
The binned $P(\slx|\mstar,z)$\ value is calculated 
as the model $P(\slx|\mstar,z)$\ value scaled by 
a factor of $N_{\mathrm{obs}}/N_{\mathrm{mdl}}$, 
while the corresponding uncertainty is the $1\sigma$
Poisson error of $N_{\mathrm{obs}}$\ scaled by the 
same factor.
In the bin of $z=0.4\text{--}0.7$ and 
$\log\mstar=11.5\text{--}12$ where no AGNs are 
detected, we derive the $1\sigma$ Poisson upper 
limits using the approach in \citet[][]{gehrels86}.
The model $P(\slx|\mstar,z)$\ is generally consistent 
with the observed values.
However, we note that at $z\lesssim 1$ and 
$\log\mstar \gtrsim 11.5$, the constraints on 
$P(\slx|\mstar,z)$ are weak. 
In this case, the $P(\slx|\mstar,z)$ is largely based
on model extrapolation and the uncertainties are 
likely underestimated.  

We also compare our best-fit $P(\slx|\mstar,z)$\ with those 
derived in some previous studies in 
Fig.~\ref{fig:P_Lsx_fit}.\footnote{We do not compare with the 
studies that investigate $\lam$ distributions 
\citep[e.g.,][]{hopkins09, weigel17}, since it is not 
feasible to convert their $\lam$ distribution to 
$P(\slx|\mstar,z)$.
}
\citet{aird12}, \citet{bongiorno12}, and \citet{wang17} 
adopted simple power-law models, and the slopes are (almost) 
independent of $\mstar$\ and redshift, similar to our results 
(see Tab.~\ref{tab:best_par}).
\citet{aird12} and \citet{wang17} probed 
$\log(\slx) \lesssim -0.5$; their slopes are $\approx 0.4$\
and $\approx 0.6$, respectively, similar to the low-$\slx$\ 
slope in our model.
Their normalizations are dependent on redshift but not $\mstar$,
while our normalization depends on both factors.
Therefore, their normalizations should be considered as 
averaged over the $\mstar$\ range of their samples. 
The relatively flat power-law shape at low-to-moderate $\slx$ 
is also confirmed by two recent studies \citep[][]{aird17, 
georgakakis17} that adopt non-parametric modeling for 
$P(\slx|\mstar,z)$.\footnote{The results of $P(\slx|\mstar,z)$
in \hbox{\citet{aird17}} and \hbox{\citet{georgakakis17}} are 
not publicly 
available. Therefore, their $P(\slx|\mstar,z)$ curves are not shown
in Fig.~\ref{fig:P_Lsx_fit}.}
These two studies also identified a sharp drop of $P(\slx|\mstar,z)$
above $\log\slx\approx0$, similar to our results. 
\citet{bongiorno12} obtained a steeper slope 
($\approx 1.0$); the reason could be that they probed 
$\slx$\ above the break $\slx$, where the slope is steep
($\approx 2.5$ in our model; see Tab.~\ref{tab:best_par}).
Thus, it is understandable that their slope is between our 
low-$\slx$\ and high-$\slx$\ slopes.
\citet{bongiorno16} modeled the bivariate distribution function 
of $\mstar$\ and redshift, i.e., $\Psi(\slx, \mstar, z)$, and 
we convert it to $P(\slx|\mstar,z)$\ by dividing it by 
our adopted SMF (\S\ref{sec:smf_xlf}).
The resulting $P(\slx|\mstar,z)$\ shape is also a double 
power law. 
Their low-$\slx$\ slope depends on redshift and is steeper
than our low-$\slx$\ slope at $z\lesssim 1.5$.
From our data, it appears \citet{bongiorno16}
overestimated the low-$\slx$\ slope. 
A possible reason for this difference might be that 
they did not include the ultradeep fields (\goodss\ and 
\goodsn) that are critical in constraining the low-$\slx$\ 
slope. 

\begin{figure*}
\begin{center}
\includegraphics[width=\linewidth]{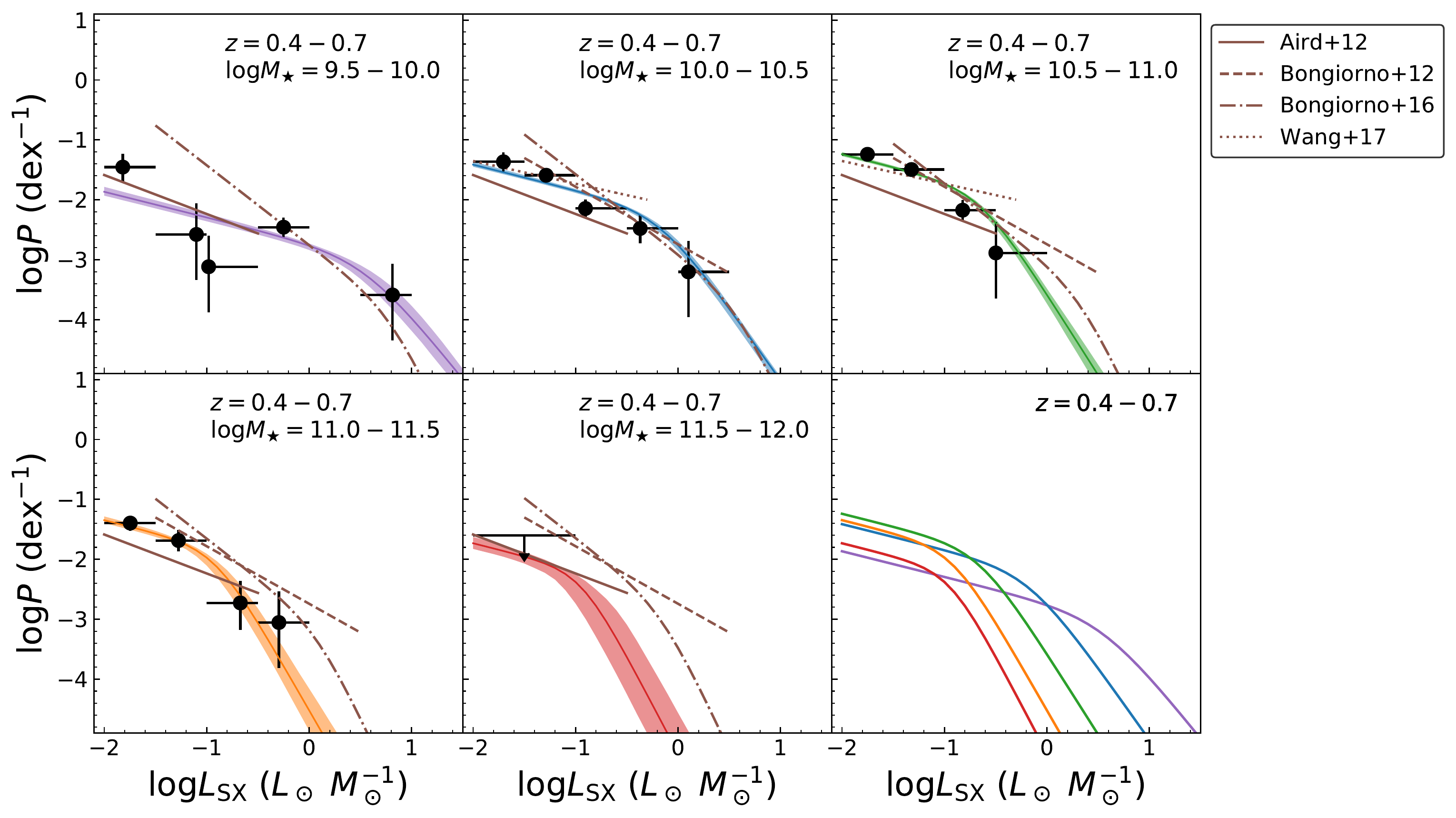}
\caption{$P(\slx|\mstar,z)$\ for different $\mstar$\ and 
redshift ranges.
The solid curves with blue, orange, green, red, and purple 
colors indicate our best-fit $P(\slx|\mstar,z)$\ model, 
and the shaded regions indicate the $1\sigma$\ confidence intervals.
For each of the seven redshift ranges, the first five panels 
display the probability distributions for five $\mstar$ ranges;
the last panel compares our best-fit $P(\slx|\mstar,z)$ for 
these $\mstar$ ranges.
The data points indicate the binned $P(\slx|\mstar,z)$\ values 
from our sample.
The error bars (upper limit) indicate the Poisson 1$\sigma$ 
confidence level.
The brown curves represent $P(\slx|\mstar,z)$\ from the 
literature. 
Our best-fit model is in agreement with the data. 
The brown lines indicate $P(\slx|\mstar,z)$\ from different 
studies.
Here, we only present the figures for $z=0.4\text{--}0.7$ and
$z=1.0\text{--}1.5$ for display 
purposes; the entire set of 6 figures for the other redshift ranges
is available in the online version of the journal.
}
\label{fig:P_Lsx_fit}
\end{center}
\end{figure*}

\begin{figure*}
\begin{center}
\includegraphics[width=\linewidth]{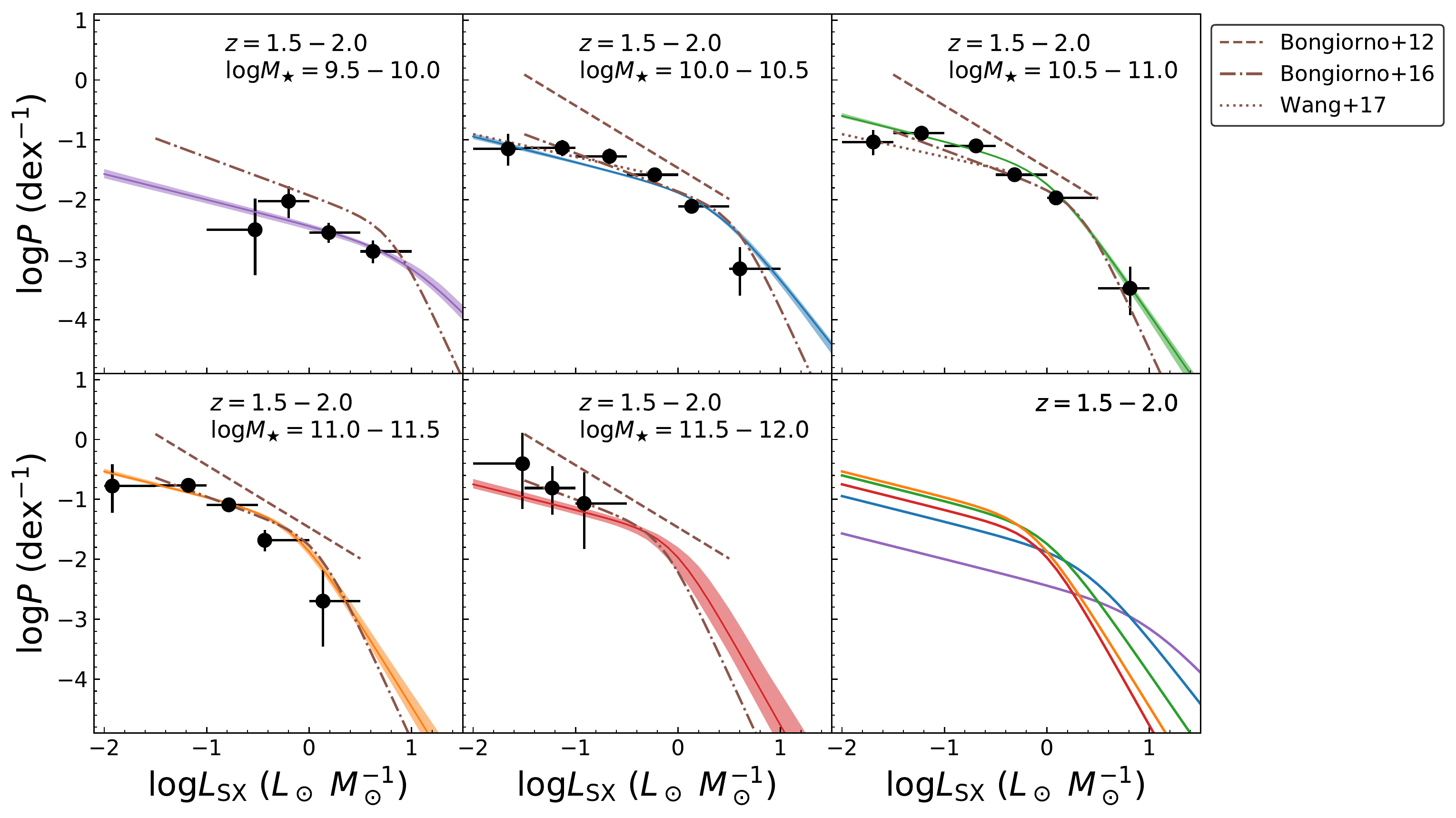}
\contcaption{}
\end{center}
\end{figure*}

%
%


\bsp	
\label{lastpage}
\end{document}